\documentclass[twocolumn,aps,prd,showpacs,nofootinbib,superscriptaddress,preprintnumbers,floatfix,10pt]{revtex4-1}

\usepackage[english]{babel} 
\usepackage[utf8]{inputenc}
\usepackage[T1]{fontenc}
\usepackage{lmodern}

\usepackage{amsmath}
\usepackage{amsfonts}
\usepackage{amssymb}
\usepackage{slashed}
\usepackage{bbm}

\usepackage{graphicx}
\usepackage{color}
\usepackage{multirow}

\usepackage{hyperref}

\newcommand{\I}{\mathrm{i}}

\newcommand{\tr}{\mathrm{tr}}

\newcommand{\im}{\mathrm{Im}}
\newcommand{\re}{\mathrm{Re}}

\newcommand{\rmA}{\mathrm{A}}

\newcommand{\SUN}{\mathrm{SU}(N)}
\newcommand{\SU}{\mathrm{SU}}
\newcommand{\U}{\mathrm{U}}
\newcommand{\rmO}{\mathrm{O}}

\newcommand{\rmT}{\mathrm{T}}
\newcommand{\Sp}{\mathrm{Sp}}
\newcommand{\SO}{\mathrm{SO}}
\newcommand{\Ztwo}{\mathbb{Z}_{2}}

\newcommand{\rms}{\mathrm{s}}
\newcommand{\rmf}{\mathrm{f}}
\newcommand{\rmh}{\mathrm{h}}
\newcommand{\rma}{\mathrm{a}}
\newcommand{\rmsym}{\mathrm{2s}}
\newcommand{\rmasym}{\mathrm{2a}}

\newcommand{\mh}{m_{\mathrm{h}}}

\newcommand{\MA}{M_{\mathrm{A}}}

\newcommand{\phid}{\phi^{\dagger}}

\begin{document}

\title{Analytical relations for the bound state spectrum of gauge theories with a Brout-Englert-Higgs mechanism}

\author{Ren\'{e} Sondenheimer}
\email{rene.sondenheimer@uni-graz.at}
\affiliation{Institute of Physics, NAWI Graz, University of Graz, Universit\"atsplatz 5, A-8010 Graz, Austria}

\begin{abstract}
We apply the method proposed by Fr\"ohlich, Morchio, and Strocchi to analyze the bound state spectrum of various gauge theories with a Brout-Englert-Higgs mechanism. These serve as building blocks for theories beyond the standard model but also stress the exceptional role of the standard model weak group. We will show how the Fr\"ohlich-Morchio-Strocchi mechanism relates gauge-invariant bound state operators to invariant objects of the remaining unbroken gauge group after gauge fixing. In particular, this provides a strict gauge-invariant formulation of the latter in terms of the original gauge symmetry without using the terminology of spontaneous gauge symmetry breaking. 
We also demonstrate that the Fr\"ohlich-Morchio-Strocchi approach allows us to put new constraints on theories beyond the standard model by pure field-theoretical considerations. Particularly the conventional construction of grand unified theories has to be rethought.
\end{abstract}

\pacs{}
\maketitle

\section{Introduction}
Within the theoretical formulation of the electroweak sector, the Higgs field is a key ingredient as the Brout-Englert-Higgs (BEH) mechanism allows us to introduce a mass generating mechanism for the elementary gauge fields and fermions without spoiling the basic gauge invariance \cite{Englert:1964et,Higgs:1964ia,Higgs:1964pj,Guralnik:1964eu,Higgs:1966ev}. This mechanism also plays an important role in various beyond the standard model (BSM) theories which exhibit an extended gauge and Higgs sector.

In order to compute the mass spectrum of a theory with BEH mechanism, the scalar field is usually split into its vacuum expectation value (VEV) as well as a fluctuating field in case the Higgs potential has a nontrivial minimum. By inserting this split into the Lagrangian, it can be analyzed which fields obtain a nonvanishing mass parameter. However, this convenient picture of BEH physics comes with a grain of salt on a field-theoretical level as a number of subtle issues arise and have to be addressed for a solid definition of such theories. Moreover, a puzzling situation occurs in BSM scenarios.

Recent lattice simulations demonstrated that the usual framework of treating the observable particle spectrum of a theory with BEH mechanism has to be questioned. By calculating the particle spectrum in an $\SU(3)$ gauge theory coupled to a scalar field in the fundamental representation, a mismatch was discovered between the lattice results and the naive expectation of identifying $\SU(3)$ gauge-variant objects as observable quantities after spontaneous symmetry breaking \cite{Maas:2016ngo,Maas:2018xxu,Maas:2017pcw}. 
The standard approach would predict for the elementary degrees of freedom from which observable quantities can be build a massive Higgs, five massive vector bosons, as well as three gauge bosons without a mass term belonging to the unbroken $\SU(2)$ subgroup. While a scalar particle was identified in the lattice formulation with the same mass as one would expect from the notion of spontaneous gauge symmetry breaking for the Higgs, a qualitative different picture manifests for more sophisticated scalar states as well as in the vector channel. Albeit the investigated properties of some of the vector particles coincided with those properties predicted by the conventional analysis, states one would naively expect in the standard gauge-fixed formulation have not been seen. Currently it is under investigation whether additional scalar bound state particles are present in the spectrum as expected from the standard gauge-fixed treatment of the theory \cite{Maas:2018xxu}. By contrast, lattice simulations of an $\SU(2)$ theory precisely confirm the existence of a single massive Higgs boson as well as three massive vector bosons and thus substantiate the particle content of the weak-Higgs subsector of the standard model beyond perturbation theory \cite{Maas:2013aia,Maas:2014pba,Shrock:1985ur,Shrock:1985un}.

Within the conventional approach, the BRST symmetry of the gauge-fixed action is used to define the physical state space such that the elementary fields of the Lagrangian can be considered to make adequate predictions for experiments and can be identified with physical particles in the standard model. However, the assumption that the BRST mechanism takes sufficient care of gauge invariance fails beyond perturbation theory due to the Gribov-Singer ambiguity \cite{Gribov:1977wm,Singer:1978dk,Fujikawa:1982ss,DellAntonio:1991mms,vanBaal:1991zw,vanBaal:1997gu,Vandersickel:2012tz}. The BRST symmetry might be recovered by extended nonperturbative gauges at least in pure Yang-Mills theories but then all quantities carrying a gauge index, e.g., the elementary gauge bosons, are absent from the physical spectrum, see \cite{Maas:2011se} for a review. This is clearly in contrast to the perturbative description of the electroweak sector.

These statements might be mitigated in case a BEH mechanism provides mass terms for either all gauge bosons, i.e., the gauge group is fully broken in the perturbative language, or the stability group consists only of an Abelian subgroup. In these cases, it turns out that the Gribov horizons cannot be reached for sufficiently large gauge boson masses as the zero-point fluctuations get suppressed. Thus for infinitely large gauge boson masses, the Gribov-Singer ambiguity has no quantitative impact on gauge-dependent correlation functions for these type of theories \cite{Lenz:1994tb,Lenz:2000zt,Maas:2010nc,Capri:2012ah,Capri:2013oja,Capri:2013gha}. Nonetheless, it still affects the spectrum on a qualitative level because the structure of the BRST symmetry is unchanged as it depends only on the group structure and not the internal dynamics. In particular, it is expected that some remnant of the Gribov-Singer ambiguity survives for theories with a remaining non-Abelian gauge symmetry, like grand unified theories (GUTs), albeit the details have not been explored yet.

A consequence of this is also the fact that all attempts to define an observable gauge charge in a non-Abelian gauge theory failed so far. While it seems to be possible to construct an ’observable’ color charge to any order in perturbation theory via suitable dressing functions similar to the Abelian case, there is an obstruction in the nonperturbative regime due to the Gribov-Singer ambiguity \cite{Lavelle:1995ty,Ilderton:2007qy,Heinzl:2008bu}. This underlines that the physical state space of a non-Abelian gauge theory is further restricted once the nonperturbative regime is taken into account.

To bedevil the situation further, also the VEV of the Higgs field is not a reliable order parameter as both the actual direction as well as the modulus depend on the gauge. Furthermore, gauges can be constructed such that the VEV vanishes identically even if the scalar potential has nontrivial minima \cite{Maas:2012ct}, implying that the gauge bosons would be massless to any order in a perturbative description. In particular, the Higgs VEV vanishes in case no gauge fixing is performed which is generally true for any gauge-dependent object \cite{Frohlich:1981yi}. Its absence in a gauge-invariant formulation is put on solid ground by Elitzur’s theorem which proves the impossibility of local gauge symmetry breaking \cite{Elitzur:1975im}. Some of the philosophical consequences can be found in \cite{Lyre:2008af,Friederich:2011xs,Francois:2017akk}.

For the weak subsector of the standard model, it was shown by Osterwalder and Seiler as well as Fradkin and Shenker that no gauge-invariant order parameter for the BEH effect exists \cite{Osterwalder:1977pc,Fradkin:1978dv}. Thus, QCD-like physics and BEH physics are qualitatively indistinguishable and the occurrence of the different phases in a gauge-fixed setup is a pure gauge artifact. In addition, it was shown that no unique transition line between the two phases exists in case an order parameter is constructed from a residual global symmetry group after gauge fixing. In principle such residual global symmetries can be broken. However, different gauge choices can have different remaining global symmetries. Therefore, this strategy is not able to define a unique transition which physically distinguishes between the two phases as this depends on the different global symmetries \cite{Caudy:2007sf,Greensite:2008ss}. Suggestions to circumvent this fact involve nonlocal constructions \cite{Greensite:2017ajx,Greensite:2018ebg,Greensite:2018mhh,Matsuyama:2019lei}.

The proof of the Osterwalder-Seiler--Fradkin-Shenker argument only works for theories in which all elementary fields receive a mass within the conventional picture as for the $\SU(2)$ case with one fundamental scalar field. For instance, in gauge theories with an adjoint Higgs field, different phases can be distinguished by the invariant Casimir operators of the gauge group \cite{Olynyk:1984pz,Baier:1986ni}. However, the VEV of the elementary Higgs field is still a gauge-dependent object and not a well-suited quantity to analyze the gauge-invariant properties of the theory. These examples demonstrate in various ways that a proper definition which treats all these subtle field theoretical issues in an appropriate way is rather sophisticated and a rethinking of the observables in terms of gauge-charge singlets appears to be mandatory \cite{Frohlich:1981yi,Frohlich:1980gj}. This statement is supported by several lattice calculations in different models \cite{Lee:1985yi,Maas:2012tj,Maas:2013aia,Maas:2016ngo}, see also \cite{Maas:2017wzi} for a detailed review.

In order to put the successful phenomenological description of the electroweak standard model mathematically on solid ground, various attempts have been performed. These include a reformulation of the bosonic subsector of the Glashow-Weinberg-Salam model by a change of variables such that the action depends only on $\SU(2)$ gauge-invariant fields by factoring out the local $\SU(2)$ symmetry \cite{Chernodub:2008rz,Masson:2010vx,Ilderton:2010tf}, as well as an investigation of the equations of motion of gauge-invariant degrees of freedom obtained by nonlinear field redefinitions \cite{Farhi:1995pt}. In addition, a gauge-invariant description was proposed by Fr\"ohlich, Morchio, and Strocchi (FMS) which investigates gauge-invariant bound state operators in analogy to QCD \cite{Frohlich:1981yi,Frohlich:1980gj}. While the former approaches directly rely on the particular structure of the $\SU(2)$ group, the latter can easily be generalized to other BEH theories as they appear in many BSM contexts. This was demonstrated for $\SUN$ gauge theories with a Higgs field in the fundamental or adjoint representation \cite{Maas:2017xzh}. In addition, the FMS approach explains why the lattice spectroscopy confirms the perturbative prediction of the particle content in the standard model but yields different results in a gauge theory with larger gauge group.

Generally, the FMS approach provides a convenient tool to define the spectrum of a theory with BEH mechanism in a gauge-invariant manner and reveals how the properties of the strictly gauge-invariant bound state operators can be extracted from the correlators of gauge-variant objects. We will use this powerful tool in the following to predict the nonperturbative bound state spectrum in the $0^{+}$ and $1^{-}$ channel for various representations of the Higgs field for $\SO(N)$ and $\SU(N)$ gauge theories and identify those states that can be described according to the usual lines of the BEH framework.

This paper is organized as follows: 
In Sec.~\ref{sec:FMS}, we discuss some of the generic properties of the FMS approach for an arbitrary BEH model and provide a brief introduction into its basics. In particular, we will systematically elaborate the precise duality relation between different gauge theories connected via the BEH mechanism for the first time. As this section might appear rather abstract, we recommend to read it together with Sec.~\ref{sec:ONfun}. In Sec.~\ref{sec:ON}, we construct gauge-invariant bound state operators for $\SO(N)$ gauge theories with scalar fields in the fundamental (\ref{sec:ONfun}) as well as irreducible second-rank tensor representations (\ref{sec:ON2anti} and \ref{sec:ON2sym}). In Sec.~\ref{sec:SUN}, we use the FMS formalism to discuss the gauge-invariant spectrum of $\SUN$ gauge theories. After a brief preparatory work in Sec.~\ref{sec:SUNpre}, we study in Secs.~\ref{sec:SUN2sym} and \ref{sec:SUN2asym} the symmetric and antisymmetric second-rank tensor representation, respectively. Finally, we outline the implications of the FMS formulation for general GUT-like structures in Section~\ref{sec:GUT}. A summary can be found in Sec.~\ref{sec:Conclusions}.

\section{General properties of the FMS mechanism}
\label{sec:FMS}

The seminal work of Fr\"ohlich, Morchio, and Strocchi illustrates a convenient way to compute observables in the electroweak sector of the standard model in a strict gauge-invariant manner \cite{Frohlich:1980gj,Frohlich:1981yi}. A generalization to other gauge groups and representations is straightforward and the analysis of Ref.~\cite{Maas:2017xzh} has shown that the computation of the gauge-invariant bound state spectrum via the FMS approach sheds a new light on the particle content of BSM models with a BEH mechanism. In order to obtain a better understanding of the underlying mechanism and under which circumstances the conventional gauge-fixed treatment of the particle spectrum is adequate for such theories, we systematically extend this analysis.

In general, we consider a theory with gauge group $\mathcal{G}$ that breaks to a subgroup $\mathcal{H}$ due to the BEH mechanism in the conventional picture. However, from a conceptual point of view only those Green's functions can be nonvanishing which are invariant under local gauge transformations of the original gauge group $\mathcal{G}$ and it is impossible to have spontaneous gauge symmetry breaking \cite{Elitzur:1975im}. The symmetry breaking is merely imposed by gauge fixing, implying that gauge-dependent correlation functions become nonvanishing \cite{Frohlich:1980gj,Frohlich:1981yi}.

The trivial fact that a physical observable has to be gauge-invariant by definition implies that any observable has to be formulated in a $\mathcal{G}$-invariant way, i.e., it has to transform as a singlet under transformations of the gauge group $\mathcal{G}$. Nevertheless, under certain conditions it turns out that it is very convenient to compute properties of $\mathcal{H}$-invariant objects, i.e., to investigate properties of operators in the effective theory after gauge symmetry breaking due to gauge fixing, as they can be directly related to $\mathcal{G}$-invariant states. This duality is formally explained by the FMS mechanism.

\subsection{The FMS formalism}
The first task of the FMS procedure is the following, formulate an operator which is strictly invariant with respect to gauge transformations of $\mathcal{G}$ albeit the gauge fixed action is only invariant under $\mathcal{H}$. These operators are necessarily composite. The only trivial exception is realized in case the action already contains a field transforming as a $\mathcal{G}$-singlet. Further, we consider in the following only local (or almost local) operators that allow for a simple interpretation as potential bound state operators. Of course, it is also feasible to define other gauge invariant objects such as connecting two elementary fields at different spacetime points via a Wilson line. However, these are inherently nonlocal and thus do not allow for a straightforward interpretation as common particles, i.e., we do not expect that these type of gauge-invariant operators are able to describe localized states. Nonetheless, such operators have to be considered to gain a comprehensive picture of the investigated model but this is clearly beyond the scope of this work.

The second step is to choose a gauge in which the scalar field acquires a nonvanishing VEV which minimizes the scalar potential.  Then it is convenient to split the scalar field into its VEV and a fluctuation part and to analyze the properties of the resulting fields that are now described by an effective low energy theory being only invariant under $\mathcal{H}$ transformations as usual. Further, we are able to expand the $\mathcal{G}$-invariant operators associated to physical observables of the system which provides a mapping to $\mathcal{H}$-invariant objects. Although, each $\mathcal{H}$-invariant operator is gauge dependent with respect to the original gauge group $\mathcal{G}$, the gauge-invariant information of the $\mathcal{G}$-invariant operator is encoded at least in the sum of the $\mathcal{H}$-invariant terms of the expansion by construction. Thus, we may take either a top-down or bottom-up viewpoint. From the top-down perspective, we simply investigate via this FMS mapping as to whether $\mathcal{G}$-invariant objects can be described by possibly simpler $\mathcal{H}$-invariant ones. The bottom-up viewpoint allows to answer the question if a physical system described by an effective gauge theory $\mathcal{H}$ can be embedded into a $\mathcal{G}$-invariant description with the aid of the BEH mechanism.

Let us formulate these statements mathematically more precise by considering the classical action of a non-Abelian gauge theory with field strength tensor $F_{\mu\nu}$ coupled to a single scalar field $\phi$ in some representation of the gauge group $\mathcal{G}$, 
\begin{align*}
 \!\!S = \int_{x} \left[ -\frac{1}{2} \tr (F_{\mu\nu}F^{\mu\nu}) + (D_{\mu}\phi)^{* \bar{\mathbf{a}}}(D^{\mu}\phi)^{\mathbf{a}} - V(\phi) \right]\!,
\end{align*}
where $\mathbf{a}$ is a multi-index encoding all possible indices characterizing the given representation and $D_{\mu}$ is the covariant derivative. 
Note, that we use bared notation to indicate complex conjugate representations if necessary.

In principle, the potential $V$ can consist of any gauge-invariant term build from the scalar field but we restrict the analysis to those operators which are renormalizable by power-counting. 
Additionally, we always assume that the scalar potential has one or several nontrivial minima such that the scalar field acquires a nonvanishing VEV $\langle \phi \rangle$ in a suitable chosen gauge and perform the following split 
\begin{align}
 \phi(x) = \langle \phi \rangle  + \varphi(x).
 \label{eq:split}
\end{align}
Further, we use the notation $\langle \phi \rangle = v \phi_{0}$ for real-valued scalar fields and  $\langle \phi \rangle = \frac{v}{\sqrt{2}}\phi_{0}$ if $\phi$ is complex. $\phi_{0}$ characterizes the normalized direction of the VEV in gauge space, i.e., defines the breaking pattern, and $v$ is its modulus setting the symmetry breaking scale. 
Throughout this paper, we analyze the mass spectra for the elementary fields appearing in the action as well as for the bound states on a classical level and do not include higher-order quantum corrections. As long as also the quantum effective potential obeys the same properties (nonvanishing VEV for the scalar field with same breaking pattern), the results will vary only on a quantitative but not on a qualitative level.

At the beginning of each subsection, we will sketch the textbook computation for the mass matrices of the gauge boson and scalar fields in a fixed gauge with nonvanishing VEV. 
The simplest way to extract the mass parameters of the involved elementary fields at tree level is to perform the calculations in the unitary gauge where the would-be Goldstone modes are removed from the spectrum. Of course, analogous calculations can be done in other gauges with a nontrivial VEV as well. For instance, the results can be translated to $R_{\xi}$ gauges in a straightforward manner. The analysis on the level of the elementary fields also shows the decomposition of the fields assigned to $\mathcal{G}$ multiplets into multiplets of the remaining unbroken subgroup $\mathcal{H} \subset \mathcal{G}$. This decomposition can be formulated in a gauge-covariant but obviously not in a gauge-invariant way with respect to $\mathcal{G}$ with the aid of $\phi_{0}$.

As a simple example consider an $\SUN$ gauge theory with a scalar field $\phi$ in the fundamental representation. If the scalar field acquires a nonvanishing VEV, we have the breaking $\mathcal{G} = \SUN \to \mathcal{H} = \SU(N-1)$. 
The gauge field $A^{\mu}$ can be decomposed into an $\mathcal{H}$-singlet, $ \phid_{0}A^{\mu}\phi_{0}^{\phantom{\dagger}} \equiv A_{\rms}^{\mu} $, a field transforming as a complex fundamental vector of $\mathcal{H}$, $A^{\mu}\phi_{0}-A_{\rms}^{\mu}\phi_{0} \equiv A_{\rmf}^{\mu} $, and $(\mathbbm{1}-\phi_{0}^{\phantom{\dagger}}\phid_{0})A^{\mu}(\mathbbm{1}-\phi_{0}^{\phantom{\dagger}}\phid_{0}) \equiv A_{\rma}^{\mu}$ being the massless gauge field of the remaining unbroken gauge group $\mathcal{H}$. 
The subscripts $\rms$, $\rmf$, $\rma$ of the $\mathcal{H}$ multiplets indicate that the fields are in the singlet, fundamental, and adjoint representation of $\mathcal{H}$.%
\footnote{Note that the inhomogeneous part of a gauge transformation restricted to the $\mathcal{H}$ subgroup drops out for $A_{\rms}^{\mu}$ and $A_{\rmf}^{\mu}$ as such a transformation leaves $\phi_{0}$ invariant. Thus, these fields transform indeed as spin-one matter fields being in the singlet and fundamental representation of $\mathcal{H}$.}
Later, we will also use $\rmsym$ and $\rmasym$ to indicate the second-rank symmetric and antisymmetric tensor representations, respectively. Similar $\mathcal{G}$-covariant decompositions into multiplets of $\mathcal{H}$ can be constructed for other representations, gauge groups, and the scalar sector as well. 
A particularity of the decomposition of the scalar field $\phi$ into $\mathcal{H}$ multiplets is given by the fact that it always contains a singlet with respect to $\mathcal{H}$, irrespective of the original gauge group $\mathcal{G}$ or the representation. This singlet is given by the fluctuating field proportional to the direction of the VEV ${\sim}\re(\phi_{0}^{*\bar{\mathbf{a}}}\phi^{\mathbf{a}})$ which we will also call radial Higgs excitation throughout this paper.

In order to obtain gauge-invariant objects, a dressing as in Abelian gauge theories is not possible due to the Gribov-Singer ambiguity \cite{Lavelle:1995ty,Ilderton:2007qy,Heinzl:2008bu}. As long as no generalization of such a dressing is known, non of the elementary degrees of freedom are observable particles in a non-Abelian gauge theory. This might be obvious for the fields transforming under a nontrivial representation of $\mathcal{H}$. If $\mathcal{H}$ is non-Abelian, the gauge coupling associated to this subgroup can become large in the infrared (IR), depending on the precise field content in the gauge fixed theory. Thus, this subsector would develop a behavior similar to QCD where the quarks are replaced by bosonic degrees of freedom which become confined.

Although they are often treated in the literature as they would, also the fields being singlets with respect to the remaining gauge group $\mathcal{H}$ after symmetry breaking cannot be part of the physical state space due to the Gribov problem even though they are singlets of the perturbative BRST transformations. Strictly speaking, the same is true for the embedding of $\mathcal{H}$-invariant bound states in the context of the original gauge structure $\mathcal{G}$. Once more, a physical observable has to be formulated in a strict $\mathcal{G}$-invariant way. The term spontaneous gauge symmetry breaking is merely a figure of speech, though a quite convenient one for the standard model due to its special group theoretical structure \cite{Frohlich:1980gj,Frohlich:1981yi} but not necessarily for BSM models \cite{Maas:2016ngo,Maas:2017xzh,Maas:2018xxu,Maas:2015gma}. For this reason, we will be as conservative as possible and do formally not interpret $\mathcal{H}$-invariant objects as observables of the actual $\mathcal{G}$-gauge theory. Nevertheless, some of the $\mathcal{H}$-invariant objects can be extracted from a strict $\mathcal{G}$-invariant counter part via the FMS mechanism such that a duality between the states of the different theories can be established.

Following the FMS strategy, we will analyze the spectrum of a theory with BEH mechanism in terms of its bound states. Any gauge-invariant object can only be classified according to the global symmetries of the theory. Besides spin and parity, internal global symmetries can be used to characterize the different channels. We will restrict this discussion and the construction of bound state operators for the various theories investigated in the following to the scalar ($0^{+}$) and vector ($1^{-}$) channel but further subdivide these two channels into multiplets of possible internal global symmetries if they exist.

A simple example for the FMS approach can be formulated in the scalar channel. We can always build a gauge-invariant composite scalar operator given by $O(x) = \phi^{*\bar{\mathbf{a}}}(x)\phi^{\mathbf{a}}(x)$. Choosing a gauge with nonvanishing VEV and splitting the elementary field operators according to Eq.~\eqref{eq:split}, we obtain the FMS expansion for this scalar operator,
\begin{align}
 O = \phi^{* \bar{\mathbf{a}}}\phi^{\mathbf{a}} 
 = \langle \phi^{* \bar{\mathbf{a}}} \rangle \langle \phi^{\mathbf{a}} \rangle 
 + 2\re \big( \langle \phi^{* \bar{\mathbf{a}}} \rangle \varphi^\mathbf{a} \big) + \varphi^{* \bar{\mathbf{a}}}\varphi^{\mathbf{a}}. 
\label{eq:FMSexpansion}
\end{align}
Ignoring the unimportant constant term $\langle \phi^{* \bar{\mathbf{a}}} \rangle \langle \phi^{\mathbf{a}} \rangle$, 
the leading order term in the fluctuating field is given by the radial Higgs excitation $\re \big( \langle \phi^{* \bar{\mathbf{a}}} \rangle \varphi^\mathbf{a} \big)\sim \re(\phi_{0}^{*\bar{\mathbf{a}}}\phi^{\mathbf{a}})$. At next-to-leading order, we obtain an operator that will generate a trivial scattering state at the elastic threshold of the elementary scalar singlet and possibly additional bound states, depending on the representation and the range of validity of the FMS mechanism, see Sec.~\ref{sec:validity}.

The spectrum as well as further properties of the bound state operator $\phi^{* \bar{\mathbf{a}}}\phi^{\mathbf{a}}$ or any other operator are encoded in its $n$-point functions. In order to extract the mass, we investigate its propagator with the aid of the FMS expansion \eqref{eq:FMSexpansion},
\begin{align}
 \langle O(x)O(y) \rangle 
 &= 4 \Big\langle \re \big( \langle \phi^{* \bar{\mathbf{a}}} \rangle \varphi^\mathbf{a}(x) \big) \, \re \big( \langle \phi^{* \bar{\mathbf{b}}} \rangle \varphi^\mathbf{b}(y) \big) \Big\rangle \notag\\
 &\quad + \big\langle \big( \varphi^{* \bar{\mathbf{a}}}\varphi^{\mathbf{a}}\big)(x) \big( \varphi^{* \bar{\mathbf{b}}}\varphi^{\mathbf{b}}\big)(y) \big\rangle + \cdots.
 \label{eq:FMSpropagator}
\end{align}
The most important result is highlighted in the first line on the right-hand side. In nontrivial leading order in the FMS description, the bound state propagator behaves like the propagator of the radial Higgs excitation. Thus the masses coincide if the pole structure on the right-hand side is not altered by the higher-order $n$-point functions. 
In general, the bound state operator on the left-hand side is an inherently nonperturbative object but in case the FMS mechanism works for a theory with BEH mechanism, we have a simple approach to address its properties in a suitable gauge by investigating the properties of the radial Higgs excitation in the current example. 
The dots in the second line hide an unimportant constant term, disconnected parts, as well as three-point functions which vanish if the $n$-point functions are evaluated at trivial order in the coupling constants but give nontrivial contributions to the analytic structure of the bound state propagator at higher order. Further, the second line explicitly contains the four-point function for later illustrative purposes. We also would like to stress the importance of choosing a gauge with nonvanishing VEV. Otherwise the FMS expansion is rather trivial.

From this analysis we are able to conclude that although the radial Higgs excitation is a gauge-dependent object, it generates a state which also has a strict gauge-invariant description in terms of a $\mathcal{G}$-invariant composite operator. This latter fact justifies that this particular state is part of the spectrum of the theory. At this point one might be tempted to infer that similar constructions can be found for the other elementary degrees of freedom as well. Indeed, it is possible to find simple operators in the vector channel that expand to a single elementary gauge boson for $\SU(N)$ gauge theories with a scalar field in the fundamental or adjoint representation \cite{Maas:2017xzh}. All examples that provide a mapping between a $\mathcal{G}$-invariant bound state and a single elementary field that have been found in the literature as well as presented in the subsequent sections have one common property. The elementary fields are always singlets with respect to the remaining unbroken gauge group $\mathcal{H}$. This is consistent with the following comparatively simple group-theoretical point of view.

Due to symmetry breaking, we decompose the original $\mathcal{G}$ multiplets of the model into multiplets of the unbroken gauge group $\mathcal{H}$. The operator on the left-hand side of the FMS expression is strictly invariant under $\mathcal{G}$ transformations which implies that the resulting terms on the right-hand side are invariant with respect to $\mathcal{H}$, i.e., they transform as singlets and thus no mapping to a single elementary field that transforms in a nontrivial way according to $\mathcal{H}$ can be constructed. However, that does not mean that the resulting objects on the right-hand side cannot contain nontrivial $\mathcal{H}$ multiplets. They are rather bound into $\mathcal{H}$-invariant composite states. This is expected as in the conventional picture they are anyhow confined if $\mathcal{H}$ is non-Abelian.

For instance, consider once more the fundamental case for $\SU(N)$. The $\mathcal{G}$-invariant vector operator $\phid D_{\mu}\phi = \frac{\I}{2}gv^{2} \phid_{0} A_{\mu} \phi_{0} + \mathcal{O}(\varphi)$ expands in leading order to the $\mathcal{H}$ singlet $A_{\rms}^{\mu}$ as well as scattering states at higher order in the expansion but not to an element of the massive fundamental vector field $A_{\rmf}^{\mu}$. Nonetheless, we may consider the scalar operator $\phid D^{2} \phi = -\frac{g^{2}v^{2}}{2} \phid_{0}A^{2}\phi_{0} + \mathcal{O}(\varphi)$ where we have omitted an unimportant term proportional to the longitudinal part of $A_{\rms}^{\mu}$ at leading order. The FMS expansion projects on a particular $\mathcal{H}$-invariant combination of the massive vector fields which can be written as a superposition of two $\mathcal{H}$-invariant objects if we use the multiplet decomposition, $\phid_{0}A^{2}\phi_{0} = A_{\rmf\mu}^{\dagger}A_{\rmf}^{\mu} + A_{\rms \mu}^{\phantom{\mu}}A_{\rms}^{\mu}$. The first term $A_{\rmf\mu}^{\dagger}A_{\rmf}^{\mu}$ is precisely a composite operator one would naively investigate in the low energy effective theory valid below the scale $v$. The term ${\sim} A_{\rms}^{2}$ simply reflects the fact that we expect a cut at the level of the $n$-point functions in the scalar channel starting at twice the mass of the massive vector boson $A_{\rms}^{\mu}$. Indeed, $A_{\rms}^{\mu}$ can be used to describe a physical particle due to its nontrivial description in terms of the $\mathcal{G}$-invariant operator $\phid D^{\mu}\phi$.

Strictly speaking the FMS mechanism does formally not provide a mapping from $\mathcal{G}$-invariant bound states on states generated by the elementary field operators but defines a relation between $\mathcal{G}$- and $\mathcal{H}$-invariant states in a fixed gauge. Of course, in case $\mathcal{H}$ is trivial, all elementary degrees of freedom are $\mathcal{H}$-invariant and the $\mathcal{G}$-invariant bound states expand in leading order to elementary fields. This is indeed the case for the standard model as the weak sector has no non-Abelian structure left after gauge symmetry breaking and an exact bound-state--elementary-state duality can be established due to the custodial symmetry group, see App.~\ref{app:SUNfun}. This duality has to be revised to the notion of $\mathcal{H}$-invariant objects if a broader context is considered, emphasizing the special structure of the standard model.

\subsection{Validity of the FMS mechanism and classification of operators/states}
\label{sec:validity}
Before we start to discuss various examples, a few more statements and details are necessary. The identification provided by the FMS mechanism is nontrivial. At first glance, the expansion is an exact identity/\-rewriting of the original $\mathcal{G}$-invariant operator in terms of $\mathcal{H}$-invariant objects in a fixed gauge. Thus, if any $\mathcal{G}$-invariant composite operator indeed generates a bound state, the pole structure of its correlator is hidden in one of the terms on the right-hand side of the expansion. There, each term is individually gauge-dependent with respect to $\mathcal{G}$ and can meaningfully be addressed only in the specific chosen gauge. Nonetheless, the sum of all the terms is invariant by construction.

Let us consider first the case, where the $\mathcal{G}$-invariant operator expands to a single elementary field in leading order. At least as long as the scalar fluctuations are small compared to the VEV, it can be expected that the identification of the bound state operator with the elementary field holds. This is obviously the case for the Higgs sector of the standard model and has been tested in nonperturbative lattice calculations for the weak-Higgs subsector as well as for an $\SU(3)$ gauge theory with fundamental scalar field \cite{Maas:2017wzi,Maas:2014pba,Maas:2013aia,Maas:2016ngo,Maas:2017pcw,Maas:2018xxu}. Up to what extend it also holds in the nonperturbative regime, i.e., at which coupling strength the pole structure of the $n$-point functions on the right-hand side might get altered, is not fully explored yet. But as long as the theory is weakly coupled at the characteristic scale defined by the minimum of the (effective) potential, it is likely that it holds similar to the standard-model case.

Furthermore, it is not clear as to whether the identification also holds, if the $\mathcal{H}$ singlet on the right-hand side is not a single elementary field but a composite operator build from nontrivial $\mathcal{H}$-multiplets. Provided that some of the constituents receive a mass term due to the BEH mechanism such that the mass parameter is $\mathcal{O}(v)$, it can be assumed that the mass of the $\mathcal{H}$ bound state is also of that order, if $v \gg \Lambda_{\mathcal{H}}$ with $\Lambda_{\mathcal{H}}$ the characteristic scale of the non-Abelian gauge theory $\mathcal{H}$ where the associated coupling constant becomes large. Assuming a simple constituent model, we will approximate the mass by the sum of the mass parameters of the elementary fields in the following but keep in mind that the mass relation is rather between the $\mathcal{G}$ and $\mathcal{H}$ bound state. These type of operators do not allow for a simple perturbative treatment as in the case of single elementary field operators on the right-hand side. Nevertheless, they can be calculated by lattice simulations or via functional methods. Assuming that the bound state indeed exists in the effective theory with gauge group $\mathcal{H}$ being a gauge fixed version of the original theory with gauge group $\mathcal{G}$, it would be interesting to see if the FMS approach also allows for an identification of the bound states as the underlying group theoretical structure of the mechanism dictates.

The answer to this question might be more intricate than one would naively expect. Formally, we are able to divide the composite operators as well as their mapping and the states they produce into different classes. First, we define a single $\mathcal{H}$-invariant operator as an $\mathcal{H}$-singlet which cannot be decomposed further, i.e., it is created as a product of irreducible $\mathcal{H}$ multiplets and covariant derivatives thereof. For instance, $h$, $A_{\rms}^{\mu}$, and $A_{\rmf\mu}^{\dagger}A_{\rmf}^{\mu}$ are single $\mathcal{H}$-invariant operators while $\phid_{0}A^{2}\phi_{0}^{\phantom{\,}} = A_{\rmf\mu}^{\dagger}A_{\rmf}^{\mu} + A_{\rms \mu}^{\phantom{\mu}}A_{\rms}^{\mu}$ is not. The FMS expansion in terms of the split~\eqref{eq:split} results in a linear combination of products of $\mathcal{G}$-covariant objects which are $\mathcal{H}$ singlets. Some of these products can be decomposed further via the standard multiplet decomposition while others already express single $\mathcal{H}$-invariant operators in a unique manner. 
Therefore, we define the following:
\begin{enumerate}
\item The first class consists of those single $\mathcal{H}$-invariant operators that can be extracted in a unique way from a $\mathcal{G}$-invariant operator via the FMS expansion at some fixed order in the expansion parameter $\varphi/v$ without requiring any further multiplet decompositions.
\end{enumerate}
Correspondingly, we define a state of the first class as a state generated by an operator of the first class and similar for the later introduced classes. 
For convenience, we will also use the mathematical symbols characterizing operators (e.g., $h,\phi^{\bar{\mathbf{a}}}\phi^{\mathbf{a}},\cdots$) for both the operators themselves as well as the states generated by these operators in the following. Which interpretation is meant will be obvious from the context. States of the first class play an important role for the $\mathcal{G}$-$\mathcal{H}$ duality as sufficiently strong evidence is accumulated from lattice simulations that these are part of the $\mathcal{G}$-invariant spectrum, see Sec.~\ref{sec:tests}.

At most those operators can belong to the first class who contain at least one constituent that operates in the same subspace in which $\phi_{0}$ acts nontrivial. Straightforward examples of the first class are the cases where the FMS expansion projects on an elementary field, e.g., the radial Higgs excitation or the massive gauge singlet $A_{\rms}^{\mu}$ for an $\SU(N)$ theory with fundamental scalar field.

A nontrivial example where we get a relation between a $\mathcal{G}$-invariant operator and an $\mathcal{H}$-invariant composite object belonging to the first class is given by an vector operator with open global $\U(1)$ quantum number for an $\SU(3)$ gauge theory, see App.~\ref{app:SUNfun}. An easier example can be found by considering an $\SO(N)$ gauge theory with scalar field in the fundamental representation. Here, we obtain only one massive gauge multiplet $A^{\mu}_{\rmf} = A^{\mu}\phi_{0}$ transforming under the fundamental representation of $\SO(N-1)$ in the conventional picture of gauge symmetry breaking. From the perspective of the effective theory with broken gauge group, we would expect that the scalar meson-type operator $A_{\rmf}^{\rmT}A_{\rmf}^{\phantom{\rmT}}$ (we omit the Lorentz indices for better readability in the following) generates a possible $\mathcal{H}$-invariant bound state. 
This operator has a strict $\mathcal{G}$-invariant analog by the operator $\phi^{\rmT} D^{2} \phi = -g^{2}v^{2}  A_{\rmf}^{\rmT}A_{\rmf}^{\phantom{\mu}} + \cdots$ which expands in leading order in the FMS prescription to that meson operator.%
\footnote{Throughout this paper, we will use the term meson for any object which consists of two matter fields regardless of whether these are scalars, fermions, or massive vector particles.} 
Depending now on the precise details of the model, in particular its couplings, the propagator of the $\mathcal{G}$-invariant composite operator
\begin{align*}
 &\big\langle \big(\phi^{\rmT} D^{2} \phi\big)(x) \big(\phi^{\rmT} D^{2} \phi\big)(y) \big\rangle \\
 &\quad = \frac{v^{4}g^{4}}{4} \big\langle \big(A_{\rmf}^{\rmT}A_{\rmf}^{\phantom{\mu}}\big)(x) \big(A_{\rmf}^{\rmT}A_{\rmf}^{\phantom{\mu}}\big)(y) \big\rangle + \cdots \phantom{E = mc^{2}}
\end{align*}
may have a pole, if the four point function on the right-hand side can indeed be interpreted as the propagator of a bound state of the remaining $\mathcal{H}$-invariant gauge theory.

By contrast, the gauge field $A$ contains two massive multiplets in the $\SU(N)$ case, namely $A_{\rms}$ and $A_{\rmf}$. Thus, we obtain from the FMS expansion of $\phid D^{2} \phi$ at leading order only a projection on the superposition of $\mathcal{H}$-invariant combinations of both multiplets as we have seen previously. The contained scattering state generated by $A_{\rms}^{2}$ and the bound state operator $A_{\rmf}^{\dagger}A_{\rmf}^{\phantom{\mu}}$ can only be identified if we additionally decompose the multiplet by hand. Moreover, there exists no simple $\mathcal{G}$-invariant operator that expands unambiguously to $A_{\rmf}^{\dagger}A_{\rmf}^{\phantom{\mu}}$ such that it is not contained in the first class and can only be extracted if we use the gauge-variant decomposition. Therefore, we assign it to the second class which we define as follows:
\begin{enumerate}
 \item[2.] The second class is defined as the union of those single $\mathcal{H}$-invariant operators that can only be extracted from a $\mathcal{G}$-invariant operator if we allow for a decomposition of the $\mathcal{G}$ into $\mathcal{H}$ multiplets. 
\end{enumerate}
Some further straightforward examples can be obtained from those $\mathcal{G}$-invariant operators which have a trivial expansion in terms of \eqref{eq:split} such that we rely on the decomposition of the multiplets to decompose the $\mathcal{G}$-invariant operators into  $\mathcal{H}$-invariant objects anyway. For instance consider the $\mathcal{G}$-invariant scalar glueball operator $\tr F^{2}$. In the conventional gauge-fixed language, we would decompose $\tr F^{2} = \tr F_{\rma}^{2} + 2|F_{\rmf}|^{2} + A_{\rmf}^{\dagger}F_{\rma}^{\phantom{\,}}A_{\rmf}^{\phantom{\,}} + \cdots$ and interpret the states generated by these three $\mathcal{H}$-invariant operators as observable quantities. However, we will sketch in Sec.~\ref{sec:tests} that states of the second class will likely not be part of the observable spectrum. The dots indicate that we omitted further terms of the decomposition. All of them describe trivial scattering states in the simple constituent model, e.g., $F_{\rms}^{2}$ , $|A_{\rmf}|^{2}A_{\rms}^{2}$, $A_{\rmf}^{\dagger\mu}A_{\rmf}^{\nu}A_{\rms\mu}A_{\rms\nu}$.\footnote{We have chosen to decompose the Yang-Mills degrees of freedom at the level of the gauge potential $A$. This decomposition does generally not translate to the corresponding field strength tensors except for accidental cases. For instance, for the massive vector multiplet for the $\SO(N)$ fundamental case, $A_{\rmf} = A\phi_{0}$, and its field strength tensor $F_{\rmf}$, we obtain indeed $F^{\mu\nu}\phi_{0} \equiv F_{\rmf}^{\mu\nu} = D^{\mu}_{\mathcal{H}}A_{\rmf}^{\nu} - D^{\nu}_{\mathcal{H}}A_{\rmf}^{\mu}$ with $D^{\mu}_{\mathcal{H}} = \partial^{\mu} + g A_{\rma}^{\mu}$. In case there is only one massive vector multiplet, the projection on this field will likely also extract the corresponding field strength tensor if it is applied on the $\mathcal{G}$ field strength tensor for all possible gauge theories. However, one always obtains additional terms from the commutator $[A^{\mu},A^{\nu}]$ if we consider the subspace of the unbroken gauge sector, e.g., $(\mathbbm{1}-\phi_{0}^{\phantom{\,}}\phi_{0}^{\rmT})F^{\mu\nu}(\mathbbm{1}-\phi_{0}^{\phantom{\,}}\phi_{0}^{\rmT}) = F_{\rma}^{\mu\nu} + g (A_{\rmf}^{\mu}A_{\rmf}^{\rmT\nu}-A_{\rmf}^{\nu}A_{\rmf}^{\rmT\mu})$ for $\SO(N)\to\SO(N-1)$. We obtain additional terms also for theories with more than one massive vector multiplet, e.g., $\phid_{0}F\phi_{0}^{\phantom{\,}}\neq F_{\rms}$ and $F\phi_{0} - F_{\rms}\phi_{0} \neq F_{\rmf}$ for $\SU(N)$ with fundamental scalar.}

Note, that there is no $\mathcal{G}$-invariant operator build by powers of the elementary $\mathcal{G}$-multiplets that expands to either $\tr F_{\rma}^{2}$ or $F_{\rmf}^{\dagger}F_{\rmf}^{\phantom{\rmT}}$ according to the requirements of the first class. Although we can define a projection operator which is nonlinear in the scalar field, $\mathbbm{1}-\frac{\phi \phid}{\phid\phi}$, and thus can build a $\mathcal{G}$-invariant operator $\tr \big( (\mathbbm{1}-\frac{\phi \phid}{\phid\phi})F^{2} \big)$ where the constituent multiplet $(\mathbbm{1}-\frac{\phi \phid}{\phid\phi})F^{\mu\nu}$ expands in leading order to the desired $\mathcal{H}$ multiplet, it is a straightforward exercise to show that the gauge-invariant operator is just a rewriting of the operators $\tr F^{2}$ and $\phid F^{2}\phi$. Thus, we have to assume that the decomposition for the gauge-invariant spectrum holds to find a strict gauge-invariant description of a possible glueball state $\tr F_{\rma}^{2}$. This feature is generic for all glueball operators associated to the unbroken subgroup independent of the gauge group or the representation of the scalar field.

The fact that some $\mathcal{H}$-invariant states follow only from a decomposition of the multiplets holds not only for those $\mathcal{G}$-invariant operators that have a trivial FMS expansion. This circumstance can appear to any order in the FMS expansion at which we do not obtain a single $\mathcal{H}$-invariant operator. We have already seen this effect at leading order for $A_{\rmf}^{\dagger}A_{\rmf}^{\vphantom{f}}$. An example where this effect appears at higher orders in the FMS expansion is the scalar operator defined in Eq.~\eqref{eq:FMSexpansion} at $\mathcal{O}(\varphi^{2})$. Suppose $\phi$ contains further non--would-be-Goldstones besides the radial Higgs excitation which are assigned to some $\mathcal{H}$ multiplets as it is generally the case for tensor representations. Then, $\varphi^{*\bar{a}}\varphi^{a}$ might contain not only the trivial scattering state of the radial Higgs excitation but also possible bound state operators whose propagators are encoded in the four-point function in the second line of Eq.~\eqref{eq:FMSpropagator} if the gauge-variant multiplet decomposition is meaningful. The FMS formalism does not provide a further ordering of the $\mathcal{H}$-invariant terms at this order in the expansion. If no further $\mathcal{G}$-invariant operator can be found that expands to such a single $\mathcal{H}$-invariant bound state operator, they belong to the second class. This seems to be the case at least for all investigated examples in which the additional Higgs fields are not singlets with respect to $\mathcal{H}$.

Though the first two classes cover a wide range of possible states, there are still some states left which are not related via a duality relation. Therefore, we define the third class in the following way:
\begin{enumerate}
\item[3.] The third class contains those operators generating states which are well defined observables from the perspective of a theory with gauge group $\mathcal{H}$ but cannot be defined in a $\mathcal{G}$-invariant way.
\end{enumerate}
We provide a simple example for such states at the end of Sec.~\ref{sec:ONfun}. In particular, they will play an important role for BSM model building which we outline in the context of grand unified scenarios in Sec.~\ref{sec:GUT}.

Finally, we would like to mention that also a fourth class of states is conceivable. These are states appearing in the spectrum of the theory with gauge group $\mathcal{G}$ but cannot be addressed in terms of $\mathcal{H}$-invariant operators and are thus generated by inherently nonperturbative effects of the $\mathcal{G}$-invariant composite operators. As such states cannot be addressed within the FMS framework, we will ignore this possibility in the following. If they indeed exist, they can only be revealed by other nonperturbative tools. Currently, no such state has been seen by lattice simulations.

\subsection{Nonperturbative tests of the FMS mechanism}
\label{sec:tests}
It is an interesting but challenging task for lattice or functional tools to show that the FMS predictions are valid beyond the examined case where the $\mathcal{G}$-invariant operators expand to elementary $\mathcal{H}$-singlet fields, i.e., for the case bound state to bound state mapping. The only preparatory work in this direction is the investigation of an $\SU(3)$ gauge theory with a single scalar field in the fundamental representation \cite{Maas:2018xxu}. There, the validity of the FMS expansion has been shown for the mapping to elementary fields being singlets of $\mathcal{H}$ in a first step. Further, nontrivial bound state operators  were considered in the vector channel with open U(1) quantum number that expand to $\SU(2)$-invariant bound states belonging to the first class. Due to resources, the $\mathcal{G}$-invariant bound state spectrum was investigated in this channel but the pole structure of the $\mathcal{H}$-invariant bound state counterpart was not. Nonetheless, it could be shown that the ground state mass of the $\mathcal{G}$-invariant spectrum is in accordance with the simple constituent model one would apply to the $\mathcal{H}$-invariant bound states as its assumptions are fulfilled. This provides evidence that the $\mathcal{G}$-$\mathcal{H}$ duality operates for all operators of the first class. Not only for those that expand to simple elementary fields but also for those who map to more sophisticated objects. Thus, states of this class can be investigated along the usual lines within the BEH framework. In particular, masses of $\mathcal{G}$-invariant states that can be mapped onto elementary fields can be computed in a standard perturbative setting in the weak coupling regime.

An additional nontrivial test can be done within the $\SU(3)$ model by considering the $\SU(2)$-invariant operators $A_{\rmf}^{\dagger}A_{\rmf}^{\phantom{f}}$ or $F_{\rmf}^{\dagger}F_{\rmf}^{\phantom{\,}}$. These might be encoded in different $\SU(3)$-invariant operators but belong to the second class as we have seen in the previous discussion. Interestingly, these operators have not been seen on the lattice yet as no operator included in the variational analysis seems to have substantial overlap with these states if they are part of the spectrum \cite{Maas:2018xxu}. This might have several reasons. 
The only investigated operator that may have some overlap with the state $F_{\rmf}^{\dagger}F_{\rmf}^{\phantom{\,}}$ is the $\mathcal{G}$-invariant glueball operator $\tr F^{2}$ if the decomposition of the symmetry breaking viewpoint transfers in a straightforward manner. 
It seems mandatory to extend the variational analysis by other operators such as $\phid F^{2} \phi$ to make a decisive statement. 
Unfortunately there is another obstacle. The parameter sets of the lattice simulation studied so far imply that the mass of the hybrid bound states  $F_{\rmf}^{\dagger}F_{\rmf}^{\phantom{\,}}$ is close to the mass of the radial Higgs excitation by accident, making the identification of this state difficult for the current data sets. The same is true for the meson-type state $A_{\rmf}^{\dagger}A_{\rmf}^{\phantom{\,}}$ which would be included in $\phid D^{2} \phi$ if states of the second class belong to the $\mathcal{G}$-invariant spectrum.

In case the $\mathcal{G}$-$\mathcal{H}$ duality is applicable for operators of the second class, we expect that not only the states $A_{\rmf}^{\dagger}A_{\rmf}^{\phantom{i}}$ and $F_{\rmf}^{\dagger}F_{\rmf}^{\phantom{i}}$ are part of the $\SU(3)$-invariant spectrum but also the $\SU(2)$ glueball. However, the state generated by this operator is also not seen. To extend the lattice analysis into that direction is challenging. The mass of this operator will be set by the scale $\Lambda_{\SU(2)}$, where the gauge coupling of the unbroken subsector becomes large. This scale is usually far separated from the characteristic scale set by the Higgs sector $v$, if the original gauge theory is not already strongly coupled at that scale. Such a large separation by several orders of magnitude is technically not accessible yet. 
At least the massless elementary gauge bosons behave perturbatively as one would expect at the investigated scales in a gauge-fixed set up on the lattice. 
One may conclude that this can be viewed as a hint that such a glueball operator exists. 
However, it would then manifest in the volume dependence of the ground state if the operator $\tr F^{2}$ has sufficient overlap. This is not seen in the actual lattice calculations \cite{Maas:2016ngo,Maas:2018xxu}. Also in this case a larger operator basis for the variational analysis will settle the question as to whether the $\SU(2)$ glueball state is contained in the spectrum. 
Alternatively one may perform a lattice study at almost strong coupling at the scale $v$. However, it will then become an intricate task to disentangle effects coming from BEH physics from those generated by the large gauge coupling similar to a QCD system. Thus, we conclude that it seems reasonable to first focus on the states generated by $A_{\rmf}^{\dagger}A_{\rmf}^{\phantom{i}}$ or $F_{\rmf}^{\dagger}F_{\rmf}^{\phantom{i}}$ to examine as to whether the state duality holds for states of the second class.

In the following, we will analyze the bound state spectrum of various BEH models via the FMS mechanism and highlight the particularities of the spectra of the different models. We will mainly focus on operators belonging to the first class. At least for those, we have sufficient lattice support to trust the nontrivial FMS mapping. We also consider operators of the second class to demonstrate the maximal possible overlap between the states of the theories with gauge group $\mathcal{G}$ and $\mathcal{H}$. However, one should keep in mind that for these operators the results have to be treated with caution as current lattice simulations provide hints that the duality cannot be extend to operators of this class. Thus, we assume at this point that states of the second class are not present in the $\mathcal{G}$-invariant spectrum, although one would naively expect them from the conventional perspective of spontaneous gauge symmetry breaking. This might be traced back to the following difference. Technically both, the multiplet decomposition as well as the FMS decomposition defined by the split~\eqref{eq:split}, are gauge dependent and one could scrutinize the meaningfulness of the distinction between operators of the first and second class. However, the FMS expansion provides a clear ordering scheme of the occurring terms on the right-hand side in terms of $\varphi/v$ in a fixed gauge. This ordering scheme is lost for the conventional multiplet decomposition and thus for the extraction of objects assigned to the second class from $\mathcal{G}$-invariant operators.

\section{\texorpdfstring{$\SO(N)$}{SO(N)} gauge theory}
\label{sec:ON}

We start our concrete analysis by extending the investigations of the fundamental and adjoint representation for $\SU(N)$ gauge theories, see \cite{Maas:2017xzh}, to the special orthogonal group. Additionally, we include also the symmetric second-rank tensor representation to obtain a complete overview over the low dimensional representations up to second-rank tensors.

As the scalar fields are real valued for these representations, the only potential global symmetry is given by a discrete $\Ztwo$ symmetry if we consider only one scalar field, $\phi \to -\phi$ and $A_{\mu} \to A_{\mu}$. Therefore, we will classify all $\SO(N)$-invariant observable states into $\Ztwo$-even and $\Ztwo$-odd states. The global symmetry may be broken explicitly, e.g., by a cubic coupling for the symmetric second-rank tensor, due to dynamical effects, or is in some cases already part of the gauge transformations. Regardless of whether this is the case, we will provide examples for both $\Ztwo$-even and $\Ztwo$-odd operators. If the global symmetry is broken, transitions between the two channels are allowed without preserving the $\Ztwo$ quantum number, e.g., a decay of a $\Ztwo$-odd state into two $\Ztwo$-even states is possible if kinematically allowed. From the perspective of the spectroscopy, we just have enlarged the number of operators of the considered $J^{P}$ channel and the distinction is not necessary but can be done. If the symmetry is intact, the global quantum number has to be conserved in a decay process. In particular, we obtain two different ground states in every $J^{P}$ channel distinguished by the global quantum number. Depending on the precise details of the theory, in particular the mass ratios of the states, the ground states of the different channels can be bound states, resonances, or only scattering states.

\subsection{Fundamental representation}
\label{sec:ONfun}

In order to warm up for the more involved tensor representations, we start with the fundamental representation of the $\SO(N)$ group. In this case, the scalar field transforms as a vector $\phi \to R \phi$, where $R$ is an element of the orthogonal group, i.e., $R^{\rmT}R = \mathbbm{1}$. The covariant derivative reads $D_{\mu}\phi = \partial_{\mu}\phi + g A_{\mu} \phi$ and the most general potential up to fourth order in the field is
\begin{align}
 V(\phi) = -\frac{\mu^2}{2} \phi^{\rmT}\phi + \frac{\lambda}{8} (\phi^{\rmT}\phi)^{2}.
\end{align}
Without loss of generality, we can choose $\phi_{0}^{a} = \delta^{aN}$ as the direction of the vacuum expectation value if $\mu^{2} > 0$. Thus, the field configuration that minimizes the potential is $\phi_{\mathrm{min}} = v \phi_{0} = \langle \phi \rangle$ with $\mu^{2}= \frac{1}{2}\lambda v^{2}$. All other solutions are related by an $\SO(N)$ transformation. Obviously, this solution is invariant under transformations of the subgroup $\SO(N-1)$ and the breaking pattern reads $\SO(N) \to \SO(N-1)$ due to gauge fixing. Albeit the gauge symmetry of the system is a redundancy in our description rather than an actual symmetry, this situation is often called spontaneous gauge symmetry breaking, adopting the vocabulary of spontaneous symmetry breaking for a global symmetry. The $(N-1)$ fields stored in $\phi$ which are orthogonal to $\phi_{0}$ will become would-be Goldstone bosons and are removed from the elementary spectrum in the unitary gauge. Therefore, the scalar field can be written in the unitary gauge as
\begin{align}
 \phi(x) = v \phi_{0} + \varphi(x) = \big(v+h(x)\big) \phi_{0}
\end{align}
with one real-valued scalar degree of freedom $h(x) = \phi_{0}^{\rmT}\varphi$. The mass of the radial Higgs excitation $h$ is given by $\mh^{2} = \lambda v^{2}$. The mass matrix of the gauge bosons $M_{\rmA}^{2}$ can be obtained from the kinetic term of the scalar field as usual and we obtain, 
\begin{align*}
 \frac{1}{2} (M_{\rmA}^{2})_{ij} A_{i \mu} A_{j}^{\mu} = g^2v^2 \phi_{0}^{\rmT}A^{\rmT}_{\mu}A^{\mu}\phi_{0} 
\end{align*}
with $A^{\mu} = A^{\mu}_{i}T_{i}^{\phantom{\,}}$. Normalizing the generators $T_{i}$ to fulfill $\tr (T_{i}^{\rmT}T_{j}) =\frac{1}{2} \delta_{ij}$, $(N-1)$ gauge fields obtain a mass parameter $m_\mathrm{A_{f}}^{2} = \frac{1}{2}g^2v^2$. Note that we use lower indices starting from $i,j,k,\cdots$ to indicate objects in the adjoint representation while upper indices starting from $a,b,c,\cdots$ are fundamental indices. The massive vector bosons transform as a fundamental vector of $\SO(N-1)$ and can be extracted from the $\SO(N)$ gauge field by $A_{\rmf} = A\phi_{0}$. The remaining $\frac{1}{2}(N-1)(N-2)$ gauge bosons encoded in $A_{\rma} = (\mathbbm{1}-\phi_{0}^{\phantom{}}\phi_{0}^{\rmT}) A (\mathbbm{1}-\phi_{0}^{\phantom{}}\phi_{0}^{\rmT})$ are the massless gauge fields of the unbroken $\SO(N-1)$ subgroup.

Ignoring Elitzur's theorem and the Gribov-Singer ambiguity, we would expect the following spectrum in the conventional picture for $N\geq 4$: In the scalar channel the states generated by operators with lowest field content are the radial Higgs excitation $h$, a scalar glueball state $\tr(F_{\rma}^{2})$, and a scalar meson-type operator where the constituents are the massive vector bosons $A_{\rmf}^{\rmT}A_{\rmf}^{\phantom{i}}$. In the vector channel, one would expect a hybrid state containing two massive vector bosons and a massless gauge boson, e.g., $A_{\rmf\nu}^{\rmT}D^{\nu}A_{\rmf}^{\mu}$. Furthermore, we have possible states generated by composite operators involving more elementary fields as the scalar hybrid operator $F_{\rmf \mu\nu}^{\rmT} F_{\rmf}^{\mu\nu}$. The case $N<4$ is analyzed at the end of this subsection.

In order to discuss the gauge-invariant bound state spectrum of the original $\SO(N)$ theory, we carefully distinguish between the groups with even $N=2K$ and odd $N=2K+1$. If we consider the special orthogonal group for odd $N$, the action of the model obeys a discrete $\mathbb{Z}_{2}$ symmetry, $\phi \to -\phi$, such that we can distinguish the states due to this global symmetry in $\mathbb{Z}_{2}$ even and odd states. This symmetry transformation is already part of the gauge transformation for even $N$, $\phi \to -\phi = -\mathbbm{1}\phi$, as $-\mathbbm{1}\in \SO(2K)$. Therefore, we distinguish the gauge-invariant states only with respect to their spin and parity in this particular case.

The simplest gauge-invariant scalar state that can be build for all $N$ is given by the product of two fundamental scalar fields and expands in the FMS framework to a single Higgs field $h$ in nontrivial leading order.
\begin{align}
 \phi^a\phi^a = v^{2} + 2v \phi_{0}^a\varphi^a + \varphi^a\varphi^a = v^{2} + 2v h + h^{2},
 \label{eq:ONfunZeven}
\end{align}
where we have used the unitary gauge condition in the second identity. Thus, we have found a gauge-invariant description of the radial Higgs excitation also for the fundamental representation of $\SO(N)$ models, implying ${m_\mathrm{\phi^{\rmT}\phi} = \mh}$. Of course this is not a surprise as this is always possible for the radial Higgs excitation irrespective of the representation or the gauge group as we have seen in Sec.~\ref{sec:FMS}, cf. Eq.~\eqref{eq:FMSexpansion}.

Further, we investigate in this channel the $\Ztwo$-even operator 
\begin{align} 
\phi^{\rmT}D_{\mu}D^{\mu}\phi = -g^{2}v^{2}A_{\rmf \mu}^{\rmT}A_{\rmf}^{\mu} + \mathcal{O}(\varphi)
\end{align}
which expands in leading order to the $\SO(N-1)$-invariant scalar meson operator build by two massive vector bosons. We approximate its mass by the sum of its constituents $m_{\phi^{\rmT}D^{2}\phi} = m_\mathrm{A_{\rmf }^{\rmT}A_{\rmf}^{\phantom{a}}} \approx 2 m_\mathrm{A_{f}}$. At higher order in the expansion, the pole structure generated by \eqref{eq:ONfunZeven} is formally encoded in the propagator of $\phi^{\rmT}D^{2}\phi$ as well. Thus, we will obtain off-diagonal terms at the level of the propagators of $\phi^{\rmT}\phi$ and $\phi^{\rmT}D^{2}\phi$. In order to avoid this, one may equivalently consider the operator $(D_{\mu}\phi)^{\rmT}D^{\mu}\phi$. Furthermore, we obtain trivial scattering states at $2\mh$, $m_\mathrm{A_{\rmf }^{\rmT}A_{\rmf}^{\phantom{a}}} + \mh$, $m_\mathrm{A_{\rmf }^{\rmT}A_{\rmf}^{\phantom{a}}} + 2\mh$. According to the definitions of Sec.~\ref{sec:FMS}, the bound state $A_{\rmf }^{\rmT}A_{\rmf}^{\phantom{a}}$ belongs to the first class as we can extract it purely by the FMS decomposition. Thus, it is likely part of the $\SO(N)$-invariant spectrum. Also the hybrid states generated by $F_{\rmf \mu\nu}^{\rmT} F_{\rmf}^{\mu\nu}$ belong to the first class. This can be deduced from the operator $\phi^{\rmT}F_{\mu\nu}F^{\mu\nu}\phi$ and its FMS expansion. They might be encoded in $\tr F^{2}$ as well. However, the operator $F_{\rmf}^{\rmT}F_{\rmf}^{\vphantom{\rmT}}$ can only be extracted via the conventional multiplet decomposition from $\tr F^{2}$. This is also the case for $\tr F_{\rma}^{2}$ and $A_{\rmf}^{\rmT}F_{\rma}A_{\rmf}^{\phantom{\,}}$. In contrast to $F_{\rmf}^{\rmT} F_{\rmf}^{\phantom{\,}}$, the operators $\tr F_{\rma}^{2}$ and $A_{\rmf}^{\rmT}F_{\rma}A_{\rmf}^{\phantom{\,}}$ belong to the second class as no $\SO(N)$-invariant operator can be found that maps unambiguously to these two operators without using the conventional multiplet decomposition.

For $\SO(2K+1)$, the scalar spectrum contains an additional channel which is odd with respect to the global $\mathbb{Z}_{2}$ symmetry, resulting in a further potentially stable particle (depending on the mass spectrum in the other $J^{P}$ channels). A possible example of an operator with odd $\mathbb{Z}_{2}$ parity involves the invariant epsilon tensor and reads
\begin{align}
 &\epsilon^{a_1 \cdots a_{2K+1}} \phi^{a_1} (F_{\mu_1}^{\,\,\,\mu_2})^{a_2a_3} (F_{\mu_2}^{\,\,\,\mu_3})^{a_4a_5} \cdots (F_{\mu_K}^{\,\,\,\mu_1})^{a_{2K}a_{2K+1}} \notag \\
 &\,\, = (v+h)\, \epsilon^{a_1 \cdots a_{2K+1}} \phi_{0}^{a_1} (F_{\mu_1}^{\,\,\,\mu_2})^{a_2a_3} \cdots (F_{\mu_K}^{\,\,\,\mu_1})^{a_{2K}a_{2K+1}} \notag \\
 &\,\, = v \epsilon^{\dot{a}_1 \cdots \dot{a}_{2K}} (F_{\mu_1}^{\,\,\,\mu_2})^{\dot{a}_1\dot{a}_2} \cdots (F_{\mu_K}^{\,\,\,\mu_1})^{\dot{a}_{2K-1}\dot{a}_{2K}} + \mathcal{O}(\varphi) 
 \label{eq:Z2oddglueball}
\end{align}
where dotted indices indicate elements living in the unbroken $\SO(2K)$ subgroup, $\dot{a}_{i} = \{1,\cdots,2K\}$, and we have again assumed $N>3$ ($K>1$). Using the multiplet decomposition, we have $F_{\mu\nu}^{\dot{a}_1\dot{a}_2} = F_{\rma\mu\nu}^{\dot{a}_1\dot{a}_2} + g A_{\rmf\mu}^{\dot{a}_1}A_{\rmf\nu}^{\dot{a}_2} - g A_{\rmf\nu}^{\dot{a}_1}A_{\rmf \mu}^{\dot{a}_2}$. Thus, the $\SO(2K)$-invariant operator on the right-hand side of Eq.~\eqref{eq:Z2oddglueball}, appearing at leading order in the FMS expansion, decomposes in several single $\SO(2K)$-invariant operators which belong to the second class. One of the operators can be described by $K$ constituents being massless gauge bosons of the unbroken subgroup in the gauge-fixed language. These form a nontrivial scalar $\SO(2K)$ glueball operator whose mass will likely be nonvanishing and of the order $\Lambda_\mathrm{\SO(2K)}$ which can be checked by lattice or functional methods. The other operators form hybrids with an increasing number of massive vector multiplets as constituents. At next-to-leading order, we obtain the trivial scattering state of the states generated by these operators with the radial Higgs excitation $h$ or, to be more precise, with the $\Ztwo$-even scalar bound state $\phi^{\rmT}\phi$ from the perspective of the actual unbroken gauge group $\SO(2K+1)$.

Additional scalar hybrid states containing massless and massive vector fields can be described in an $\SO(2K+1)$-invariant way as well, e.g., consider 
\begin{align}
&\epsilon^{a_1 \cdots} \phi^{a_1} (F_{\mu_1}^{\,\,\,\mu_2}\phi)^{a_2} (F_{\mu_2}^{\,\,\,\mu_3}\phi)^{a_3} (F_{\mu_3}^{\,\,\,\mu_4})^{a_4a_5} \!\cdots\! (F_{\mu_{K+1}}^{\,\,\,\mu_1})^{a_{2K}a_{2K+1}} \notag\\
&\,\, = v \epsilon^{\dot{a}_1 \cdots \dot{a}_{2K}} (F_{\rmf \mu_{1}}^{\,\,\,\mu_2})^{\dot{a}_1}(F_{\rmf \mu_{2}}^{\,\,\,\mu_3})^{\dot{a}_2}(F_{\mu_{3}}^{\,\,\,\mu_4})^{\dot{a}_3\dot{a}_4} \!\cdots\! + \mathcal{O}(\varphi)
\label{eq:Z2oddscalarhybrid}
\end{align} 
for spacetime dimensions larger than two, otherwise the operator vanishes. All $\SO(2K)$-invariant operators are assigned to the second class also in this case. The only operator we have found in the $\Ztwo$-odd scalar channel that belongs to the first class is given by a hybrid containing $2K$ massive constituents $A_{\rmf}$,
\begin{align}
&\epsilon^{a_1 \cdots a_{2K+1}} \phi^{a_1} (F_{\mu_1}^{\,\,\,\mu_2}\phi)^{a_2} (F_{\mu_2}^{\,\,\,\mu_3}\phi)^{a_3} \cdots (F_{\mu_{2K}}^{\,\,\,\mu_1}\phi)^{a_{2K+1}} \notag\\
&\,\, = v \epsilon^{\dot{a}_1 \cdots \dot{a}_{2K}} (F_{\rmf \mu_{1}}^{\,\,\,\mu_2})^{\dot{a}_1} \cdots (F_{\rmf \mu_{2K}}^{\,\,\,\mu_1})^{\dot{a}_{2K}} + \mathcal{O}(\varphi).
\label{eq:Z2oddscalarhybrid2}
\end{align} 
However both, the $\SO(2K+1)$-invariant and the resulting $\SO(2K)$-invariant operator are only nonvanishing for $4K\leq d(d-1)$ where $d$ is the spacetime dimension. 
Further note, that this construction principle for $\mathbb{Z}_{2}$-odd operators via the epsilon tensor does not work for even $N$, as the elementary building blocks that transform in a covariant manner are the scalar field $\phi^{a}$, its covariant derivative $(D_{\mu}\phi)^{a}$, or the field strength tensor $F_{\mu\nu}^{ab}$. Thus, any operator always contains an even number of scalar fields for $\SO(2K)$.

In contrast to $\SUN$ gauge theories with a scalar field in the fundamental representation, the elementary vector spectrum does not contain a singlet with respect to the unbroken subgroup. This fact can be translated to the gauge-invariant spectrum as it is not possible to build a vector operator that expands to a single gauge field. As the generators of the Lie algebra are antisymmetric, we have $\phi^{\rmT}F_{\mu\nu}\phi = 0$ and $\phi^{\rmT}D_{\mu}\phi = \phi^{\rmT}\partial_{\mu}\phi$. The latter expression reflects the fact that the gauge-invariant operator $\phi^{\rmT}\partial_{\mu}\phi = \frac{1}{2}\partial_{\mu}(\phi^{\rmT}\phi)$ generates a state that mixes with other states in the vector channel due to its quantum numbers but does not necessarily give rise to an additional vector particle as it is build by a partial derivative of the scalar degrees of freedom. Though, internal excitations of the scalar meson states with $J>0$ might be conceivable, we do not extend the analysis into this direction. Thus, the simplest operator that transforms as a vector will contain at least three covariant derivatives $\phi^{\rmT}D^{\mu}D^{\nu}D_{\nu}\phi$. For simplicity, we antisymmetrize the first two indices and obtain 
\begin{align}
 \phi^{\rmT}F^{\mu\nu}D_{\nu}\phi &= v^{2}g\, \phi_{0}^{\rmT}F^{\mu\nu}A_{\nu}\phi_{0} + \mathcal{O}(\varphi) \notag \\
 &= -v^{2}g\, F_{\rmf}^{\rmT\mu\nu}A_{\rmf\nu}^{\phantom{a}} + \mathcal{O}(\varphi).
\label{eq:ONfunVec}
\end{align}
The FMS expansion reveals that this vector operator provides a gauge-invariant description of the $\SO(N-1)$-invariant vector hybrid operator build by two massive elementary vector bosons that transform as fundamental vectors of the $\SO(N-1)$ subgroup and a massless adjoint gauge boson. In order to make this more transparent, we notice that $F^{\mu\nu}\phi_{0} \equiv F_{\rmf}^{\mu\nu} = D^{\mu}A_{\rmf}^{\nu}-D^{\nu}A_{\rmf}^{\mu}$ such that $F_{\rmf}^{\rmT\mu\nu}A_{\rmf\nu}^{\phantom{a}} = \frac{1}{2}\partial^{\mu}(A_{\rmf\nu}^{\rmT}A_{\rmf}^{\nu}) - A_{\rmf\nu}^{\rmT}D^{\nu}A_{\rmf}^{\mu}$. The latter term describes the proposed vector hybrid operator in an $\SO(N-1)$-invariant fashion. The first term is simply a derivative of the already discussed scalar meson operator. Of course also this state might have nontrivial internal excitations such that we obtain an additional resonance in the vector channel. Similar conclusions can be drawn from the general form $\phi^{\rmT}D^{\mu}D^{\nu}D_{\nu}\phi$.

For $\SO(2K+1)$, we get additional $\mathbb{Z}_{2}$-odd states. To explore this channel, we first investigate the following operator and its FMS expansion which leads to an $\SO(2K)$-invariant operator of the first class, 
\begin{align}
&\epsilon^{a_1 \cdots a_{2K+1}} \phi^{a_1}  (D^{\mu}\phi)^{a_2}  (F_{\mu_1}^{\,\,\,\mu_2}\phi)^{a_3} \cdots (F_{\mu_{2K-1}}^{\,\,\,\mu_1}\phi)^{a_{2K+1}} \notag \\ 
 &= g v^{2K} \epsilon^{a_1 \cdots a_{2K+1}} \phi_{0}^{a_1}  (A_{\rmf}^{\mu})^{a_2}  (F_{\rmf\mu_1}^{\,\,\,\mu_2})^{a_3} \cdots (F_{\rmf\mu_{2K-1}}^{\,\,\,\mu_1})^{a_{2K+1}} \notag \\
 &\quad + \mathcal{O}(\varphi).
\label{eq:Z2oddhybrid}
\end{align}
As for the scalar operator defined in Eq.~\eqref{eq:Z2oddscalarhybrid2}, this operator is nonvanishing only for sufficient small $K$, here $4K \leq d(d-1)+2$. 
An operator that can be defined for all $N=2K+1$ reads
\begin{align}
&\epsilon^{a_1 \cdots a_{2K+1}} \phi^{a_1}  (D^{\mu}\phi)^{a_2}  (F_{\mu_1}^{\,\,\,\mu_2}\phi)^{a_3} \notag \\ 
 &\quad \times (F_{\mu_2}^{\,\,\,\mu_3})^{a_4a_5} \cdots (F_{\mu_K}^{\,\,\,\mu_1})^{a_{2K}a_{2K+1}} \notag \\
 &= g v^{3} \epsilon^{a_1 \cdots a_{2K+1}} \phi_{0}^{a_1}  (A^{\mu}\phi_{0})^{a_2}  (F_{\mu_1}^{\,\,\,\mu_2}\phi_{0})^{a_3} \notag \\
 &\quad \times (F_{\mu_2}^{\,\,\,\mu_3})^{a_4a_5} \cdots (F_{\mu_K}^{\,\,\,\mu_1})^{a_{2K}a_{2K+1}} + \mathcal{O}(\varphi).
\label{eq:Z2oddhybrid2}
\end{align}
and generates various $\SO(2K)$-invariant hybrid states containing at least two massive vector bosons and multiple massless gauge bosons or massive vector bosons. All these states belong to the second class.

So far, we investigated various $\SO(N)$-invariant operators that have a nontrivial expansion in terms of $\varphi/v$ to address possible states in the different channels of the model. This provided a mapping to single $\SO(N-1)$-invariant operators either directly via the FMS decomposition or by using the multiplet decomposition additionally. We concentrated on operators with least or almost minimal field content in every channel as we assume that those operators have the largest overlap with the ground state as well as possible next-level excitations. 
As expected, none of the operators assigned to the first class generated a state which is described by an operator only containing fields living in the orthogonal subspace to the direction of the VEV in the broken formulation of the theory. The simplest representative of such an operator would be the $\SO(N-1)$ scalar glueball $\tr F_{\rma}^{2}$. If this operator or any other operator belonging to the second class can indeed be described by an $\SO(N)$-invariant operator, e.g., $\tr F^{2}$, a strict duality can be established between the bound state operators of the $\SO(N)$ gauge theory and the $\SO(N-1)$ model for $N>3$. Again, nonperturbative calculations indicate that states of the second class are not part of the $\mathcal{G}\big({=}\SO(N)\big)$-invariant spectrum but more investigations into this direction are needed to make a final statement.


\begin{table*}[t!]
\begin{center}

\begin{tabular}{c|| ll | ll || lcc}
\toprule

& \multicolumn{2}{c|}{\quad $\SO(N)$ invariant \quad} & \multicolumn{2}{c||}{\quad $\SO(N-1)$ singlets \quad} & \multicolumn{3}{c}{$\SO(N-1)$ multiplets} \\
\hline
$J^P$ & $\Ztwo$\footnote{$\Ztwo$ odd states exist only for $N=2K+1$, see main text.}  &  Operator                & 1. Class  & [2. Class]                 & Field & DOF &  $m_{\mathrm{Field}}^{2}$ \\
\hline

$0^+$ & $+$ & $\phi^{\rmT}\phi$ & $h$ &   & $h$ & 1 & $\lambda v^{2}$ \cr
& $+$ & $\phi^{\rmT}D^{2}\phi$        & $A_{\rmf \mu}^{\rmT}A_{\rmf}^{\mu}$, $h$ &            & & & \cr
& $+$ & $\phi^{\rmT}F^{2}\phi$        & $F_{\rmf \mu\nu}^{\rmT}F_{\rmf}^{\mu\nu}$ &            & & & \cr
& $+$ & $\tr F^{2}$        & [$F_{\rmf \mu\nu}^{\rmT}F_{\rmf}^{\mu\nu}$] & $[\tr F_{\rma}^{2}]$, [$A_{\rmf}^{\rmT}F_{\rma}A_{\rmf}^{\phantom{\,}}$]           & & & \cr
\cline{2-5}
& $-$ & see Eq.\eqref{eq:Z2oddglueball}        &  & [see Eq.\eqref{eq:Z2oddglueball}]           & & & \cr
& $-$ & see Eq.\eqref{eq:Z2oddscalarhybrid}        &  & [see Eq.\eqref{eq:Z2oddscalarhybrid}]           & & & \cr
& $-$ & see Eq.\eqref{eq:Z2oddscalarhybrid2}        & see Eq.\eqref{eq:Z2oddscalarhybrid2}   &         & & & \cr
\hline
$1^-$ & $+$ & $\phi^{\rmT}F^{\mu\nu}D_{\nu}\phi$             & $A_{\rmf\nu}^{\rmT}D^{\nu}A_{\rmf}^{\mu}$ &              & $A_{\rma}^{\mu}$ & $\frac{(N-1)(N-2)}{2}$ & 0 \cr
\cline{2-5}
 & $-$ & see Eq.\eqref{eq:Z2oddhybrid}             & see Eq.\eqref{eq:Z2oddhybrid}  &              & $A_{\rmf}^{\mu}$ & $N-1$ & $\frac{g^{2}v^{2}}{2}$ \cr
 & $-$ & see Eq.\eqref{eq:Z2oddhybrid2}             &   & [see Eq.\eqref{eq:Z2oddhybrid2}]             &  &  &  \cr
 \toprule
\end{tabular}
\caption{Particle content of an $\SO(N>3)$ gauge theory with scalar field in the fundamental representation. 
Left: Comparison between operators/states that are strict invariant with respect to $\SO(N)$ transformations, i.e., observables, and operators/states that one would predict from the conventional but gauge-variant viewpoint of spontaneous gauge symmetry breaking ($\SO(N-1)$ singlets). Trivial scattering states are ignored. Right: Properties of the elementary building blocks obtained from the standard multiplet decomposition after gauge fixing which are used to construct $\SO(N-1)$ singlets.}

\label{tab:SOfun}
\end{center}
\end{table*}


In either case, we obtain nontrivial implications. If it turns out that operators of the second class cannot be described in an $\SO(N)$-invariant way, we have a mismatch between the spectra of the $\SO(N)$ and $\SO(N-1)$ gauge theory, showing that the heuristic picture of spontaneous gauge symmetry breaking is not adequate for BSM scenarios. In the other case, we may obtain nontrivial phenomenological modifications. For instance consider the $\SO(2K)$ scalar glueballs $\tr F_{\rma}^{2}$ and $\epsilon^{\dot{a}_1 \cdots \dot{a}_{2K}} (F_{\rma\mu_1}^{\,\,\,\mu_2})^{\dot{a}_1\dot{a}_2} (F_{\rma\mu_2}^{\,\,\,\mu_3})^{\dot{a}_3\dot{a}_4} \cdots (F_{\rma\mu_K}^{\,\,\,\mu_1})^{\dot{a}_{2K-1}\dot{a}_{2K}}$. 
While from the perspective of the broken theory one state might decay into the other if the mass ratio is sufficiently large, the situation is different and more constrained from the perspective of the unbroken theory. There, the latter glueball operator is described by an $\SO(2K+1)$-invariant operator with odd $\Ztwo$ parity defined in expression \eqref{eq:Z2oddglueball} such that the conservation of an additional quantum number has to be fulfilled in a possible decay process.

The results of the present investigation are summarized in Tab.~\ref{tab:SOfun}. The left part of the table lists $\mathcal{G}$- as well as $\mathcal{H}$-invariant operators and has the intention to provide a transparent presentation of the duality between the two different gauge theories in both directions. From the top down perspective, it can be read off which states are present in the spectrum of the $\mathcal{G}$($=\SO(N)$) gauge theory. From the bottom up perspective, it demonstrates which states of an $\mathcal{H}$($=\SO(N-1)$) gauge theory can be embedded into a model with larger gauge group. In the right part, we list the properties of the elementary fields obtained from the multiplet decomposition. These serve as building blocks for $\mathcal{H}$-invariant quantities. The column DOF lists the number of nontrivial independent real degrees of freedom of the associated multiplet with respect to the internal gauge group but not with respect to the Lorentz group.\footnote{Note that we use a $\mathcal{G}$ covariant embedding of the $\mathcal{H}$ multiplets throughout this paper. For instance, the $\SU(N-1)$ multiplet $A_{\rmf} = A\phi_{0}$ contains $N-1$ nontrivial (Lorentz-vector) degrees of freedom but is described by an $N$ component $\SU(N)$ covariant object.} Thus, a field (either scalar or vector) in a complex representation with multiplicity $N$ will be listed as $2N$.

In the left column, we sort the $\mathcal{G}$-invariant bound state operators with respect to their global quantum numbers. In the next column, we list $\mathcal{H}$-invariant operators and distinguish as to whether they can be extracted from the $\mathcal{G}$-invariant operator in the same row according to the definitions of the different classes defined in Sec.~\ref{sec:FMS}. 
We mainly focus on identifying operators of the first class. For illustration, we list operators of the second class with small field content (two constituents as well as some examples containing three and four constituents), unless for $\Ztwo$ odd or open $\U(1)$ channels where the $\mathcal{G}$-invariant operators as well as their expansion are involved and may contain a larger field content. For all models considered in this work, any $\mathcal{H}$-singlet which consists of finitely many fields can be extracted via the multiplet decomposition. However, this is not generically true, e.g., for theories with scalar fields in three and higher index representations or theories with fermions where the fermions are in lower rank tensor representations than the scalar field.

Due to the indeterminate role of operators of the second class, we put them in square brackets to indicate that they are likely not present in the $\mathcal{G}$-invariant spectrum according to present lattice simulations \cite{Maas:2016ngo,Maas:2018xxu,Maas:2017pcw}. Furthermore, we put an operator of the first class into square brackets if we can extract it from the $\mathcal{G}$-invariant operator in the same row only via the standard gauge-dependent multiplet decomposition. Of course such an operator of the first class has to appear at some other place in the table without square brackets where it can be obtained via a clean projection at a fixed order in the FMS prescription. Otherwise it would not be an operator of the first class. As an example, consider the row of the $\SO(N)$ glueball operator $\tr F^{2}$. Using the conventional decomposition, we obtain the scalar hybrids $F_{\rmf}^{\rmT}F_{\rmf}^{\phantom{a}}$ and $A_{\rmf}^{\rmT}F_{\rma}A_{\rmf}^{\phantom{\,}}$ as well as the glueball $\tr F_{\rma}^{2}$. All of them would be operators of the second class. However, another $\SO(N)$-invariant operator, $\phi^{\rmT}F^{2}\phi$, expands unambiguously to the scalar hybrid $F_{\rmf}^{\rmT}F_{\rmf}^{\phantom{\,}}$ at leading order in the FMS procedure such that the states generated by $F_{\rmf}^{\rmT}F_{\rmf}^{\phantom{\,}}$ are states of the first class. What kind of states are actually generated by an operator like $\tr F^{2}$ is currently unknown in case the $\mathcal{G}$-$\mathcal{H}$ duality of states does not hold for the second class. Although the conventional decomposition is not meaningful in this case, it could be that it still has overlap with the formally contained state of the first class. Equally likely is that it might generate a state of the fourth class which has no dual description in terms of the $\mathcal{H}$ gauge theory.


\begin{table*}[t!]
\begin{center}

\begin{tabular}{c|| ll | lll || lcc}
\toprule

& \multicolumn{2}{c|}{$\SO(3)$ invariant} & \multicolumn{3}{c||}{$\SO(2)$ singlets} & \multicolumn{3}{c}{$\SO(2)$ multiplets}  \\
\hline
$J^P$ &  $\Ztwo$ \,  &  Operator                & 1.Class & & 3.Class                &  Field & DOF &  $m_{\mathrm{Field}}^{2}$   \\
\hline

$0^+$ & $+$ & $\phi^{\rmT}\phi$ & $h$ &  &               & $h$ & 1 & $\lambda v^{2}$  \cr
& $+$ & $\phi^{\rmT}D^{2}\phi$        & $W^{+\mu}W^{-}_{\mu}$, $h$ &            & & \cr
\hline
 $1^-$ & $-$ & $\partial^{\nu}(\epsilon^{a_1a_2a_3}\phi^{a_1}F_{\mu\nu}^{a_2a_3})$         & $A_{\U(1)\perp\mu}$ &            & & $A_{\U(1)\mu}$ & 1 & 0 \cr
\cline{2-6}
 \quad & &         &  &  & $\mathcal{D}W^{\pm}_{\mu}$            & $W^{\pm}_{\mu}$ & $2$ & $\frac{g^{2}v^{2}}{2}$ \cr
\toprule
\end{tabular}
\caption{Particle content of an $\SO(3)$ gauge theory with scalar field in the fundamental representation. 
Left: Comparison between operators/states that are strict invariant with respect to $\SO(3)$ transformations, i.e., observables, and operators/states that one would predict from the conventional but gauge-variant viewpoint of spontaneous gauge symmetry breaking ($\SO(2)$ singlets). Note that states of the third class appear in this specific model and we have removed the column '[2.Class]' as it is not relevant here. The symbol $\mathcal{D}$ indicates a suitable dressing with a Dirac phase factor to obtain observable $\SO(2)$-invariant vector bosons. Trivial scattering states are ignored. Right: Properties of the elementary building blocks obtained from the standard multiplet decomposition after gauge fixing.}

\label{tab:SOfun2}
\end{center}
\end{table*}


At this point, we also want to emphasize that the above given operators for the various channels are merely simple examples to extract information of the ground and excited states of the various models. Their construction is based on the assumption that the operator with least field content has some overlap with the ground and first excited states. By no means this allows for a comprehensive analysis of the mass spectrum which is beyond the scope of this paper. As a first investigation, we will rather try to identify and obtain a first glance on the properties of the low-lying observable states of each channel.

Before we go on to more involved representations of the $\SO(N)$ group, we will briefly discuss a few particularities of the groups $\SO(2)$ and $\SO(3)$. These two groups are special for the following reasons. The group $\SO(2) \cong \U(1)$ is Abelian. Thus, there is no need to consider bound state operators as the transversal part of the gauge field is already gauge invariant and a physical scalar field is given by the elementary scalar field dressed with a nonlocal photon cloud $\exp(-\I g\frac{\partial_{\mu}}{\partial^{2}}A^\mu)$ similar to Dirac's physical electron \cite{Dirac:1955uv,Lavelle:1993wy}. Nonetheless, also the FMS mechanism may be used to describe the physical particle spectrum. It predicts a single particle in the scalar channel described by the properties of the elementary Higgs field as well as a single massive vector particle with mass proportional to $gv$, consistent with the standard description. Also this result is supported by nonperturbative lattice calculations \cite{Woloshyn:2017rhe,Lewis:2018srt}.

The group $\SO(3)$ is particular with respect to the above analysis because it breaks to an Abelian subgroup after gauge fixing. By contrast, the unbroken subgroup is always non-Abelian for $N>3$. Furthermore, $\SO(3) \cong \SU(2)/\mathbb{Z}_{2}$ and thus we expect to get the same gauge-invariant spectrum as in case of the adjoint representation of $\SU(2)$ which contains a massless vector degree of freedom, cf. \cite{Maas:2017xzh}. Indeed, we can construct a gauge-invariant $\SO(3)$ vector operator which expands to the massless gauge boson of the unbroken $\SO(2)$. This is easiest displayed at the level of the field-strength tensors where $\epsilon^{a_1a_2a_3}\phi^{a_1}F_{\mu\nu}^{a_2a_3}$ expands in leading order to $F_{\U(1)}^{\mu\nu}$, the field strength tensor of the unbroken Abelian subgroup, and a $\U(1)$-neutral Lorentz-tensor state given by two massive vector fields.\footnote{Note that both operators can be assigned to the first class as the field strength tensor decomposes naturally to these to terms via the FMS projection and no further multiplet decomposition is required in this particular case.} Here, we used the $\SO(3)$ property that the fundamental representation can be mapped to the adjoint representation, i.e., the adjoint field $\phi^{a_1a_2} = \epsilon^{a_1a_2a_3}\phi^{a_3}$ is the dual field of the fundamental vector and vice versa. The corresponding massless $\SO(3)$-invariant vector particle can be described by $\partial^{\nu}(\epsilon^{a_1a_2a_3}\phi^{a_1}F_{\mu\nu}^{a_2a_3}) \sim A_{\U(1)\perp \mu}$ which maps on the $\U(1)$-invariant transversal part of $A_{\U(1)}$.\footnote{We also obtain a term $\partial^{\nu}(W_{\mu}^{\pm}W_{\nu}^{\mp})$ from the commutator of $F_{\mu\nu}$ that mixes with the vector states. However, the propagator of this object does not show the characteristics of a common vector particle but might describe some nontrivial internal excitation of the corresponding tensor state. As for $\partial_{\mu}\phi$, we do not extend the analysis into that direction.} Thus, $\SO(3)$ is the only non-Abelian $\SO(N)$ group for which an operator can be constructed that expands to a single gauge field. For all other groups this is not possible as no gauge singlet with respect to the unbroken subgroup exists.

At this point we would like to emphasize another particularity of the breaking pattern $\SO(3)\to\SO(2)$. From the conventional perspective of gauge symmetry breaking, all elementary fields become observable particles. This is obvious for the radial Higgs excitation $h$ and the massless $\U(1)$ gauge field $A_{\U(1)}^{\mu}$. The only two remaining elementary fields $A_{\rmf}^{1\mu}$ and $A_{\rmf}^{2\mu}$ are the massive vector fields which are charged under the remaining Abelian subgroup. For convenience we combine them to a complex massive vector boson $W^{\pm} = (A_{\rmf}^{1} \mp \I A_{\rmf}^{2})/\sqrt{2}$ in analogy to the standard model nomenclature. Due to the Abelian nature of the charge, we can define physical states via suitable dressings from the $\SO(2)$ perspective. Nonetheless, there is no gauge-invariant description of these charged particles in a strict $\SO(3)$-invariant formulation. As the FMS mechanism can only provide a mapping to singlets of the unbroken subgroup $\mathcal{H}$, we find only mappings to charge-neutral bound state operators ($\phi^{\rmT}D^{2}\phi \sim A_{\rmf}^{\rmT}A_{\rmf}^{\phantom{a}} = \big((A_{\rmf}^{1})^{2} + (A_{\rmf}^{2})^{2}\big) = 2W^{+}W^{-}$) of the latter two fields but not to a single $\U(1)$-charged particle. Thus, the two $\SO(3)$ gauge-variant states generated by $W^{\pm}$ belong to the third class and have no $\SO(3)$-invariant counterpart in the embedding theory such that they cannot be part of the $\SO(3)$ bound state spectrum. We summarize the particle content of this model in Tab.~\ref{tab:SOfun2}.

\subsection{Antisymmetric second-rank tensor (adjoint) representation}
\label{sec:ON2anti}

After we have analyzed the simplest nontrivial representation for the $\SO(N)$ group, we now focus on the second-rank tensor representations. 
It is convenient to introduce matrix notation, $(\phi)^{ab} = \phi^{ab}$, such that the transformation property of the scalar field reads ${\phi \to R \phi R^{\rmT}}$. 
For the covariant derivative we have $D_{\mu}\phi = \partial_{\mu} \phi + g [A_{\mu},\phi]$. 
In case the scalar field acquires a nonvanishing VEV, we split it into its VEV and a part containing the fluctuations as usual
\begin{align}
 \phi(x) = v \phi_{0} + \varphi(x),
\label{eq:split2ON}
\end{align}
where we use the normalization $\tr (\phi_{0}^{\rmT}\phi_{0}^{\phantom{\rmT}}) =\frac{1}{2}$ and obtain for the mass matrix of the gauge bosons 
\begin{align}
 \frac{1}{2}(\MA^{2})_{ij} A_{i\mu}^{\phantom{m}}A_{j}^{\mu} = g^{2}v^{2}\, \tr \big( [A_{\mu},\phi_0^{\rmT}]  [A^{\mu},\phi_0] \big).
 \label{eq:massmatrix2ndrankON}
\end{align}
Of course, any second-rank $\SO(N)$ tensor can be decomposed into two nontrivial irreducible representations, an antisymmetric tensor and a symmetric traceless tensor (as well as a trace part which is a singlet). 
First, we investigate the antisymmetric second-rank tensor representation, i.e., $\phi^{\rmT} = -\phi$, which coincides with the adjoint for $\SO(N)$.

We consider the following fourth-order scalar potential for the scalar self-interactions
\begin{align}
 V(\phi) &= - \mu^2 \phi^{ab}\phi^{ab} + \frac{\lambda}{2} (\phi^{ab}\phi^{ab})^2 + \frac{\tilde\lambda}{2} \phi^{ab}\phi^{bc}\phi^{cd}\phi^{da} \notag \\
 &= \mu^2 \tr \phi^2 + \frac{\lambda}{2} (\tr \phi^2)^2 + \frac{\tilde\lambda}{2} \tr \phi^4.
 \label{eq:pot2aON}
\end{align}
Note that a possible cubic term $\tr \phi^{3}$ vanishes due to the antisymmetry of $\phi$. 
The field configuration which minimizes the potential \eqref{eq:pot2aON} can always be transformed to a block diagonal standard form as $\phi$ is real and antisymmetric. 
The breaking pattern is governed by the non-isotropic coupling $\tilde\lambda$ \cite{Li:1973mq}.

\subsubsection{\texorpdfstring{$\tilde\lambda > 0$}{\tilde\lambda > 0}}

If $\tilde\lambda > 0$, the direction of the VEV in gauge space is 
\begin{align}
 \phi_{0} = \frac{1}{\sqrt{4K}} \begin{pmatrix} \varepsilon & & \\ & \ddots & \\ & & \varepsilon   \end{pmatrix}
\end{align}
for $N = 2K$ with block-diagonal elements $\varepsilon = \begin{pmatrix} 0 & 1 \\ -1 & 0 \end{pmatrix}$ and all off-diagonal elements vanish, and
\begin{align}
 \phi_{0} = \frac{1}{\sqrt{4K}} \begin{pmatrix} \varepsilon & & & \\ & \ddots & & \\ & & \varepsilon & \\ & & & 0   \end{pmatrix}
\end{align}
for $N = 2K+1$, implying the breaking patterns ${\SO(2K) \to \U(K)}$ and $\SO(2K+1) \to \U(K)$, respectively \cite{Li:1973mq}. Due to the minimization of the potential, we further obtain $\mu^{2} = \frac{1}{4K}(2K\lambda + \tilde\lambda)v^{2}$.

Indeed, the mass matrix of the gauge bosons \eqref{eq:massmatrix2ndrankON} contains $K^{2}$ vanishing eigenvalues. We further have $K(K-1)$ degenerate eigenvalues $m_\mathrm{A_{2a}}^2 = \frac{g^2v^2}{K}$ for all $N$ as well as additional $2K$ gauge bosons with mass $m_\mathrm{A_\rmf}^2 = \frac{g^2v^2}{4K}$ forming a $U(K)$ fundamental vector for $N=2K+1$.
In order to extract the massive vector fields, we can use the operator $A\phi_{0}-\phi_{0}A$. For even $N$, this simple form is sufficient as we have only one massive multiplet given by a complex antisymmetric second-rank tensor of the remaining unbroken $\U(K)$ subgroup, $A_{2\rma} = A\phi_{0}-\phi_{0}A$ where the subscript $2\rma$ indicates second-rank antisymmetric. For odd $N$, we have to define more sophisticated projections to distinguish the different subspaces corresponding to the different multiplets with nonvanishing mass parameter. We use $A_{2\rma} = 16K^{2} \phi_{0}^{2}[A,\phi_{0}]\phi_{0}^{2} = 16K^{2} [\phi_{0}^{2}A\phi_{0}^{2},\phi_{0}]$ and $A_{\rmf} = [A,\phi_{0}] - 16K^{2} \phi_{0}^{2}[A,\phi_{0}]\phi_{0}^{2}$ to project on the fields transforming as the antisymmetric second-rank tensor and the fundamental vector of $\U(K)$, respectively. Note, that $[\phi_{0}^{2}A\phi_{0}^{2},\phi_{0}] \sim [A,\phi_{0}]$ for $N=2K$.
The massless gauge bosons transforming under the adjoint representation of the remaining unbroken $\U(K)$ gauge group can be extracted by considering the anticommutator $\{\phi_{0}A\phi_{0},\phi_{0}\}$. This expression simplifies to $\{A,\phi_{0}\}$ for even $N$. The massless field proportional to the direction of the VEV is the gauge boson associated to the $\U(1)$ generator while the remaining degrees of freedom form the gauge bosons of $\SU(K)$, $A_{\rma} = \{\phi_{0}A\phi_{0},\phi_{0}\} - 2\tr(\phi_{0}A)\phi_{0}$.

Correspondingly, the elementary scalar spectrum contains either $K^2-K$ ($N=2K$) or $K^2-K+2K$ ($N=2K+1$) would-be Goldstone bosons stored in $[\varphi,\phi_{0}]$ which we remove by the unitary gauge condition. Further, we have the radial scalar excitation $h(x) = 2\,\tr (\phi_{0}^{\rmT}\varphi)$ which is proportional to the direction of the VEV and transforms as a singlet with respect to the unbroken subgroup $\U(K)$ with mass $\mh^2 = \big(\lambda + \frac{\tilde\lambda}{2K}\big) v^2$. The remaining $K^{2}-1$ components of $\varphi$ are degenerate with mass parameter $m_\mathrm{h_{\rma}}^2 = \frac{\tilde\lambda}{2K} v^2$ and transform according to the adjoint representation of the non-Abelian part of the unbroken subgroup $\SU(K)$. They can be extracted from $h_{\rma} = \{\varphi-h\phi_{0},\phi_{0}\}$ in the unitary gauge.

In the conventional picture of spontaneous gauge symmetry breaking, observable states of the model are described with the aid of $\SU(K)$-invariant operators. Thus, we would expect a singlet Higgs $h$, a massless gauge boson $A_{\U(1)} = 2\tr(\phi_{0}A)$ of the unbroken $\U(1)$, and meson operators build from two Higgs fields transforming as adjoint fields $\tr (h_{\rma}^{2})$ or two massive vector fields $\tr (A_{2\rma}^{2})$ for all $N$. Further, we have the scalar $\SU(K)$ glueball $\tr (F_{\rma}^{2})$, hybrid states which read schematically $\tr(F_{\rma}h_{\rma})$, $\tr(F_{\rmasym}A_{\rmasym})$, $\tr(F_{\rmasym}^{2})$, and $\tr(A_{2\rma}F_{\rma}A_{2\rma})$, as well as baryonic-like operators containing three massive matter fields, e.g., $\tr(A_{2\rma}h_{\rma}A_{2\rma})$, as well as others. For odd $N$, we have additional mesons, hybrids, and baryons containing the massive vector field $A_{\rmf}$, e.g., $\tr(A_{\rmf}^{2})$, $\tr(A_{\rmf}F_{\rma}A_{\rmf})$, $\tr(F_{\rmf}A_{\rmf})$, $\tr(A_{\rmf}h_{\rma}A_{\rmf})$, and $\tr(A_{\rmf}A_{\rmasym}A_{\rmf})$. Once more, these states are gauge-variant quantities with respect to the original gauge symmetry $\SO(N)$ and can meaningfully be defined only in a suitable chosen gauge with nonvanishing VEV.

In order to classify the strict gauge-invariant spectrum of the $\SO(N)$ gauge theory without the notation of gauge symmetry breaking, we first note that the action of the model has a global $\mathbb{Z}_{2}$ symmetry. 
Again, it is straightforward to write down a scalar bound state operator that expands to the Higgs singlet with respect to the $\SU(K)$ subgroup in the FMS approach. This operator is $\Ztwo$ even and reads
\begin{align}
 \tr (\phi^{\rmT}\phi) &= \frac{v^2}{2} + v h + \tr(\varphi^{\rmT}\varphi) \notag \\
 &= \frac{v^2}{2} + v h + \frac{1}{2}h^2 - \tr h_{\rma}^{2}.
\label{eq:ZevenscalarON}
\end{align}
The FMS expansion on the right-hand side shows, that the gauge-invariant operator $\tr (\phi^{\rmT}\phi)$ can be described by the elementary field $h$ in leading order. At next-to-leading order in the fluctuation fields an intricate situation appears which we have already sketched in Sec.~\ref{sec:validity}. Usually, we would expect a trivial scattering behavior from the term $\tr\varphi^{2}$. However, using the decomposition of $\varphi$ into $\U(K)$ multiplets, i.e., $h$ and $h_{\rma}$, which we provided in the second line of Eq.~\eqref{eq:ZevenscalarON}, the propagator is described by the trivial scattering state at twice the mass of $h$ as well as the propagator of the $\U(K)$ invariant meson operator $\tr (h_{\rma}^{2})$ at next-to-leading order.

Unfortunately, more sophisticated nonperturbative methods are required to make a final statement under which circumstances the mass of the $\U(K)$-invariant meson operator $\tr (h_{\rma}^{2})$ is indeed contained in the spectrum of the scalar channel of the $\SO(N)$ gauge theory. According to the classification of operators of Sec.~\ref{sec:FMS}, this operator formally belongs to the second class. There is no $\SO(N)$-invariant operator build from $\phi$ that expands solely to $\tr (h_{\rma}^{2})$ on the right-hand side via the split \eqref{eq:split2ON} which would provide a natural ordering principle. We can only conclude that the nontrivial next-to-leading order of $\tr\phi^{2}$ produces a formal superposition of $h^{2}$ and $\tr (h_{\rma}^{2})$ which cannot be disentangled as there is no meaningful definition of the latter from the perspective of the original $\SO(N)$ gauge theory. Thus, we would only get a scattering state.

Further, we can investigate various other $\SO(N)$-invariant operators that have the same quantum numbers but provide a mapping to different $\U(K)$-invariant states. 
A careful distinction between even and odd $N$ is required in order to classify the states as to whether they follow unambiguously from the FMS expansion due to the expansion in $\varphi/v$ or can only be obtained by a decomposition of the involved multiplets.

As a first example consider the operator 
\begin{align}
 \tr (\phi D^{2}\phi) = -g^{2}v^{2} \tr ([A,\phi_{0}]^{2}) +\mathcal{O}(\varphi). 
\label{eq:ZevenscalarON2}
\end{align}
For $N=2K$, we obtain in leading order an $\SO(N)$-invariant description of the $\U(K)$ meson operator $\tr (A_{2\rma}^{2})$. At next-to-leading order, also the pole structure generated by the radial Higgs excitation will be generated at the level of the propagator of $\tr (\phi D^{2}\phi)$ as in the fundamental case. However, the state $h$ can only be extracted via the multiplet decomposition. Whether $\phi^{\rmT}D^{2}\phi$ has overlap with $h$, which can be extracted from another operator without requiring the multiplet decomposition, is an open issue. Further, we find several scattering states at $\mathcal{O}(\varphi)$ and $\mathcal{O}(\varphi^{2})$ as well as operators of the second class, e.g., containing an adjoint Higgs field and two massive multiplets, $\tr(A_{\rmasym}h_{\rma}\bar{A}_{\rmasym})$ where ${\bar{A}_{\rmasym/\rmf} \equiv [\phi_{0},A_{\rmasym/\rmf}]}$.\footnote{Note that we used the freedom to express the fields $A_{\rmasym/\rmf}$ which transform according to a complex representation of $\U(K)$ in terms of multiplets of real fields embedded in an $\SO(N)$ covariant way for convenience. The presence of the bared field within an operator as $\tr(A_{\rmasym}h_{\rma}\bar{A}_{\rmasym})$ ensures that the real components get combined in an appropriate way to form a meaningful $\U(K)$-invariant composite operator. Nonetheless, the bar notation here should not be confused with the Hermitian conjugate of the corresponding operator when retranslating to complex $\U(K)$ multiplets as both operations are not equivalent.}

For odd $N$, the situation is more intricate as the field operator $[A_{\mu},\phi_{0}]$ contains two different massive multiplets. Thus, we rely on the standard multiplet decomposition to identify the resulting two different meson operators $\tr (A_{2\rma}^{2})$ and $\tr (A_{\rmf}^{2})$ as they are interlinked in the strict gauge-invariant formulation and cannot be disentangled by using only the split \eqref{eq:split2ON}. 
Nonetheless, it is possible to find a strict gauge-invariant formulation of the former. The operator $\tr\big( \phi(D_{\mu}\phi)\phi D^{\mu}\phi \big)$ provides an unambiguous mapping to the $\U(K)$-invariant meson operator $\tr (A_{2\rma}^{2})$ at leading order in the expansion such that it belongs to the first class. Such a construction is not possible for the meson operator $\tr (A_{\rmf}^{2})$ as the components of $A_{\rmf}$ live in an orthogonal subspace to $\phi_{0}$ such that we always have to rely on the standard multiplet decomposition to find an $\SO(N)$-invariant description. Analyzing the operator $\tr\big( \phi(D_{\mu}\phi)\phi D^{\mu}\phi \big)$ at next-to-leading order in the FMS expansion, we also find the $\U(K)$-invariant operator generating the baryonic-type state $\tr(A_{\rmasym}h_{\rma}\bar{A}_{\rmasym})$. We assign this operator to the second class as we can only extract it from $\tr\big( \phi(D_{\mu}\phi)\phi D^{\mu}\phi \big)$ or any other $\SO(N)$-invariant operator we have investigated if we use the conventional multiplet decomposition. Otherwise, we cannot disentangle it from the trivial scattering state given by $h$ and $\tr(A_{\rmasym}^{2})$. 
This is not a surprise as $h_{\rma}$ is orthogonal to $\phi_{0}$ such that every operator containing this field and two massive vector fields can only be extracted via the multiplet decomposition.

In order to extend a variational analysis of the model, it might be useful to consider another $\Ztwo$ even operator, e.g., $\tr(\phi^{2}(D_{\mu}\phi)D^{\mu}\phi)$. It contains the meson $\tr A_{\rmasym}^{2}$ at leading order and the baryonic-type states $\tr(A_{\rmf}h_{\rma}A_{\rmf})$ (only for odd $N$) and $\tr(A_{\rmasym}h_{\rma}A_{\rmasym})$ at next-to-leading order. All can only be extracted via the multiplet decomposition at the respective orders of the FMS expansion.

The scalar $\SU(K)$ glueball operator $\tr  F_{\rma}^{2}$ can be extracted from $\SO(N)$-invariant operators only via the multiplet decomposition. Thus, it belongs always to the second class. As usual, we might consider the operator $\tr F^{2} $ which contains $\tr F_{\rmasym}^{2}$ and $\tr(A_{\rmasym}F_{\rma}A_{\rmasym})$ for all $N$ as well as $\tr F_{\rmf}^{2}$, $\tr(A_{\rmf}F_{\rma}A_{\rmf})$, $\tr(A_{\rmf}\bar{F}_{\rmasym}A_{\rmf})$, and $\tr(F_{\rmf}A_{\rmf}\bar{A}_{\rmasym})$ for odd $N$ as well. In order to enlarge the operator basis, it is worthwhile to consider further operators, e.g., $\tr\big((D_{\mu}\phi)F^{\mu\nu}D_{\nu}\phi\big)$. At leading order, the FMS expansion yields an $\U(K)$-invariant operator that decomposes into the hybrid $\tr(A_{\rmasym}F_{\rma}A_{\rmasym})$ as well as scattering states such as $A_{\U(1)}$ and $\tr A_{\rmasym}^{2}$ for even $N$. Note that for $K=2$ the hybrid operator $\tr(A_{\rmasym}F_{\rma}A_{\rmasym})$ vanishes as the second-rank antisymmetric tensor representation transforms trivial with respect to the non-Abelian subgroup. For odd $N$, we additionally obtain the hybrid $\tr(A_{\rmf}F_{\rma}A_{\rmf})$ as well as more exotic operators, e.g., $\tr(A_{\rmf}\bar{F}_{\rmasym}A_{\rmf})$ at leading order via the multiplet decomposition.

Studying the vector channel, it is straightforward to find a gauge-invariant description in terms of the original gauge symmetry for the only vector singlet state in the elementary spectrum after gauge fixing. This is the massless gauge boson associated with the $\U(1)$ generator of $\SU(K)\times \U(1)$. 
Considering the $\Ztwo$ odd operator $\tr (F^{\mu\nu}\phi) = v F^{\mu\nu}_{\U(1)} + gv\, \tr([A_{\tilde{m}}^{\mu},A_{\tilde{m}}^{\nu}]\phi_{0}) + \tr (F^{\mu\nu}\varphi)$ where $\tilde{m} \in \{\rmasym\}$ for $N=2K$ and $\tilde{m} \in \{\rmasym, \rmf\}$ for $N=2K+1$ indicates the massive multiplet(s), we find that the leading order contribution is precisely the field strength tensor whose generator is proportional to $\phi_{0}$ and thus the associated gauge boson is massless for any orthogonal group.\footnote{As in the case of a fundamental scalar field for $N=3$, the leading order contribution is a linear combination of a standard Abelian field strength tensor and the commutator of the massive vector multiplet(s) without relying on the multiplet decomposition. Thus, $F_{\U(1)}$ belongs to the first class. For $N=2K$, we can assign $\tr([A_{\rmasym}^{\mu},A_{\rmasym}^{\nu}]\phi_{0}) = \tr(\bar{A}_{\rmasym}^{\mu}A_{\rmasym}^{\nu})$ to the first class as well. For $N=2K+1$, we need the multiplet decomposition to identify $\tr(\bar{A}_{\tilde{m}}^{\mu}A_{\tilde{m}}^{\nu}) = \tr(\bar{A}_{\rmasym}^{\mu}A_{\rmasym}^{\nu}) + \tr(\bar{A}_{\rmf}^{\mu}A_{\rmf}^{\nu})$ and both operators belong to the second class. Nevertheless, they will mix only with the vector channel due to the quantum numbers of $\partial^{\mu}$, thus describing at most internal excitations of the tensor states.} As $\tr(D_{\mu}\phi) = 0$, the simplest vector operator and its FMS expansion reads
\begin{align}
 \partial_{\nu} \tr (F^{\nu\mu}\phi) &= v \partial^{2}  A^{\mu}_{\perp \U(1)} + v \partial_{\nu}\tr(A_{\tilde{m}}^{\nu}\bar{A}_{\tilde{m}}^{\mu}) +\mathcal{O}(\varphi)
\label{eq:Zevena2ONvector}
\end{align}
where $A^{\mu}_{\perp \U(1)}$ is the transversal part of the gauge field $A^{\mu}_{\U(1)}$. 
Thus, the FMS formalism predicts a massless state in the $\Ztwo$-odd vector channel. Such a result of a massless vector bound state can also be obtained in case the scalar field is in the adjoint representation of an $\SU(N)$ gauge theory, see \cite{Maas:2017xzh} or App.~\ref{app:SUNadj}, or the fundamental case for $\SO(3)$. Nonetheless, it is unexpected. Investigating the long-range effective degrees of freedom of a non-Abelian gauge theory, i.e., bound states, one would naively expect that they are massive at least due to radiative corrections unless a symmetry dictates that a mass term vanishes. Due to the duality of states of the $\SO(N)$ with the $\U(\lfloor N/2 \rfloor)$ gauge theory, the mass term of the $\SO(N)$ vector operator~\eqref{eq:Zevena2ONvector} vanishes as the mass of the Abelian gauge boson is protected by the remaining unbroken gauge structure.

How this fact translates into a pure gauge-invariant description is unexplored. It might be the case that the $\SO(N)$-invariant low energy effective theory of the bound states\footnote{Not to be confused with the $\SO(N)$-variant gauge theory $\U(\lfloor N/2 \rfloor)$ after gauge fixing.} obeys an emergent gauge symmetry and the operator~\eqref{eq:Zevena2ONvector} turns out to be the corresponding gauge boson. This could be realized if the system has an intact $\Ztwo$ symmetry and states with the same mass appear in both, $\Ztwo$ even and odd, channels. The corresponding two independent operators can be combined to form a complex object which might couple to the massless vector operator in the correct way to form an Abelian gauge theory. To check whether such a scenario is possible is clearly beyond the scope of this work. At least, we will see in the following that the FMS mechanism predicts composite states with identical constituents at leading order in different $\Ztwo$ channels. Furthermore, the nontrivial prediction of the FMS mechanism that the vector channel contains indeed a massless state for the $\SU(2)$ adjoint case is supported by lattice simulations \cite{Lee:1985yi,Afferrante:2019vsr}.

Considering the operator~\eqref{eq:Zevena2ONvector} at next-to-leading order in the FMS expansion, we may use the conventional multiplet decomposition to extract the trivial scattering state at $\mh + m_\mathrm{A_{\U(1)}} = \mh$ and the mixing with possible hybrid bound state formed by the massless non-Abelian gauge bosons $A_{\rma}$ and the Higgs fields $h_{\rma}$, $\tr (F_{\rma}h_{\rma})$.

Further $\Ztwo$ odd operators in the vector channel and their FMS expansion read,
\begin{align}
 \tr (F^{\mu\nu}D_{\nu}\phi) &= gv\, \tr (F^{\mu\nu}[A_{\nu},\phi_{0}]) +\mathcal{O}(\varphi), 
 \label{eq:Zevena2ONvector2} \\
 \tr (\phi F^{\mu\nu}\phi D_{\nu}\phi) &= gv^{3}\, \tr (F^{\mu\nu}A_{\rmasym \nu}^{\phantom{a}}) +\mathcal{O}(\varphi).
 \label{eq:Zevena2ONvector3}
\end{align}
The operators defined on the left-hand side of Eq.~\eqref{eq:Zevena2ONvector2} and Eq.~\eqref{eq:Zevena2ONvector3} expand unambiguously to the vector hybrid $\tr (\bar{F}_{\rmasym}^{\mu\nu}A_{\rmasym \nu}^{\phantom{a}})$ at leading order in the FMS formalism for even $N$. Therefore, $\tr (\bar{F}_{\rmasym}^{\mu\nu}A_{\rmasym \nu}^{\phantom{a}})$ belongs to the first class. Assuming that the constituent model is valid for this hybrid bound state, the mass will be approximately $2 m_\mathrm{A_{2\rma}} = 2gv/\sqrt{K}$. For $N$ odd, we have to use the standard multiplet decomposition to get $\tr (F^{\mu\nu}[A_{\nu},\phi_{0}]) = \tr (\bar{F}_{\rmasym}^{\mu\nu}A_{\rmasym \nu}^{\phantom{\,}}) + \tr (\bar{F}_{\rmf}^{\mu\nu}A_{\rmf \nu}^{\phantom{\,}})$ in Eq.~\eqref{eq:Zevena2ONvector2}. The mass of the latter can be approximated by $2 m_\mathrm{A_{\rmf}} =  gv/\sqrt{K}$ within the simple constituent model. We also need the multiplet decomposition to obtain single $\U(K)$-invariant operators in Eq.~\eqref{eq:Zevena2ONvector3} for odd $N$, $\tr(F^{\mu\nu}A_{\rmasym \nu}^{\phantom{a}}) = (\bar{F}_{\rmasym}^{\mu\nu}A_{\rmasym \nu}^{\phantom{\,}}) + 2g \tr(A_{\rmf}^{\mu}A_{\rmf}^{\nu}A_{\rmasym\nu})$. As we do not find an $\SO(2K+1)$-invariant operator from which we obtain $\tr(\bar{F}_{\rmasym}^{\mu\nu}A_{\rmasym \nu}^{\phantom{\,}})$ or $\tr(\bar{F}_{\rmf}^{\mu\nu}A_{\rmf \nu}^{\phantom{\,}})$ via the FMS expansion, we assign both operators to the second class. The situation is different for $\tr(A_{\rmf}^{\mu}A_{\rmf}^{\nu}A_{\rmasym\nu})$. An $\SO(2K+1)$-invariant operator with three covariant derivatives can be found that contains this baryonic state at leading order in the expansion.

The simplest operator containing three covariant derivatives acting on three scalar fields, $\tr \big((D^{\mu}\phi)(D^{\nu}\phi)D_{\nu}\phi \big)$, vanishes due to the antisymmetry of the adjoint representation. However, considering
\begin{align}
 &\tr \big(\phi^{2}(D^{\mu}\phi)(D^{\nu}\phi)D_{\nu}\phi \big) \notag \\
 &\quad = g^{3}v^{5}\, \tr (\phi_{0}^{2}[A^{\mu},\phi_{0}][A^{\nu},\phi_{0}][A_{\nu},\phi_{0}]) +\mathcal{O}(\varphi) \notag \\
 &\quad = -\frac{g^{3}v^{5}}{4K} \tr (A_{\rmf}^{\mu} A_{\rmf}^{\nu} A_{\rmasym \nu}) +\mathcal{O}(\varphi),
\end{align}
we find an $\SO(2K+1)$-invariant description of the baryonic-type operator $\tr (A_{\rmf}^{\mu} A_{\rmf}^{\nu} A_{\rmasym \nu})$ at leading order. For $N=2K$ the leading order contribution vanishes.

In order to investigate the $\Ztwo$ even vector channel, we first consider the operator
\begin{align}
  \tr (\phi F^{\mu\nu}D_{\nu}\phi)
  = g v^{2} \tr( \phi_{0} F^{\mu\nu} [A_{\nu},\phi_{0}]) + \mathcal{O}(\varphi).
\label{eq:Zeven2aON}
\end{align}
For even $N$, the leading order contribution is governed by two massive vector fields of the $\SU(K)$ antisymmetric second-rank tensor and a massless adjoint $\SU(K)$ gauge boson forming the hybrid $\tr (F_{\rmasym}^{\mu\nu}A_{\rmasym \nu}^{\phantom{\mu}})$. This hybrid belongs to the first class as it follows directly from the split \eqref{eq:split2ON}. Further, we get several higher order excitations given by $\U(K)$ invariant composite operators which contain three, four, and five elementary fields that may form more involved hybrids as well as scattering states. 
We would like to emphasize that the field content of the hybrids $\tr (F_{\rmasym}^{\mu\nu}A_{\rmasym \nu}^{\phantom{\,}})$ and $\tr (\bar{F}_{\rmasym}^{\mu\nu}A_{\rmasym \nu}^{\phantom{\,}})$ is the same but the internal dynamics how the $\U(K)$-invariant operator is formed is different.\footnote{Retranslating $A_{\tilde{m}}$ to complex $\U(K)$ multiplets denoted by $a_{\tilde{m}}$ with field strength $f_{\tilde{m}}$, the operators $\tr (F_{\tilde{m}}^{\mu\nu}A_{\tilde{m} \nu}^{\phantom{\,}})$ and $\tr (F_{\tilde{m}}^{\mu\nu}\bar{A}_{\tilde{m} \nu}^{\phantom{\,}})$ are expressed as $\tr (f_{\tilde{m}}^{\dagger\mu\nu}a_{\tilde{m} \nu}^{\phantom{\,}} + a_{\tilde{m} \nu}^{\dagger}f_{\tilde{m}}^{\mu\nu})$ and $\tr (f_{\tilde{m}}^{\dagger\mu\nu}a_{\tilde{m} \nu}^{\phantom{\,}} - a_{\tilde{m} \nu}^{\dagger}f_{\tilde{m}}^{\mu\nu})$, i.e., the real and imaginary part of $\tr (f_{\tilde{m}}^{\dagger\mu\nu}a_{\tilde{m} \nu}^{\phantom{\,}})$, respectively.} Thus, we find indeed that $\SO(N)$-invariant operators with different global $\Ztwo$ quantum numbers expand in leading order to $\U(K)$-invariant states with the same field content. Furthermore, these two states in the different channels have the same mass as they are indeed the real and imaginary part of the corresponding $\U(K)$-invariant operator and similar for operators with larger field content.

For $N$ odd, the $\Ztwo$-even operator \eqref{eq:Zeven2aON} projects on composite $\U(K)$-invariant operators containing all massive gauge multiplets which can only be disentangled if we assume that the $\mathcal{G}$-$\mathcal{H}$ duality can be applied to operators of the second class. Using the standard decomposition, we obtain the hybrid states $\tr(F_{\rmasym}^{\mu\nu}A_{\rmasym \nu}^{\phantom{\mu}})$, $\tr(F_{\rmf}^{\mu\nu}A_{\rmf \nu}^{\phantom{\mu}})$, and $\tr(A_{\rmf}^{\mu}A_{\rmf}^{\nu}\bar{A}_{\rmasym\nu})$ at leading order. 
Likewise we rely on the multiplet decomposition if we want to extract single $\U(K)$-invariant operators from the $\Ztwo$ even version of the operator defined in Eq.~\eqref{eq:Zevena2ONvector3},
\begin{align}
 \tr (\phi F^{\mu\nu}\phi^{2} D_{\nu}\phi) = gv^{4}\, \tr (\phi_{0} F^{\mu\nu}\phi_{0}^{2} [A_{\nu},\phi_{0}]) + \cdots.
\end{align}
The leading order term is governed by $\tr(F_{\rmasym}^{\mu\nu}A_{\rmasym \nu}^{\phantom{\,}})$ and $\tr (A_{\rmf}^{\mu} A_{\rmf}^{\nu} \bar{A}_{\rmasym \nu})$ after using the multiplet decomposition. 
However, considering the operator 
\begin{align}
 \! \tr \big(\phi(D^{\mu}\phi)(D^{\nu}\phi)D_{\nu}\phi \big) = g^{3}v^{4}\, \tr (A_{\rmf}^{\mu} A_{\rmf}^{\nu} \bar{A}_{\rmasym \nu}) +\!\cdots\!,
\end{align}
we find that at least the baryonic operator $\tr (A_{\rmf}^{\mu} A_{\rmf}^{\nu} \bar{A}_{\rmasym \nu})$ belongs to the first class for $N=2K+1$.\footnote{For even $N$ the leading order contribution vanishes.} Note that an operator with equivalent constituents and belonging to the first class is also present in the $\Ztwo$ odd vector channel.


\begin{table*}[t!]
\begin{center}

\begin{tabular}{c|| ll | ll || lcc}

\toprule

& \multicolumn{2}{c|}{\quad $\SO(2K)$ invariant \quad} & \multicolumn{2}{c||}{\quad $\U(K)$ singlets \quad} & \multicolumn{3}{c}{$\U(K)$ multiplets} \\
\hline
$J^P$ & $\Ztwo$ \, & Operator                & 1. Class  & [2. Class]                &  Field  & DOF &  $m_{\mathrm{Field}}^{2}$ \\
\hline

$0^+$ & $+$ & $\tr\phi^{2}$              & $h$ & [$\tr h_{\rma}^{2}$]                                                                           & $h$  & 1 & $(\lambda + \frac{\tilde\lambda}{2K})v^{2}$ \cr
& $+$ & $\tr (\phi D^{2}\phi)$        & $\tr A_{\rmasym}^{2}$, [$h$]  & [$\tr(A_{\rmasym}h_{\rma}\bar{A}_{\rmasym})$]             & $h_{\rma}$ & $K^{2}-1$  & $\frac{\tilde\lambda}{2K}v^{2}$ \cr
& $+$ & $\tr\big( \phi(D_{\mu}\phi)\phi D^{\mu}\phi \big)$ 		& $\tr A_{\rmasym}^{2}$ & [$\tr(A_{\rmasym}h_{\rma}\bar{A}_{\rmasym})$]            &  & & \cr
& $+$ & $\tr F^{2}$        &  & $[\tr F_{\rma}^{2}$], [$\tr F_{\rmasym}^{2}$], [$\tr(A_{\rmasym}F_{\rma}A_{\rmasym})$]          & & & \cr
& $+$ & $\tr\big((D_{\mu}\phi)F^{\mu\nu}D_{\nu}\phi\big)$        &  & [$\tr(A_{\rmasym}F_{\rma}A_{\rmasym})$]           & & & \cr
\cline{2-5}
& $-$ & $\tr\big( (D_{\mu}F^{\mu\nu})D_{\nu}\phi \big)$        &  & [$\tr(A_{\rmasym \nu}^{\vphantom{\mu}}D_{\mu}\bar{F}_{\rmasym}^{\mu\nu})$]            & & & \cr
\hline
$1^-$ & $+$ & $\tr (\phi F^{\mu\nu}D_{\nu}\phi)$             & $\tr(F_{\rmasym}^{\mu\nu}A_{\rmasym \nu}^{\phantom{\mu}})$ & [$\tr(F_{\rma}^{\mu\nu}D_{\nu}h_{\rma}^{\vphantom{\mu}})$]             & $A_{\rma}^{\mu}$ & $K^{2}-1$ & 0 \cr
 \cline{2-5}
 & $-$ & $\partial_{\nu} \tr (F^{\mu\nu}\phi)$        & $A^{\mu}_{\perp \U(1)}$   &            & $A_{\U(1)}^{\mu}$ & 1 & 0\cr
 & $-$ & $\tr (F^{\mu\nu}D_{\nu}\phi)$        & $\tr(\bar{F}_{\rmasym}^{\mu\nu}A_{\rmasym \nu}^{\vphantom{\mu}})$ & [$\tr(F_{\rma}^{\mu\nu}D_{\nu}h_{\rma}^{\vphantom{\mu}})$]         & $A_{\rmasym}^{\mu}$ & $K(K-1)$ & $\frac{g^{2}v^{2}}{K}$ \cr
\toprule
\end{tabular}\\[0.5cm]

\begin{tabular}{c|| ll | ll || lcc}

\toprule

& \multicolumn{2}{c|}{\quad $\SO(2K+1)$ invariant \quad} & \multicolumn{2}{c||}{\quad $\U(K)$ singlets \quad} & \multicolumn{3}{c}{$\U(K)$ multiplets} \\
\hline
$J^P$ &  $\Ztwo$ \,  & Operator \,               & 1. Class \, & [2. Class] \,                &  Field  & DOF &  $m_{\mathrm{Field}}^{2}$ \\
\hline

$0^+$ & $+$ & $\tr\phi^{2}$              & $h$ & [$\tr h_{\rma}^{2}$]                                                                           & $h$  & 1 & \hspace{-0.2cm}$\big(\lambda + \frac{\tilde\lambda}{2K}\big) v^2$ \cr
& $+$ & $\tr (\phi D^{2}\phi)$        & [$\tr A_{\rmasym}^{2}$], [$h$] &  [$\tr A_{\rmf}^{2}$], [$\tr(A_{\rmasym}h_{\rma}\bar{A}_{\rmasym})$], [$\tr(A_{\rmf}h_{\rma}\bar{A}_{\rmf})$]            & $h_{\rma}$ & $K^{2}-1$  & $ \frac{\tilde\lambda}{2K} v^{2}$ \cr
& $+$ & $\tr\big( \phi(D_{\mu}\phi)\phi D^{\mu}\phi \big)$ 		& $\tr A_{\rmasym}^{2}$ & [$\tr(A_{\rmasym}h_{\rma}\bar{A}_{\rmasym})$]            &  & & \cr
& $+$ & $\tr(\phi^{2}(D_{\mu}\phi)D^{\mu}\phi)$ 			& [$\tr A_{\rmasym}^{2}$]  & [$\tr A_{\rmf}^{2}$], [$\tr(A_{\rmasym}h_{\rma}\bar{A}_{\rmasym})$], [$\tr(A_{\rmf}h_{\rma}\bar{A}_{\rmf})$]            &  & & \cr
& $+$ & $\tr F^{2}$        &  & [$\tr F_{\rma}^{2}$], [$\tr F_{\rmasym}^{2}$], [$\tr F_{\rmf}^{2}$], [$\tr (A_{\rmasym} F_{\rma} A_{\rmasym})$],           & & & \cr
& &        &  & [$\tr (A_{\rmf} F_{\rma} A_{\rmf})$], [$\tr(F_{\rmf}A_{\rmf}\bar{A}_{\rmasym})$], [$\tr(A_{\rmf}A_{\rmf}\bar{F}_{\rmasym})$]           & & & \cr
& $+$ & $\tr\big((D_{\mu}\phi)F^{\mu\nu}D_{\nu}\phi\big)$        &  & [$\tr(A_{\rmf}F_{\rma}A_{\rmf})$], [$\tr(A_{\rmasym}F_{\rma}A_{\rmasym})$], [$\tr(A_{\rmf}A_{\rmf}\bar{F}_{\rmasym})$]           & & & \cr
\cline{2-5}
& $-$ & $\tr\big( (D_{\mu}F^{\mu\nu})D_{\nu}\phi \big)$        &  & [$\tr(A_{\rmasym \nu}^{\vphantom{\mu}}D_{\mu}\bar{F}_{\rmasym}^{\mu\nu})$], [$\tr(A_{\rmf \nu}^{\vphantom{\mu}}D_{\mu}\bar{F}_{\rmf}^{\mu\nu})$]            & & & \cr
& $-$ & $\tr\big( \phi F^{\mu\nu}(D_{\mu}\phi)D_{\nu}\phi \big)$         &   & [$\tr(A_{\rmf}A_{\rmf}F_{\rmasym})$], [$\tr(F_{\rmf}A_{\rmf}A_{\rmasym})$]         & & & \cr
\hline
$1^-$ & $+$ & $\tr (\phi F^{\mu\nu}D_{\nu}\phi)$             & [$\tr (A_{\rmf}^{\mu} A_{\rmf}^{\nu\vphantom{\mu}} \bar{A}_{\rmasym \nu}^{\vphantom{\mu}})$] & [$\tr(F_{\rmasym}^{\mu\nu}A_{\rmasym \nu}^{\vphantom{\mu}})$], [$\tr(F_{\rmf}^{\mu\nu}A_{\rmf \nu}^{\vphantom{\mu}})$], [$\tr(F_{\rma}^{\mu\nu}D_{\nu}h_{\rma}^{\vphantom{\mu}})$]              & $A_{\rma}^{\mu}$ & $K^{2}-1$ & 0 \cr
 & $+$ & $\tr (\phi F^{\mu\nu}\phi^{2}D_{\nu}\phi)$             & [$\tr (A_{\rmf}^{\mu} A_{\rmf}^{\nu\vphantom{\mu}} \bar{A}_{\rmasym \nu}^{\vphantom{\mu}})$]  & [$\tr(F_{\rmasym}^{\mu\nu}A_{\rmasym \nu}^{\vphantom{\mu}})$]               & $A_{\U(1)}^{\mu}$ & 1 & 0  \cr       
  & $+$ & $\tr \big(\phi(D^{\mu}\phi)(D^{\nu}\phi)D_{\nu}\phi \big)$        & $\tr (A_{\rmf}^{\mu} A_{\rmf}^{\nu\vphantom{\mu}} \bar{A}_{\rmasym \nu}^{\vphantom{\mu}})$ & [$\tr (A_{\rmasym}^{\mu} (D_{\nu}h_{\rma}) \bar{A}_{\rmasym}^{\nu\vphantom{\mu}})$]         & $A_{\rmasym}^{\mu}$ & $K(K-1)$ & $\frac{g^{2}v^{2}}{K}$ \cr
 \cline{2-5}
  & $-$ & $\partial_{\nu} \tr (F^{\mu\nu}\phi)$        & $A^{\mu}_{\perp \U(1)}$ &             & $A_{\rmf}^{\mu}$ & $2K$ & $\frac{g^{2}v^{2}}{4K}$ \cr
 & $-$ & $\tr (F^{\mu\nu}D_{\nu}\phi)$        &   & [$\tr(\bar{F}_{\rmasym}^{\mu\nu}A_{\rmasym \nu}^{\vphantom{\mu}})$], [$\tr(\bar{F}_{\rmf}^{\mu\nu}A_{\rmf \nu}^{\vphantom{\mu}})$], [$\tr(F_{\rma}^{\mu\nu}D_{\nu}h_{\rma}^{\vphantom{\mu}})$]         &  &  &  \cr
  & $-$ & $\tr (\phi F^{\mu\nu}\phi D_{\nu}\phi)$        & [$\tr (A_{\rmf}^{\mu} A_{\rmf}^{\nu} A_{\rmasym \nu}^{\vphantom{\mu}})$]  & [$\tr(\bar{F}_{\rmasym}^{\mu\nu}A_{\rmasym \nu}^{\vphantom{\mu}})$]          &  &  &  \cr
  & $-$ & $\tr \big(\phi^{2}(D^{\mu}\phi)(D^{\nu}\phi)D_{\nu}\phi \big)$        & $\tr (A_{\rmf}^{\mu} A_{\rmf}^{\nu} A_{\rmasym \nu}^{\vphantom{\mu}})$ & [$\tr (A_{\rmasym}^{\mu} (D_{\nu}h_{\rma}) A_{\rmasym}^{\nu\vphantom{\mu}})$]          &  &  &  \cr
\toprule
\end{tabular}

\caption{Particle content of an $\SO(N>3)$ gauge theory with scalar field in the second-rank antisymmetric tensor (adjoint) representation and $\tilde\lambda>0$. The upper table contains the spectrum for even $N$ while the lower table summarizes the spectrum for odd $N$. 
Left: Comparison between operators/states that are strict invariant with respect to $\SO(N)$ transformations, i.e., observables, and operators/states that one would predict from the conventional but gauge-variant viewpoint of spontaneous gauge symmetry breaking ($\SU(K)$ singlets). Trivial scattering states are ignored. In case the contraction is obvious, we suppress Lorentz indices in the scalar channel for the $\U(K)$ singlets for better readability. Right: Properties of the elementary building blocks obtained from the standard multiplet decomposition after gauge fixing which are used to construct $\U(K)$ singlets.}

\label{tab:SOadj}
\end{center}
\end{table*}


Finally, we have to consider the scalar $\Ztwo$-odd channel. In general, it is not possible to construct a gauge-invariant operator which contains only scalar fields due to the antisymmetric property of $\phi$, implying $\tr(\phi^{2n+1}) = 0$. Further, it can be shown that every operator build from the $\epsilon$ tensor and $\phi$ either vanishes or is $\Ztwo$ even with one particular exception. For $N = 4K + 2$, we may consider the $\Ztwo$-odd operator $\epsilon^{a_1\cdots a_{N}} \phi^{a_1a_2}\phi^{a_3a_4} \cdots \phi^{a_{N-1}a_N}$. Its leading order term in the FMS expansion is ${\sim}v^{2K} h$, showing that the propagator of this bound state can be approximated by the propagator of the radial Higgs excitation with mass $\mh$. 
A $\Ztwo$-odd operator which exists for all $N$ is
\begin{align}
  \tr\big( (D_{\mu}F^{\mu\nu})D_{\nu}\phi \big)
  = gv \, \tr\big( (D_{\mu}F^{\mu\nu})[A_{\mu},\phi_{0}] \big) + \cdots \!.
  \label{eq:scalarOddanti2nd}
\end{align}
The dynamics of this bound state operator is governed by a composite $\U(K)$-invariant hybrid operator of two massive vector fields as well as massless gauge bosons. For even $N$, we have only one massive vector multiplet such that we obtain the $\U(K)$-invariant operator $\tr(A_{\rmasym \nu}D_{\mu}\bar{F}_{\rmasym}^{\mu\nu})$. Nevertheless, this is an operator of the second class as several scattering states are encoded in the leading order term as well. For odd $N$, we additionally obtain $\tr(A_{\rmf \nu}D_{\mu}\bar{F}_{\rmf}^{\mu\nu})$ which also belongs to the second class. 
Furthermore, we investigated the operator
\begin{align}
 2\tr\big( \phi F^{\mu\nu}(D_{\mu}\phi)D_{\nu}\phi \big) &= g^{2}v^{3}\tr\big( [\phi_{0}, F^{\mu\nu}][A_{\mu},\phi_{0}][A_{\nu},\phi_{0}] \big)  \notag \\
 &\quad  + \mathcal{O}(\varphi)
\label{eq:scalarOddanti2nd2}
\end{align}
for $N=2K+1$. From the leading order term of the FMS expansion, we obtain the hybrid states $\tr(F_{\rmasym}^{\mu\nu}A_{\rmf\mu}^{\phantom{\,}}A_{\rmf\nu}^{\phantom{\,}})$ and $\tr(F_{\rmf}^{\mu\nu}A_{\rmf\mu}^{\phantom{\,}}A_{\rmasym\nu}^{\phantom{\,}})$ which are assigned to the second class as well as various potential scattering states, $\tr(A_{\rmasym}^{\mu}A_{\rmasym}^{\nu}A_{\rmf\mu}^{\phantom{\,}}A_{\rmf\nu}^{\phantom{\,}})$, $\tr(A_{\rmasym}^{\mu}A_{\rmasym}^{\nu}A_{\rmf\nu}^{\phantom{\,}}A_{\rmf\mu}^{\phantom{\,}})$, $\tr(A_{\rmasym}^{\mu}A_{\rmasym\mu}^{\phantom{\,}}A_{\rmf}^{\nu}A_{\rmf\nu}^{\phantom{\,}})$, $\tr(A_{\rmf}^{\mu}A_{\rmf}^{\nu}A_{\rmf\mu}^{\phantom{\,}}A_{\rmf\nu}^{\phantom{\,}})$, and $\tr(A_{\rmf}^{\mu}A_{\rmf\mu}^{\phantom{\,}})^{2}$. 
The particle content for even and odd $N$ is summarized in Tab.~\ref{tab:SOadj}.


\begin{table*}[t!]
\begin{center}

\begin{tabular}{c|| ll | lll || lcc}
\toprule

& \multicolumn{2}{c|}{\quad $\SO(N)$ invariant \quad} & \multicolumn{3}{c||}{\quad $\U(1)\times\SO(N-2)$ singlets \quad} & \multicolumn{3}{c}{$\U(1)\times\SO(N-2)$  multiplets} \\
\hline
$J^P$ &  $\Ztwo$ \,  & Operator                & 1. Class  & [2. Class]    & 3. Class             &  Field  & DOF &  $m_{\mathrm{Field}}^{2}$ \\
\hline

$0^+$ & $+$ & $\tr\phi^{2}$              & $h$ & [$\tr h_{\rma}^{2}$]   &                                                                         & $h$  & 1 & $\big(\lambda+\frac{1}{2}\tilde\lambda\big)v^{2}$ \cr
& $+$ & $\tr (\phi D^{2}\phi)$        & $\tr (A_{\rmf +}A_{\rmf -})$, [$h$] & [$\tr(A_{\rmf +}h_{\rma}A_{\rmf -})$]  &            & $h_{\rma}$ & $\frac{(N-2)(N-3)}{2}$  & $\frac{1}{4}|\tilde\lambda |v^{2}$ \cr
& $+$  & $\tr F^{2}$       &  & [$\tr F_{\rma}^{2}$], [$\tr(F_{\rmf +}F_{\rmf -})$],   &          & & & \cr
&  &       &  & [$\tr(A_{\rmf +}F_{\rma}A_{\rmf -})$]  &          & & & \cr
& $+$  & $\tr (\phi^{2}F^{2})$       &  & [$\tr(F_{\rmf +}F_{\rmf -})$]  &          & & & \cr
& $+$  & $\tr \big((D_{\mu}\phi)F^{\mu\nu}D_{\nu}\phi\big)$       &  & [$\tr(A_{\rmf +}F_{\rma}A_{\rmf -})$]  &          & & & \cr

\cline{2-6}
& $-$ & $\tr\big( (D_{\mu}F^{\mu\nu})D_{\nu}\phi \big)$        & $\tr \big( (D_{\mu}F_{\rmf \pm}^{\mu\nu}) A_{\rmf \mp \nu}^{\vphantom{\mu}} \big)$  & [$\tr\big( (D_{\mu}F_{\rma}^{\mu\nu})D_{\nu}h_{\rma}^{\vphantom{\mu}} \big)$]            & & & \cr

\cline{2-6}
&   &       &  &   & $\tr (\mathcal{D}A_{\rmf \pm})^{2}$           & & & \cr
&   &       &  &   & $\tr (\mathcal{D}F_{\rmf \pm})^{2}$          & & & \cr
&   &       &  &   & $\tr(A_{\rmf \pm}F_{\rma}A_{\rmf \pm})$          & & & \cr

\hline
$1^-$ & $+$ & $\tr (\phi F^{\mu\nu}D_{\nu}\phi)$             & $\tr (F_{\rmf \pm} A_{\rmf \mp })$  &   &             & $A_{\rma}^{\mu}$ & $\frac{(N-2)(N-3)}{2}$ & 0 \cr 
\cline{2-6}

& $-$ & $\partial_{\nu} \tr (F^{\mu\nu}\phi)$        & $A^{\mu}_{\perp \U(1)}$ &     &            & $A_{\U(1)}^{\mu}$ & 1 & 0  \cr       
& $-$ & $\tr (F^{\mu\nu}D_{\nu}\phi)$         & $\tr (F_{\rmf \pm} A_{\rmf \mp })$ & [$\tr(F_{\rma}^{\mu\nu}D_{\nu}h_{\rma}^{\vphantom{a}})$]   &          & $A_{\rmf +}^{\mu}$ & $N-2$ & $\frac{1}{4}g^{2}v^{2}$  \cr
\cline{2-6}
  &  &         &  &    & $\tr(\mathcal{D}F_{\rmf \pm}^{\mu\nu} \mathcal{D}A_{\rmf \pm \nu}^{\vphantom{\mu}})$          & $A_{\rmf -}^{\mu}$ & $N-2$ & $\frac{1}{4}g^{2}v^{2}$ \cr

\toprule
\end{tabular}

\caption{Particle content of an $\SO(N>3)$ gauge theory with scalar field in the second-rank antisymmetric tensor (adjoint) representation and $\tilde\lambda <0$.  
Left: Comparison between operators/states that are strict invariant with respect to $\SO(N)$ transformations, i.e., observables, and operators/states that one would predict from the conventional but gauge-variant viewpoint of spontaneous gauge symmetry breaking ($\SO(N-2)$ singlets). Trivial scattering states are ignored. In case the contraction is obvious, we suppress Lorentz indices for better readability. The symbol $\mathcal{D}$ indicates a suitable dressing with a Dirac phase factor to obtain observable $\U(1)$-invariant states. Right: Properties of the elementary building blocks obtained from the standard multiplet decomposition after gauge fixing which are used to construct $\SO(N-2)$ singlets.}
\label{tab:SOadjneg}
\end{center}
\end{table*}


\subsubsection{\texorpdfstring{$\tilde\lambda < 0$}{\tilde\lambda < 0}}

If the nonisotropic coupling $\tilde\lambda$ is negative, the field configuration 
\begin{align}
 \phi_{0} = \frac{1}{2} \begin{pmatrix} \varepsilon & & & \\ & 0 & & \\ & & \ddots & \\ & & & 0 \end{pmatrix}
\end{align}
minimizes the potential~\eqref{eq:pot2aON} where $\mu^{2} = \frac{1}{4}(2\lambda + \tilde\lambda)v^{2}$ and the breaking pattern reads $\SO(N) \to \U(1)\times \SO(N-2)$ \cite{Li:1973mq}. 
Note that the coupling $\tilde\lambda$ is restricted by the condition $|\tilde\lambda| < 2\lambda$  in order to fulfill stability criteria for the potential (throughout this paper, we always assume $\lambda>0$).
The elementary spectrum consists of $(N-2)(N-3)/2$ massless gauge bosons $A_{\rma}^{\mu}$ as well as a further massless vector degree of freedom $A_{\U(1)}^{\mu}$ associated to the unbroken subgroups $\SO(N-2)$ and $\U(1)$, respectively. Further, we have $2(N-2)$ degenerated gauge bosons with mass $m_{\rmA_{\rmf}}^{2} = \frac{1}{4}g^{2} v^{2}$ being two fundamental vectors of $\SO(N-2)$ and charged under the $\U(1)$ which we denote by $A_{\rmf +}^{\mu}$ and $A_{\rmf -}^{\mu}$. In the scalar sector we have $2(N-2)$ would-be Goldstones, the radial Higgs excitation $h$ being a singlet with respect to $\U(1)\times \SO(N-2)$ with mass $\mh^{2} = (\lambda + \tilde\lambda/2)v^{2}$ as well as an antisymmetric 2nd-rank ($\equiv$ adjoint) $\SO(N-2)$ tensor field $h_{\rmasym} \equiv h_{\rma}$ with mass parameter $m_{\mathrm{h}_{\rma}}^{2} = \frac{1}{4}|\tilde\lambda|v^{2}$. The $\U(1)\times \SO(N-2)$ multiplets can be obtained in a covariant manner from $\phi$ and $A$ similar to the previous case.

Again, we start the analysis of the strict gauge-invariant spectrum in the scalar channel. 
The FMS expansion of the $\Ztwo$-even scalar operator given in the first line of Eq.~\eqref{eq:ZevenscalarON} defines an $\SO(N)$-invariant description of the radial Higgs excitation $h$ in leading order as usual. Thus, it generates a state at $\mh$. Allowing for the decomposition of the potential scattering contribution $\varphi^{ab}\varphi^{ab}$ at next-to-leading order into $\U(1)\times \SO(N-2)$-multiplets,
\begin{align}
 \tr( \phi^{\rmT}\phi) = \frac{v^2}{2} + v h + \frac{1}{2}h^2 - \tr (h_{\rma}^{2}),
\end{align}
we indeed obtain a trivial scattering state at $2\mh$ as well as a state of the second class described by the dynamics of the $\SO(N-2)$ meson operator $\tr(h_{\rma}^{2})$ with mass ${2 m_{\mathrm{h}_{\rma}} = \sqrt{|\tilde\lambda|} v}$.

Considering the operator $\tr(\phi D^{2}\phi)$, we find a direct mapping on the scalar meson operator with two massive vector fields as constituents $\tr(A_{\rmf +}A_{\rmf -})$, implying that a further possible bound state at approximately $2m_{\rmA_{\rmf}} = g v$ exists in the $\Ztwo$-even scalar channel. This meson operator is invariant under both unbroken subgroups, $\U(1)$ and $\SO(N-2)$. We would expect further bound states being singlets with respect to the non-Abelian subgroup $\SO(N-2)$ but with open $\U(1)$ quantum numbers, i.e., $\tr(A_{\rmf +}A_{\rmf +})$ and $\tr(A_{\rmf -}A_{\rmf -})$, from the conventional perspective of spontaneous gauge symmetry breaking. These can be dressed via suitable Dirac phase factors for the Abelian part. However, there is no $\SO(N)$-invariant analog of these states as an $\SO(N)$-invariant operator can only be expanded in singlets of the full unbroken subgroup, see Sec.~\ref{sec:validity}. Thus, the operators $\tr(A_{\rmf +}A_{\rmf +})$ and $\tr(A_{\rmf -}A_{\rmf -})$ belong to the third class. 
At next-to-leading order, $\tr(\phi D^{2}\phi)$ contains scattering states as well as the baryonic operator $\tr(A_{\rmf +} h_{\rma} A_{\rmf -})$. The latter belongs to the second class.

Furthermore, the scalar glueball $\tr F_{\rma}^{2}$ and the hybrids $\tr (F_{\rmf +}F_{\rmf -})$ and $\tr (A_{\rmf +}F_{\rma}A_{\rmf -})$ belong to the second class as can be deduced from the operators $\tr F^{2}$, $\tr(\phi^{2}F^{2})$, and $\tr \big((D_{\mu}\phi)F^{\mu\nu}D_{\nu}\phi\big)$. In analogy to the discussion of the meson states, the hybrids with open $\U(1)$ quantum number, $\tr F_{\rmf +}^{2}$, $\tr F_{\rmf -}^{2}$, $\tr(A_{\rmf +}F_{\rma}A_{\rmf +})$, and $\tr(A_{\rmf -}F_{\rma}A_{\rmf -})$, can be assigned to the third class.

The FMS expansion of the simplest operator in the scalar $\Ztwo$-odd channel reads,
\begin{align}
 \tr\big( (D_{\mu}F^{\mu\nu})D_{\nu}\phi \big) &= \tr \big( (D_{\mu}F_{\rmf +}^{\mu\nu}) A_{\rmf - \nu} + (D_{\mu}F_{\rmf -}^{\mu\nu}) A_{\rmf + \nu} \big) \notag \\
 &\quad + \mathcal{O}(\varphi),
\label{eq:scalarOddanti2ndneg}
\end{align}
describing a nontrivial $\U(1)\times \SO(N-2)$-invariant hybrid. By contrast, the operator defined on the left-hand side of Eq.~\eqref{eq:scalarOddanti2nd2} contains only trivial scattering states.

The analysis of the vector channel follows the same strategy as in the previous case. Basic $\Ztwo$ even and odd operators are provided in Tab.~\ref{tab:SOadjneg} which also summarizes the scalar sector and lists further states of the third class. We highlight that we obtain again states with identical field content and mass in the $\Ztwo$ even and odd vector channel giving rise to a potential emergent $\U(1)$ gauge structure at the bound state level.

As in the case of a fundamental scalar field, we find a mismatch between the conventional investigation of the spectrum and the strict gauge-invariant formulation for an $\SO(N)$ gauge theory with scalar field in the adjoint representation. Although some states are identical in both descriptions and the FMS mechanism provides a field theoretical tool to establish this relation, i.e., states of the first class, the situation is unclear for many composite objects as they belong to the second class. Even more critical is the situation for states of the third class for which no $\SO(N)$-invariant formulation can be found. That such states arise can be traced back to the explicit presence of the Abelian $\U(1)$ subgroup in the gauge-fixed formulation.

\subsection{Symmetric second-rank tensor representation}
\label{sec:ON2sym}

The analysis of the irreducible symmetric second-rank tensor representation ($\phi^{\rmT}=\phi$, $\tr\phi = 0$) is structurally similar to the adjoint representation of the $\SUN$ group. 
The most general fourth order potential of the scalar sector reads
\begin{align}
 V(\phi) =  -\mu^2 \tr \phi^2 + \frac{\gamma}{3} \tr \phi^{3} + \frac{\lambda}{2} (\tr \phi^2)^2 + \frac{\tilde\lambda}{2} \tr \phi^4.
\label{eq:potSONsym}
\end{align}
If the cubic coupling $\gamma$ vanishes, the action is invariant under a $\Ztwo$ symmetry. 
In case the field $\phi$ acquires a nonvanishing VEV, different breaking patterns can be realized. In fact it can be shown that the field configuration that minimizes the potential has at most two different eigenvalues, implying $\SO(N) \to \mathrm{S}(\rmO(P){\times}\rmO(N-P))$ where $P < N$ \cite{Li:1973mq}. Thus,
\begin{align}
 \phi_{0} = \frac{1}{\sqrt{2NP(N-P)}} \begin{pmatrix} (N-P)\mathbbm{1}_\mathrm{P} & \\  & -P\mathbbm{1}_\mathrm{N-P} \end{pmatrix}
\label{eq:minSONsym}
\end{align}
is a convenient parametrization of the VEV where $\mathbbm{1}_\mathrm{x}$ is the $x\times x$ identity matrix and $\mu^{2} = \frac{1}{2}\lambda v^{2}+2\tilde\lambda v^{2}\tr\phi_{0}^{4} + \gamma v \tr\phi_{0}^{3}$.
The actual global minimum is determined by the two non-isotropic couplings $\gamma$ and $\tilde\lambda$. If $\tilde\lambda < 0$, there is only one breaking pattern with little group ${\rmO(N-1)}$. If $\tilde\lambda \geq 0$, $\gamma$ and $\tilde\lambda$ pull in opposite directions, such that $|2P-N|$ becomes as large/small as possible. For example, we have $P= \lfloor N/2 \rfloor$ for $\gamma = 0$, where $\lfloor x \rfloor$ is the floor function, and $P = N-1$ for $\tilde\lambda = 0$ \cite{ORaifeartaigh:1986agb}. Without loss of generality, we will restrict the following discussion to the case $P \leq N/2$. The case $P>N/2$ is included by a simple relabeling of the occurring fields.

The mass matrix for the gauge bosons is defined in Eq.~\eqref{eq:massmatrix2ndrankON} and we obtain $P(N-P)$ degenerate massive gauge bosons $A_{\rmf \otimes \rmf}^{\mu} = [A^{\mu},\phi_{0}]$ with mass parameter $m_{\rmA_{\rmf \otimes \rmf}}^{2} = \frac{N}{2P(N-P)}g^2v^2$, transforming as $(P,N-P)$-multiplet. The subscript $\mathrm{r_{P}} \otimes \mathrm{r_{N-P}}$ indicates that the object transforms according to representation $\mathrm{r_{P}}$ and $\mathrm{r_{N-P}}$ of $\SO(P)$ and ${\SO(N-P)}$, respectively. In addition, we have the massless gauge fields $A_{\rma \otimes \rms}^{\mu}$ and $A_{\rms \otimes \rma}^{\mu}$. Of course, $A_{\rma \otimes \rms}^{\mu}$ vanishes if $P=1$, i.e., the stability group is $\rmO(N-1)$. Then, the massive vector fields simply transform under the fundamental representation of $\rmO(N-1)$, $A_{\rmf \otimes \rmf}^{\mu} \equiv A_{\rmf}$, with mass parameter $m_{\rmA_{\rmf}}^{2} = \frac{N}{2(N-1)}g^2v^2$ and we have only one kind of massless gauge bosons $A_{\rma}$.

The elementary scalar spectrum contains a singlet $h$, a symmetric tensor $h_{\rmsym \otimes \rms}$ transforming according to $\big(P(P+1)/2-1,1\big)$, and a tensor $h_{\rms \otimes \rmsym}$ transforming as a singlet with respect to the remaining $\SO(P)$ group and a symmetric second-rank tensor of $\SO(N-P)$, i.e., $\big(1,(N-P)(N-P+1)/2-1\big)$. Unless $P=1$ where $h_{\rmsym \otimes \rms}$ vanishes. The subscript $\rmsym$ indicates symmetric second-rank tensor. The mass parameters of these fields read
\begin{align}
 \mh^2 =& \lambda v^2 + \frac{P^3 + (N-P)^{3}}{N^2P(N-P)} \tilde\lambda v^{2} + \frac{N-2P}{\sqrt{8NP(N-P)}}\gamma v, \notag \\
 m_{\mathrm{h_{\rmsym \otimes \rms}}}^{2} &= \sqrt{\frac{N}{8P(N-P)}} \gamma v + \frac{2(N-P)-P}{2P(N-P)}\tilde\lambda v^2, \notag \\
 m_{\mathrm{h_{\rms \otimes \rmsym}}}^{2} &= -\sqrt{\frac{N}{8P(N-P)}} \gamma v + \frac{2P-(N-P)}{2P(N-P)}\tilde\lambda v^2.
 \label{eq:masshsymSO}
\end{align}

For the bound state spectrum of the $0^{+}$ channel, we investigate the following operators and their FMS expansion
\begin{align}
 \tr \phi^{2} &= \frac{v^{2}}{2} + v h + \tr \varphi^{2} \notag \\
 &= \frac{v^{2}}{2} + v h + \tr h_{\rmsym\otimes\rms}^{2} + \tr h_{\rms\otimes\rmsym}^{2} ,
 \label{eq:scalarSymZ2even} \\
 \tr \phi^{3} &= \big(v^{3} + 3v^{2} h\big) \tr\phi_{0}^{3} + 3v\, \tr(\phi_{0}\varphi^{2}) + \tr\varphi^{3} \notag \\
 &= \frac{N-2P}{\sqrt{8NP(N-P)}}(v^3 + 3v^2 h + 3v h^2) \notag \\
 &\quad + \frac{N-P}{\sqrt{2NP(N-P)}} \tr h_{\rmsym\otimes\rms}^{2} \notag \\
 &\quad - \frac{P}{\sqrt{2NP(N-P)}} \tr h_{\rms\otimes\rmsym}^{2} + \mathcal{O}(\varphi^3)
 \label{eq:scalarSymZ2odd}
\end{align}
which are $\Ztwo$-even and odd, respectively. While the nontrivial leading order contribution projects on the radial Higgs excitation $h$ by performing the split \eqref{eq:split2ON}, we have also provided the standard decomposition of the next-to-leading order terms in the second line of Eq.~\eqref{eq:scalarSymZ2even} and Eq.~\eqref{eq:scalarSymZ2odd}, giving the usual scattering term at $2\mh$ as well as possible bound state contributions if the $\mathcal{G}$-$\mathcal{H}$ duality extends to operators of the second class.

In case the global $\Ztwo$ symmetry would be intact, the $0^{+}$ channel would be divided into states with even and odd $\Ztwo$ symmetry and the FMS expansion of the operators \eqref{eq:scalarSymZ2even} and \eqref{eq:scalarSymZ2odd} reveals that both channels would have the same mass spectrum in a first order approximation. For $\gamma \neq 0$, however, the global symmetry is explicitly broken and the distinction is superfluous. Then, both operators have overlap with the same states of the scalar channel. But even if $\gamma = 0$, we expect that the global $\Ztwo$ symmetry is broken. In contrast to the fundamental and adjoint case, we find this time an $\Ztwo$-odd scalar operator which acquires a nonvanishing VEV, namely $\tr \phi^{3}$.\footnote{In principle, this is also possible for the adjoint case for $N=4K+2$. In this particular case, we can define the nonvanishing operator $\epsilon^{a_1\cdots a_{N}} \phi^{a_1a_2} \cdots \phi^{a_{N-1}a_N}$ which can be used as an order parameter for the global $\Ztwo$ symmetry.} Thus we conclude that the global symmetry is broken spontaneously by the internal dynamics of the microscopic degrees of freedom.

Of course, we concentrated only on those states which purely consists of elementary scalar degrees of freedom so far. Another contribution to the scalar channel is given by the operator~\eqref{eq:ZevenscalarON2} which expands in leading order to the scalar meson operator $\tr(A_{\rmf\otimes\rmf}A_{\rmf\otimes\rmf})$ in an unambiguous way such that it belongs to the first class. That such a state is also part of the $\Ztwo$-odd channel can be deduced from the operator $\tr(\phi^{2}D^{2}\phi)$.

Further scalar states from the $\mathrm{S}(\rmO(P){\times}\rmO(N-P))$-invariant effective theory may be formulated in a strict $\SO(N)$-invariant way if the $\mathcal{G}$-$\mathcal{H}$ duality works at the decomposition level.
For instance, the glueball states $\tr(F_{\rma\otimes\rms}^{2})$ and $\tr(F_{\rms\otimes\rma}^{2})$ as well as the hybrids $\tr(F_{\rmf\otimes\rmf}^{2})$, $\tr(A_{\rmf\otimes\rmf}^{\rmT}F_{\rma\otimes\rms}A_{\rmf\otimes\rmf})$, and $\tr(A_{\rmf\otimes\rmf}F_{\rms\otimes\rma}A_{\rmf\otimes\rmf}^{\rmT})$ are encoded in $\tr(F^{2})$, $\tr(F^{2}\phi)$, or $\tr(F^{2}\phi^{2})$. Note that the hybrid $\tr(F_{\rmf\otimes\rmf}^{2})$ is not contained in the $\Ztwo$-odd operator $\tr(\phi F^{2})$ if $P=N-P$. The next-to-leading order contribution of the FMS expansion of the operators $\tr(F^{2}\phi)$ and $\tr(F^{2}\phi^{2})$ contain also hybrid operators involving massless gauge fields and the corresponding second-rank tensor Higgs field transforming either nontrivial with respect to the $\SO(P)$ subgroup and being singlets of $\SO(N-P)$ or vice versa, i.e., $\tr(F_{\rma\otimes\rms}^{2}h_{\rma\otimes\rms}^{\phantom{2}})$ and $\tr(F_{\rms\otimes\rma}^{2}h_{\rms\otimes\rma}^{\phantom{2}})$. Furthermore, we obtain the hybrids $\tr(F_{\rmf\otimes\rmf}^{2}h_{\rmsym\otimes\rms}^{\phantom{2}})$ and $\tr(F_{\rmf\otimes\rmf}^{2}h_{\rms\otimes\rmsym}^{\phantom{2}})$.

As no elementary vector particle exists which transforms as a singlet, we do not expect that it is possible to find a gauge-invariant bound state operator that expands to an elementary vector field. Indeed, we have $\tr F^{\mu\nu} \phi^{n} = 0$ and $\tr \phi^{n}D_{\mu}\phi = \tr \phi^{n}\partial_{\mu}\phi$ due to the symmetry properties of $A_{\mu}$ and $\phi$. Also, $\tr (F^{\mu\nu}D_{\nu}\phi) = 0$. Thus, the simplest operator to investigate the bound state spectrum in the vector channel has to contain at least three elementary gauge fields. 
Analyzing the $\Ztwo$-even operator
\begin{align}
 \tr (\phi F^{\mu\nu}D_{\nu}\phi) = gv^{2} \tr (\phi_{0}F^{\mu\nu} [A_{\nu},\phi_{0}]) + \mathcal{O}(\varphi),
\end{align}
we predict that a possible state of the vector channel is generated by the $\mathrm{S}(\rmO(P){\times}\rmO(N-P))$-invariant vector hybrid operator $\tr (F_{\rmf\otimes\rmf}^{\mu\nu} A_{\rmf\otimes\rmf\nu}^{\phantom{2}})$ which belongs to the first class. Similar results can be obtained by investigating the following $\Ztwo$-odd vector operator
\begin{align}
 \tr (\phi^{2}F^{\mu\nu}D_{\nu}\phi) &= gv^{3} \tr (\phi_{0}^{2}F^{\mu\nu} A_{\rmf\otimes\rmf\nu}) + \mathcal{O}(\varphi). 
\label{eq:vectorSymZ2odd}
\end{align}
At leading order in the FMS expansion, this operator has overlap with the vector hybrid state $\tr (F_{\rmf\otimes\rmf}^{\mu\nu} A_{\rmf\otimes\rmf\nu}^{\phantom{2}})$ with mass $2m_{\rmA_{\rmf\otimes\rmf}}$ in the simple constituent model as well. At higher order in the FMS expansion, we obtain the hybrids $\tr(F_{\rmf\otimes\rmf}^{\mu\nu}A_{\rmf\otimes\rmf\nu}^{\vphantom{\mu}}h_{\rmsym\otimes\rms}^{\vphantom{\mu}})$ and $\tr(F_{\rmf\otimes\rmf}^{\mu\nu}A_{\rmf\otimes\rmf\nu}^{\vphantom{\mu}}h_{\rms\otimes\rmsym}^{\vphantom{\mu}})$ belonging to the second class. As usual, we provide a summary table comparing the different spectra, see Tab.~\ref{tab:SOsym}.


\begin{table*}[t!]
\begin{center}

\begin{tabular}{c|| ll | ll || lcc}

\toprule

& \multicolumn{2}{c|}{\quad $\SO(N)$ invariant \quad} & \multicolumn{2}{c| |}{\quad $\mathrm{S}(\rmO(P){\times}\rmO(N-P))$ singlets \quad} & \multicolumn{3}{c}{$\mathrm{S}(\rmO(P){\times}\rmO(N-P))$ multiplets} \\
\hline
$J^P$ &  $\Ztwo$ \, &  Operator                &  1. Class  & [2. Class]                 & Field  & DOF &  $m_{\mathrm{Field}}^{2}$ \\
\hline

$0^+$ & $+$ & $\tr\phi^{2}$ &		 $h$ & [$\tr h_{\rms\otimes\rmsym}^{2}$],[$\tr h_{\rmsym\otimes\rms}^{2}$]    & $h$ & 1 & see Eq.~\eqref{eq:masshsymSO} \cr
& $+$ & $\tr(\phi D^{2}\phi)$        & $\tr A_{\rmf\otimes\rmf}^{2}$, [$h$]  & [$\tr (A_{\rmf\otimes\rmf}^{2}h_{\rms\otimes\rmsym})$],[$\tr (A_{\rmf\otimes\rmf}^{2}h_{\rmsym\otimes\rms})$]            & $h_{\rmsym \otimes \rms}$ & $\frac{P(P+1)}{2}-1$ & see Eq.~\eqref{eq:masshsymSO} \cr
& $+$ & $\tr F^{2}$        &  & [$\tr(F_{\rma\otimes\rms}^{2})$], [$\tr(F_{\rms\otimes\rma}^{2})$], [$\tr(F_{\rmf\otimes\rmf}^{2})$],           & $h_{\rms \otimes \rmsym}$ & $\frac{(N-P)(N-P+1)}{2}-1$ & see Eq.~\eqref{eq:masshsymSO} \cr
&  &        &   & [$\tr(A_{\rmf\otimes\rmf}F_{\rma\otimes\rms}A_{\rmf\otimes\rmf})$], [$\tr(A_{\rmf\otimes\rmf}F_{\rms\otimes\rma}A_{\rmf\otimes\rmf})$]            & & & \cr

& $+$ & $\tr(\phi^{2}F^{2})$        &   & [$\tr(F_{\rma\otimes\rms}^{2})$],[$\tr(F_{\rms\otimes\rma}^{2})$], [$\tr(F_{\rmf\otimes\rmf}^{2})$],           & & & \cr
&  &        &   & [$\tr(A_{\rmf\otimes\rmf}^{\rmT}F_{\rma\otimes\rms}A_{\rmf\otimes\rmf})$], [$\tr(A_{\rmf\otimes\rmf}F_{\rms\otimes\rma}A_{\rmf\otimes\rmf}^{\rmT})$],            & & & \cr
&  &        &   & [$\tr(F_{\rma\otimes\rms}^{2}h_{\rmsym\otimes\rms}^{\phantom{2}})$], [$\tr(F_{\rms\otimes\rma}^{2}h_{\rms\otimes\rmsym}^{\phantom{2}})$],            & & & \cr
&  &        &   & [$\tr(F_{\rmf\otimes\rmf}^{2}h_{\rmsym\otimes\rms}^{\phantom{2}})$], [$\tr(F_{\rmf\otimes\rmf}^{2}h_{\rms\otimes\rmsym}^{\phantom{2}})$]            & & & \cr
\cline{2-5}
& $-$ & $\tr\phi^{3}$        & $h$   & [$\tr h_{\rms\otimes\rmsym}^{2}$], [$\tr h_{\rmsym\otimes\rms}^{2}$], [$\tr h_{\rms\otimes\rmsym}^{3}$], [$\tr h_{\rmsym\otimes\rms}^{3}$]            & & & \cr
& $-$ & $\tr(\phi^{2} D^{2}\phi)$        & $\tr A_{\rmf\otimes\rmf}^{2}$\footnote{This state is not present for ${P=N-P}$}, [$h$]  & [$\tr (A_{\rmf\otimes\rmf}^{2}h_{\rms\otimes\rmsym}^{\vphantom{2}})$],[$\tr (A_{\rmf\otimes\rmf}^{2}h_{\rmsym\otimes\rms}^{\vphantom{2}})$]            &  &  &  \cr
& $-$ & $\tr(\phi F^{2})$        &   & [$\tr(F_{\rma\otimes\rms}^{2})$], [$\tr(F_{\rms\otimes\rma}^{2})$], [$\tr(F_{\rmf\otimes\rmf}^{2})$]\footnotemark[1],            & & & \cr
&  &        &   & [$\tr(A_{\rmf\otimes\rmf}F_{\rma\otimes\rms}A_{\rmf\otimes\rmf})$], [$\tr(A_{\rmf\otimes\rmf}F_{\rms\otimes\rma}A_{\rmf\otimes\rmf})$],            & & & \cr
&  &        &   & [$\tr(F_{\rma\otimes\rms}^{2}h_{\rmsym\otimes\rms}^{\vphantom{2}})$], [$\tr(F_{\rms\otimes\rma}^{2}h_{\rms\otimes\rmsym}^{\phantom{2}})$],            & & & \cr
&  &        &   & [$\tr(F_{\rmf\otimes\rmf}^{2}h_{\rmsym\otimes\rms}^{\vphantom{2}})$], [$\tr(F_{\rmf\otimes\rmf}^{2}h_{\rms\otimes\rmsym}^{\phantom{2}})$]            & & & \cr
\hline
$1^-$ & $+$ & $\tr(\phi F^{\mu\nu}D_{\nu}\phi)$             & $\tr (F_{\rmf\otimes\rmf} A_{\rmf\otimes\rmf})$  & [$\tr(F_{\rmf\otimes\rmf}A_{\rmf\otimes\rmf}h_{\rmsym\otimes\rms})$], [$\tr(F_{\rmf\otimes\rmf}A_{\rmf\otimes\rmf}h_{\rms\otimes\rmsym})$]              & $A_{\rma \otimes \rms}^{\mu}$ & $\frac{P(P-1)}{2}$ & 0 \cr
\cline{2-5}
 & $-$ & $\tr(\phi^{2}F^{\mu\nu}D_{\nu}\phi)$             & $\tr (F_{\rmf\otimes\rmf} A_{\rmf\otimes\rmf})$   & [$\tr(F_{\rmf\otimes\rmf}A_{\rmf\otimes\rmf}h_{\rmsym\otimes\rms})$], [$\tr(F_{\rmf\otimes\rmf}A_{\rmf\otimes\rmf}h_{\rms\otimes\rmsym})$]                & $A_{\rms \otimes \rma}^{\mu}$ & $\frac{(N-P)(N-P-1)}{2}$ & $0$ \cr
  &  &              &    &               &  $A_{\rmf\otimes\rmf}^{\mu}$ & $P(N-P)$ & $\frac{N}{2P(N-P)}g^2v^2$ \cr
 \toprule
\end{tabular}
\caption{Particle content of a non-Abelian $\SO(N)$ gauge theory with scalar field in the irreducible symmetric second-rank tensor representation. 
Left: Comparison between operators/states that are strict invariant with respect to $\SO(N)$ transformations, i.e., observables, and operators/states that are considered from the conventional but gauge-variant viewpoint of spontaneous gauge symmetry breaking ($\mathrm{S}(\rmO(P){\times}\rmO(N-P))$ singlets). Trivial scattering states are ignored. We suppress Lorentz indices for better readability for the $\mathrm{S}(\rmO(P){\times}\rmO(N-P))$ singlets. Right: Properties of the elementary building blocks obtained from the standard multiplet decomposition after gauge fixing which are used to construct $\mathrm{S}(\rmO(P){\times}\rmO(N-P))$ singlets. Note that the $\Ztwo$ distinction is superfluous as the global symmetry is broken.}

\label{tab:SOsym}
\end{center}
\end{table*}


\section{\texorpdfstring{$\SUN$}{SU(N)} gauge theory}
\label{sec:SUN}

After the analysis of the low dimensional representations up to second-rank tensors for $\SO(N)$ gauge theories, we investigate the same representations for $\SU(N)$ theories now. The fundamental and adjoint representation were already discussed in the literature \cite{Maas:2017xzh}. We give a brief summary of the fundamental representation in App.~\ref{app:SUNfun} to put the results of this model into the broader context of this paper and perform a classification of states for the first time. In App.~\ref{app:SUNadj}, we generalize the results of Ref.~\cite{Maas:2017xzh} for the adjoint representation to the case of a scalar potential with cubic term for arbitrary $\SU(N)$ theories. In the main text, we focus on second-rank tensor representations.

\subsection{Second-rank tensor representations - Preliminaries}
\label{sec:SUNpre}
Before we start to analyze the different irreducible second-rank tensor representations in detail, we list some of their general properties and common features in order to prepare the ground.  
For the second-rank tensor representations, the scalar field transforms according to
$\phi^{ab} \to U^{a\bar{c}} U^{b\bar{d}} \phi^{cd}$ where $U \in \SUN$.
The potential reads 
\begin{align}
 V(\phi) &= \mu^{2} \phi^{* \, \bar{a}\bar{b}} \phi^{ab} + \frac{\lambda}{2} \big(\phi^{* \, \bar{a}\bar{b}} \phi^{ab} \big)^{2} + \frac{\tilde\lambda}{2} \phi^{* \, \bar{a}\bar{b}} \phi^{bc} \phi^{* \, \bar{c}\bar{d}} \phi^{da} \notag\\
 &= \mu^{2} \tr (\phid\phi) + \frac{\lambda}{2} \big(\tr (\phid\phi)\big)^{2} + \frac{\tilde\lambda}{2} \tr (\phi^{\dagger}\phi)^2
\label{eq:potential2ndrank}
\end{align}
where we have introduced a convenient matrix notation $(\phi)^{ab} = \phi^{ab}$ with $\phi \to U \phi U^{\rmT}$ in the second line in analogy to the $\SO(N)$ case. 
Expanding the potential in terms of the fluctuation field $\varphi$, the mass terms for the Higgs fields can be read off from the quadratic terms. 
From the kinetic term of the scalar field with covariant derivative 
$(D_{\mu}\phi)^{ab} = \partial_{\mu} \phi^{ab} + \I g A_{\mu}^{a\bar{c}}\phi^{cb} + \I g A_{\mu}^{b\bar{c}}\phi^{ac}$, 
we extract the mass matrix $\MA^{2}$ for the gauge fields. In case the field $\phi$ is either completely symmetric or antisymmetric, i.e., transforms according to an irreducible $2$nd-rank tensor representation, we obtain 
\begin{align}
 \frac{1}{2}{\MA^{2}}_{ij} A_{i\mu}^{\phantom{m}}A_{j}^{\mu} = g^{2}v^{2} \tr \big( \phi_0^{\dagger} A^{\mu}(A_{\mu}\phi_0 + \phi_0 A_{\mu}^{\rmT}) \big)
 \label{eq:massmatrix2ndrank}
\end{align}
with gauge fields $A^{\mu} = A^{\mu}_{i}T_{i}$ and $T_i$ the generators of the $\SUN$ Lie algebra.

The gauge-invariant bound state operators in the scalar and vector channel can be further divided according to possible flavor symmetries. The model is invariant under global $\U(1)$ transformations if the scalar field transforms as a complex second-rank tensor and $N>2$. Thus, we further classify the states in $\U(1)$ singlets, i.e., particles which have no observable $\U(1)$ charge, and states with an open $\U(1)$ quantum number which describe $\U(1)$ charged particles.
In case $N=2$, it is straightforward to adapt the following analysis of the bound state spectrum for the symmetric second-rank tensor. It is important to keep in mind that the dual field $\tilde\phi^{ab} = \epsilon^{ac}\epsilon^{bd}\phi^{*\bar{c}\bar{d}}$ also transforms as a symmetric tensor for $N=2$, reflecting the fact that the fundamental representation is pseudo real. Therefore, the operator basis is enlarged to construct gauge-invariant operators in this particular case. By contrast, the $\SU(2)$ antisymmetric second-rank tensor transforms as a singlet such that the scalar and pure Yang-Mills sector decouple. In the following, we focus only on the $N>2$ case.

The simplest gauge-invariant bound state operator in the scalar channel which is a $\U(1)$ singlet is given by two elementary Higgs fields and its FMS expansion reads,
\begin{align}
\phi^{* \bar{a}\bar{b}}\phi^{ab} &= \frac{v^2}{2} + \sqrt{2}v\, \re (\phi_{0}^{* \bar{a}\bar{b}}\varphi^{ab}) + \varphi^{* \bar{a}\bar{b}}\varphi^{ab}.
\label{eq:scalarSinglet}
\end{align}
Thus, this operator always expands in leading order to the single elementary Higgs field being proportional to the direction of the VEV, irrespective of the details of the action or which representation is considered. Further $\U(1)$-singlet operators in the scalar channel are
\begin{align}
 \tr (\phid D^{2}\phi) = -g^{2}v^{2} \tr \big( \phi_0^{\dagger} A^{\mu}(A_{\mu}\phi_0 + \phi_0 A_{\mu}^{\rmT}) \big) + \mathcal{O}(\varphi)
 \label{eq:scalarSinglet2}
\end{align}
or $ \tr \big((D^{\mu}\phi)^{\dagger} D_{\mu}\phi \big)$ expanding to meson operators containing massive vector multiplets, as well as $\tr F^{2}$ which can be decomposed into the corresponding $\mathcal{H}$ glueball and hybrids containing massive vector multiplets and massless gauge bosons.

A charged scalar bound state can be constructed via the epsilon tensor and thus will contain at least $N$ elementary scalar fields
\begin{align}
 \frac{1}{N!} \epsilon^{\bar{a}_1 \cdots \bar{a}_N} \epsilon^{\bar{b}_1 \cdots \bar{b}_N} \phi^{a_1b_1} \cdots \phi^{a_Nb_N} = \det\left( \frac{v}{\sqrt{2}}\phi_{0} + \varphi \right).
\label{eq:scalarNonSinglet}
\end{align}
The charge conjugated operator is given by $\det(\phi^{\dagger})$ and describes the corresponding antiparticle. 
As to whether this operator expands in leading order to a single elementary scalar field or has a more complex bound state dynamic depends on the rank $k$ of the matrix $\phi_{0}$. If $1 \leq k <N$, the leading order contribution is given by $N-k$ elementary Higgs fields. For $k=N$ the leading order contribution is given by a single elementary scalar field.

In the vector channel the $\U(1)$-singlet operator with least field content is given by
\begin{align}
 \I\, \tr \phid D_{\mu}\phi
 &= -gv^2 \tr \phid_{0}A_{\mu}\phi_{0} - \sqrt{8} g v\, \re(\tr \phid_{0}A_{\mu}\varphi) \notag \\
 &\quad + \I \frac{v}{\sqrt{2}} \tr \phid_{0}\partial_{\mu}\varphi - 2g\, \tr \varphi^{\dagger}A_{\mu}\varphi + \I\, \tr \varphi^{\dagger}\partial_{\mu}\varphi.
\label{eq:vectorSinglet} 
\end{align}
The FMS expansion demonstrates that this operator can be described by a single elementary vector boson as long as $\tr (\phi_{0}^{\phantom{\dagger}}\phid_{0}A_{\mu}) \neq 0$, i.e., the $\mathcal{H}$-invariant spectrum contains a vector singlet. 
Note that also a pole at the mass of the radial Higgs excitation seems to appear in the correlator of this vector operator. However, this merely reflects the mixing of $\partial_\mu$ with other operators in the vector channel and does not lead to a new particle, cf. the discussion in Sec.~\ref{sec:ONfun} above Eq.~\eqref{eq:ONfunVec}. Additionally, we investigate the $\SU(N)$-invariant vector operator
\begin{align}
 \tr (\phid F^{\mu\nu}D_{\nu}\phi) = \I g v^2 \tr [\phid_{0} F^{\mu\nu} (A_{\mu}\phi_0 + \phi_0 A_{\mu}^{\rmT})] + \mathcal{O}(\varphi)
 \label{eq:SUNvectorSinglet2}
\end{align}
which maps on possible $\mathcal{H}$-invariant vector hybrids created by the massive vector multiplets and the massless gauge bosons. 
Finally, vector operators with an open $\U(1)$ quantum number can only be constructed with the aid of the epsilon tensor as in case of scalar operators. 
In order to analyze this channel, we will investigate the following two operators
\begin{align}
 &\frac{1}{N!} \epsilon^{\bar{a}_1 \cdots \bar{a}_N} \epsilon^{\bar{b}_1 \cdots \bar{b}_N} (D^\mu\phi)^{a_1b_1} \phi^{a_2b_2} \cdots \phi^{a_Nb_N}, 
\label{eq:vectorNonSinglet} \\
 &\frac{1}{N!} \epsilon^{\bar{a}_1 \cdots \bar{a}_N} \epsilon^{\bar{b}_1 \cdots \bar{b}_N} (F^{\mu\nu}D_{\nu}\phi)^{a_1b_1}\phi^{a_2b_2} \cdots \phi^{a_Nb_N}.
 \label{eq:vectorNonSinglet2}
\end{align}
The conjugated version of these operators describes the corresponding antiparticles if the $\U(1)$ symmetry is intact. 
As their FMS expansions are involved, we discuss the relevant results in the appropriate subsections.

\subsection{Symmetric second-rank tensor representation}
\label{sec:SUN2sym}

The field configurations which minimize the potential~\eqref{eq:potential2ndrank} can always be transformed to either $\phi_{0}^{ab} = \delta^{ab}/\sqrt{N}$ for $\tilde\lambda>0$ or $\phi_{0}^{ab} = \delta^{aN}\delta^{Nb}$ if $\tilde\lambda<0$ \cite{Li:1973mq}. 
Thus, depending on the sign of $\tilde\lambda$ the breaking pattern reads either $\SUN \to \rmO(N)$ or $\SUN \to \SU(N-1)$, respectively.

\subsubsection{\texorpdfstring{$\tilde\lambda > 0$}{\tilde\lambda > 0}}
Beginning the analysis for the case $\tilde\lambda > 0$, it is useful to decompose the fluctuation field of the symmetric second-rank tensor into 3 parts which read for the particular choice $\phi_{0}^{ab} = \delta^{ab}/\sqrt{N}$, 
\begin{align*}
 \varphi = h \phi_{0} + \mathrm{Re}(\varphi - h \phi_{0}) + \I \mathrm{Im}(\varphi - h \phi_{0}).
\end{align*}
They are given by a complex scalar field proportional to the VEV, $\tr (\phi_{0}\varphi) \equiv h(x) = \frac{1}{\sqrt{2}}(h_{1}(x)+\I h_{2}(x))$, the real part of the remaining fluctuation field after the complex Higgs excitation $h$ proportional to the VEV has been subtracted, and the corresponding imaginary part. The last part, encoding $\frac{1}{2}N(N+1)-1$ real-valued scalar fields, mixes with the fields stored in $\re(A_{\mu})$ which acquire a nonvanishing mass term if $\phi_{0}^{ab} \sim \delta^{ab}$. Removing these would-be Goldstone fields by using the unitary gauge, the fluctuation field $\varphi$ contains $\frac{1}{2}N(N+1)+1$ real-valued scalar fields after symmetry breaking in the gauge-fixed formulation. Their mass spectrum is given by $\frac{1}{2}N(N+1)-1$ degenerated Higgs fields, $\frac{\sqrt{N}}{2}[(\varphi-h\phi_{0})\phi_{0} + \phi_{0}^{\dagger}(\varphi^{\dagger}-h^{*}\phid_{0})] = \mathrm{Re}(\varphi - h \phi_{0}) \equiv h_{\rmsym}/\sqrt{2}$, with mass $m^{2}_{h_{\rmsym}} = \tilde\lambda v^{2}$ transforming as a traceless symmetric second-rank tensor of $\rmO(N)$. Further, we obtain two real-valued singlets, the radial Higgs excitation $h_{1} \equiv \sqrt{2} \re\big( \tr(\phi_{0}\varphi)\big)$ with mass parameter $m_{\rmh_{1}}^{2} = (\lambda + \frac{1}{N} \tilde\lambda)v^{2}$, and a massless degree of freedom $h_{2} \equiv \sqrt{2} \im\big( \tr(\phi_{0}\varphi)\big)$.

Analyzing the mass matrix for the gauge fields given in Eq.~\eqref{eq:massmatrix2ndrank}, we obtain $\frac{1}{2}N(N+1)-1$ degenerated vector bosons stored in $A_{\rmsym}^{\mu} = \sqrt{N}(A^{\mu}\phi_{0} + \phi_{0}A^{\rmT\mu})/2$ which reduces to $\re(A^{\mu})$ for $\phi_{0}=\mathbbm{1}/\sqrt{N}$. Their mass parameter reads $m_{\rmA_{\rmsym}}^{2} = \frac{2}{N} g^{2} v^{2}$. The remaining $\frac{1}{2}N(N-1)$ massless vector bosons $A_{\rma}^{\mu} = \im(A^{\mu})$ form the gauge sector of the remaining $\rmO(N)$ gauge theory.

In order to investigate the bound state spectrum, we have to examine the global symmetries of the action which is given by a $\U(1)$ symmetry as in the fundamental case. 
In principle it is possible to classify the states according to this global quantum number. 
However, a particularity of the considered model is that the field configuration that minimizes the potential does not only break the gauge symmetry due to gauge-fixing but also spontaneously the global $\U(1)$ symmetry. In the gauge-fixed formulation this is reflected by the occurrence of the massless Higgs excitation $h_{2}$ which is the corresponding Goldstone boson. 
In order to show the breaking of the global $\U(1)$ symmetry in an $\SU(N)$-invariant fashion, a gauge-invariant order parameter can be constructed by $\langle \det(\phi) \rangle$ which is obviously invariant under an $\SU(N)$ gauge transformation. Indeed, this vacuum expectation value is nonvanishing and describes a homogeneous condensate for the field configuration that minimizes the potential. 
Therefore, we expect that also the gauge-invariant bound state spectrum contains a massless scalar particle. 
Thus, the $\U(1)$ quantum number is no longer a conserved quantity and transitions between $\U(1)$ singlets and non-singlets are possible.

The FMS expansion of the simplest gauge-invariant operator in the scalar channel \eqref{eq:scalarSinglet} (which would be a $\U(1)$ singlet) reads:
\begin{align}
\phi^{* \bar{a}\bar{b}}\phi^{ab} &= \frac{v^2}{2} + v h_{1} + \frac{1}{2} \big(h_{1}^{2}+h_2^2 + \tr(h_{\rmsym}^{2}) \big).
\label{eq:scalarSingletSym}
\end{align}
The operator expands in nontrivial leading order to the massive real-valued Higgs field proportional to the VEV. 
If we use the conventional multiplet decomposition, the next-to-leading order contribution contains not only the scattering of two radial Higgs fields $h_{1}$ but also scattering states of two massless fields $h_{2}$ and the meson state $\tr(h_{\rmsym}^{2})$ belonging to the second class, $2\varphi^{* \bar{a}\bar{b}}\varphi^{ab} = h_{1}^{2}+h_{2}^{2}+h_{\rmsym}^{ab}h_{\rmsym}^{ab}$. 
The possible scattering state $h_{2}^{2}$ produces a cut at $2m_{h_{2}} = 0$ for the correlator of $\tr(\phid\phi)$ implying that the state at $m_{\rmh_{1}}$ is merely a resonance. That the ground state of the scalar channel might indeed be given by a massless excitation as dictated by the Goldstone theorem can be further substantiated by investigating bound states with a nonvanishing $\U(1)$ quantum number. Although, this quantum number is no longer conserved, we can still formally build such operators and view them just as any other operator in the scalar channel which might have overlap with the ground state.
As the scalar field obeys more structure than in the fundamental case, it is easier to construct and expand such operators. For instance the operator~\eqref{eq:scalarNonSinglet},
\begin{align}
\det(\phi) &= \frac{v^{N}}{(2N)^{\frac{N}{2}}}  + \frac{\sqrt{N} v^{N-1}}{(2N)^{\frac{N-1}{2}}} (h_{1} + \I h_{2}) + \mathcal{O}(\varphi^{2}),
\label{eq:scalarNonSym}
\end{align}
expands in nontrivial leading order to the complex singlet ${h \sim h_{1}+\I h_{2}}$ via the FMS mechanism and similar for the charge conjugated operator, $\det(\phid){\sim} v^{N-1} h^{*} + \cdots$. Thus both scalar singlets $h$ and $h^{*}$ or equivalently their real-valued counterparts $h_{1}$ and $h_{2}$ belong to the first class. 
As the $\U(1)$ charged composite object $\det(\phi)$ acquires a nonvanishing VEV, the generated states linked by a $\U(1)$ transformation obtain different masses. We have a massless Goldstone mode $\im(\det\phi){\sim}h_{2}$ associated with the broken $\U(1)$ generator and a real-valued massive radial mode in the spectrum.

A further state in the scalar channel belonging to the first class is the meson operator $\tr A_{\rmsym}^{2}$ which can be deduced from $\tr(\phid D^{2}\phi)$. The hybrid $\tr F_{\rmsym}^{2}$ and the $\rmO(N)$ glueball $\tr F_{\rma}^{2}$ belong to the second class as they can only be extracted from $\tr F^{2}$ or $\tr(\phid F^{2}\phi)$ via the standard multiplet decomposition. In addition, $\tr F^{2}$ and $\tr(\phid F^{2}\phi)$ contain the $\rmO(N)$-invariant states $\tr(F_{\rma}A_{\rmsym}A_{\rmsym})$ and $\tr(F_{\rmsym}A_{\rmsym}A_{\rmsym})$.

We investigate the operator defined in Eq.~\eqref{eq:vectorSinglet} to get a first glance on the vector channel. Its FMS expansion, reveals that it does not expand to a single elementary vector particle as the leading order term, $\tr \phid_{0}A_{\mu}\phi_{0}^{\phantom{\,}} $, vanishes for $\phi_{0} \sim \mathbbm{1}$. At next-to-leading order we obtain the vector meson $\tr(h_{\rmsym}^{\phantom{\,}}A_{\rmsym}^{\mu})$ being invariant under $\rmO(N)$ transformations. Further next-level states which belong to the second class as well as the trivial contributions from the scalar channel are encoded in the remaining terms.
Albeit, it is not possible to construct a gauge-invariant vector operator that expands to a single vector field, it is straightforward to write down an operator that expands to an $\rmO(N)$-invariant operator containing several vector fields in the FMS prescription. For instance, the operator 
\begin{align}
 \tr (\phid F^{\mu\nu}D_{\nu}\phi) = \frac{2\I g v^2}{N} \tr ( F_{\rmsym}^{\mu\nu} A_{\rmsym \nu}^{\phantom{\,}}) + \mathcal{O}(\varphi)
\end{align}
expands unambiguously to the hybrid state $\tr ( F_{\rmsym}^{\mu\nu} A_{\rmsym \nu}^{\phantom{\,}})$ containing the massive vector multiplet. Higher orders in the FMS expansion combined with the multiplet decomposition reveal several $\rmO(N)$-invariant operators of the second class which are summarized in Tab.~\ref{tab:SUsym}.


\begin{table*}[t!]
\begin{center}

\begin{tabular}{c|| cl | ll || lcc}

\toprule

&   \multicolumn{2}{c|}{$\SU(N)$ invariant} & \multicolumn{2}{c||}{\quad $\rmO(N)$ singlets \quad} & \multicolumn{3}{c}{$\rmO(N)$ multiplets} \\
\hline
$J^P$ & $\U(1)$ &  Operator                &  1. Class \, & [2. Class] \,                & Field \, & DOF &  $m_{\mathrm{Field}}^{2}$ \\
\hline

$0^+$  & 0 & $\tr(\phid\phi)$  & $h_{1}$		  & [$\tr h_{\rmsym}^{2}$]    &		 $h_{1}$ & 1 & $(\lambda + \frac{\tilde\lambda}{N})v^{2}$  \cr
 & 0 & $\tr(\phid D^{2}\phi)$        & $\tr A_{\rmsym}^{2}$, [$h_{1}$]   & [$\tr(h_{\rmsym}^{\vphantom{\mu}}D_{\mu}A_{\rmsym}^{\mu})$], [$\tr(A_{\rmsym}^{\mu}D_{\mu}h_{\rmsym}^{\vphantom{\mu}})$], [$\tr(h_{\rmsym}^{\vphantom{\mu}}A_{\rmsym}A_{\rmsym})$]           & $h_{2}$ & 1 & 0 \cr
 & 0 & $\tr F^{2}$        &   & [$\tr F_{\rma}^{2}$], [$\tr F_{\rmsym}^{2}$], [$\tr(F_{\rma}A_{\rmsym}A_{\rmsym})$], [$\tr(F_{\rmsym}A_{\rmsym}A_{\rmsym})$]          & $h_{\rmsym}$ & $\frac{N(N+1)}{2}-1$ & $\frac{\tilde\lambda v^{2}}{N}$  \cr
 & 0 & $\tr (\phid F^{2}\phi)$        &    & [$\tr F_{\rma}^{2}$], [$\tr F_{\rmsym}^{2}$], [$\tr(F_{\rma}A_{\rmsym}A_{\rmsym})$], [$\tr(F_{\rmsym}A_{\rmsym}A_{\rmsym})$]           &  &  &  \cr
\cline{2-5}
  & $1/\bar{1}$ & $\det(\phi)/\det(\phid)$        & $h_{1}$, $h_{2}$    & [$\det h_{\rmsym}$], [$\tr h_{\rmsym}^{n}$ ($n<N$)]           &  &  &  \cr
\hline
$1^-$  & 0 & $\tr(\phid D^{\mu}\phi)$             & $\tr (A_{\rmsym}^{\mu}h_{\rmsym}^{\phantom{\mu}})$  & [$\tr(h_{\rmsym}^{2\vphantom{\mu}}A_{\rmsym}^{\mu})$], [$\tr(h_{\rmsym}D^{\mu}h_{\rmsym})$]              & $A_{\rma}^{\mu}$ & $\frac{N(N-1)}{2}$ & 0 \cr
 & 0 & $\tr(\phid F^{\mu\nu}D_{\nu}\phi)$             & $\tr ( F_{\rmsym}^{\mu\nu} A_{\rmsym \nu}^{\phantom{\mu}})$   & [$\tr(F_{\rmsym}^{\mu\nu}D_{\nu}h_{\rmsym}^{\vphantom{\mu}})$], [$\tr(F_{\rmsym}^{\mu\nu}A_{\rmsym\nu}^{\vphantom{\mu}}h_{\rmsym}^{\vphantom{\mu}})$], [$\tr(F_{\rma}^{\mu\nu}A_{\rmsym\nu}^{\vphantom{\mu}}h_{\rmsym}^{\vphantom{\mu}})$]              & $A_{\rmsym}^{\mu}$ & $\frac{N(N+1)}{2}-1$ & $\frac{2g^{2}v^{2}}{N}$ \cr
\cline{2-5}
 & $1/\bar{1}$ & see Op.~\eqref{eq:vectorNonSinglet}            & $\tr (A_{\rmsym}^{\mu}h_{\rmsym}^{\phantom{\mu}})$    & [see main text]              &  &  &  \cr
 & $1/\bar{1}$ & see Op.~\eqref{eq:vectorNonSinglet2}             & $\tr ( F_{\rmsym}^{\mu\nu} A_{\rmsym \nu}^{\phantom{\mu}})$    & [see main text]             &  &  &  \cr
 \toprule
\end{tabular}
\caption{Particle content of an $\SU(N)$ gauge theory ($N>2$) with a scalar field in the symmetric second-rank tensor representation and gauge-variant breaking pattern $\SU(N) \to \rmO(N)$.  
Left: Comparison between operators/states that are strict invariant with respect to $\SU(N)$ transformations, i.e., observables, and operators/states that are considered from the conventional but gauge-variant viewpoint of spontaneous gauge symmetry breaking ($\rmO(N)$ singlets). Trivial scattering states are ignored. In case the contraction is obvious, we suppress Lorentz indices for better readability. Albeit the global $\U(1)$ symmetry is broken, we formally divide the $\SU(N)$-invariant operators into $\U(1)$ singlets and nonsinglets. We assign a $\U(1)$ charge of $1/N$ to the scalar field $\phi$. Right: Properties of the elementary building blocks obtained from the standard multiplet decomposition after gauge fixing which are used to construct $\rmO(N)$ singlets.}

\label{tab:SUsym}
\end{center}
\end{table*}


Similar conclusions can be drawn from more involved operators. For instance, the operator~\eqref{eq:vectorNonSinglet} expands in leading order in the scalar fluctuation field to $\tr A^{\mu} = 0$ confirming the findings from the analysis of the $\U(1)$ singlet operator~\eqref{eq:vectorSinglet}. At next to leading order we obtain $\tr (A_{\rmsym}^{\mu}\varphi) = \tr (A_{\rmsym}^{\mu}h_{\rmsym}^{\phantom{\mu}})$. Therefore, we also infer from a formally $\U(1)$ charged operator\footnote{Keep in mind that the global $\U(1)$ symmetry is broken and thus the distinction is merely for practical purposes.} that this $\rmO(N)$-invariant vector meson belongs to the first class. Using the simple constituent model, we approximate its mass by $m_{\rmA_{\rmsym}} + m_{\rmh_{\rmsym}}$. Similarly, the FMS expansion of the operator~\eqref{eq:vectorNonSinglet2} predicts a state of the first class at ${\approx} 2m_{\rmA_{\rmsym}}$ as we obtain a term proportional to $\tr ( F_{\rmsym}^{\mu\nu} A_{\rmsym \nu}^{\phantom{\,}})$. This vector hybrid is also the leading order contribution of the operator $\epsilon^{\bar{a}_1 \cdots \bar{a}_N} \epsilon^{\bar{b}_1 \cdots \bar{b}_N} (F^{\mu\nu}\phi)^{a_{1}b_{1}} (D_{\nu}\phi)^{a_2b_2} \phi^{a_3b_3} \cdots \phi^{a_{N}b_{N}}$. Further states assigned to the second class and containing an increasing number of $h_{\rmsym}$ as constituents appear in these $\U(1)$ charged operators at higher orders in the expansion coefficients of the FMS series. An overview of the $\mathcal{G}$-$\mathcal{H}$ duality of the present model is depicted in Tab.~\ref{tab:SUsym}.

\subsubsection{\texorpdfstring{$\tilde\lambda < 0$}{\tilde\lambda < 0}}
Without loss of generality the field configuration that minimizes the potential~\eqref{eq:potential2ndrank} is given by $\phi_{0}^{ab} = \delta^{aN}\delta^{Nb}$ for a negative coupling $\tilde\lambda$ \cite{Li:1973mq}. Note that $|\tilde\lambda| < \lambda$ for a stable potential. It is straightforward to show that the gauge-dependent vacuum configuration remains invariant under an $\SU(N-1)$ subgroup. Inserting the minimizing field configuration into Eq.~\eqref{eq:massmatrix2ndrank}, we obtain the mass parameter matrix for the gauge bosons. We obtain a complex fundamental vector $A_{\rmf} \equiv A\phi_{0} - A_{\rms}\phi_{0}$ with $m_{\rmA_{\rmf}}^{2}= \frac{1}{2}g^{2}v^{2}$ and an $\SU(N-1)$ singlet $A_{\rms} = \tr(\phid_{0} A \phi_{0})$ being generically heavier $m_{\rmA_{\rms}}^{2}=\frac{4(N-1)}{N}m_{\rmA_{\rmf}}^{2}$. The remaining $(N-1)^2-1$ gauge bosons are the massless vector degrees of freedom of the unbroken $\SU(N-1)$ group. 
As the breaking pattern is the same as in the fundamental case, we obtain the same structural decomposition of the elementary gauge field. Nonetheless, the internal dynamics is different which manifests, e.g., in the different ratio $m_{\rmA_{\rmf}}^{2}/m_{\rmA_{\rms}}^{2}$, c.f. App.~\ref{app:SUNfun}.
The elementary scalar field $\phi$ decomposes into the real-valued radial Higgs excitation $h = \sqrt{2}\re \big(\tr(\phid_{0}\varphi) \big)$ with $m_{\rmh}^2 = (\lambda + \tilde\lambda)v^{2}$ and $N(N-1)$ degenerated complex scalars forming a complex symmetric second-rank $\SU(N-1)$ tensor field $h_{\rmsym}$ with $m_{\rmh_{\rmsym}}^2 = |\tilde\lambda| v^{2}$. The remaining massless scalar degrees of freedom mix with those gauge bosons acquiring a nonvanishing mass term and thus are would-be Goldstones which we remove by the unitary gauge condition.

In contrast to the previous case ($\tilde\lambda > 0$), the gauge-invariant spectrum can be defined according to the global $\U(1)$ quantum number of the model. 
At first sight, it seems that the field configuration $\phi_{0}^{ab} = \delta^{aN}\delta^{Nb}$ breaks the generator of the global $\U(1)$ symmetry in the gauge-fixed set up. However, this is a gauge-dependent statement as $\phi_{0}$ is a gauge-dependent order parameter. The model still obeys an unbroken global $\U(1)$ symmetry which manifests in an unbroken diagonal subgroup of $\SU(N)\times \U(1)$ in the gauge-fixed formulation. Thus, it is still possible to classify the states according to their $\U(1)$ quantum number. This statement can be made more precise in a gauge-invariant way by realizing that the gauge-invariant order parameter $\langle \det \phi \rangle$ vanishes for the particular field configuration which minimizes the potential for $\tilde\lambda < 0$. Also any other $\U(1)$ charged operator does not develop a homogeneous condensate at least within the perturbative regime.

Beginning with the $\U(1)$-singlet scalar channel, we analyze the operators $\tr(\phid\phi)$, $\tr(\phid D^{2}\phi)$, $\tr F^{2}$, and $\tr (\phid F^{2}\phi)$. From their FMS expansion, e.g., see Eq.~\eqref{eq:scalarSinglet} and Eq.~\eqref{eq:scalarSinglet2}, we conclude that only the radial Higgs excitation $h$ generates a state of the first class. All other scalar $\SU(N-1)$-singlets contained in these operators, e.g., $\tr(h_{\rmsym}^{\dagger}h_{\rmsym}^{\phantom{\,}})$, $\tr(A_{\rmf}^{\dagger}A_{\rmf}^{\phantom{\,}})$, $\tr(F_{\rmf}^{\dagger}F_{\rmf}^{\phantom{\,}})$, and $\tr F_{\rma}^{2}$, can only be obtained via the conventional multiplet decomposition as they mix with the scattering states $A_{\rms}^{2}$ and $h^{2}$. For the precise relation between the $\SU(N)$ and $\SU(N-1)$ states see Tab.~\ref{tab:SUsymneg}.


\begin{table*}[t!]
\begin{center}

\begin{tabular}{c|| cl | ll || lcc}

\toprule

&   \multicolumn{2}{c|}{$\SU(N)$ invariant} & \multicolumn{2}{c||}{\quad $\SU(N-1)$ singlets \quad} & \multicolumn{3}{c}{$\SU(N-1)$ multiplets} \\
\hline
$J^P$ & $\U(1)$ &  Operator \,               &  1. Class \, & [2. Class] \,                & Field \, & DOF &  $m_{\mathrm{Field}}^{2}$ \\
\hline

$0^+$  & 0 & $\tr(\phid\phi)$  & $h$		  & [$\tr(h_{\rmsym}^{\dagger}h_{\rmsym}^{\vphantom{\dagger}})$]    &		 $h$ & 1 & $(\lambda + \tilde\lambda)v^{2}$  \cr
 & 0 & $\tr(\phid D^{2}\phi)$        & [$h$]    & [$\tr(A_{\rmf}^{\dagger}A_{\rmf}^{\vphantom{\dagger}})$], [$\tr(h_{\rmsym}^{*}A_{\rmf}^{\vphantom{\rmT}}A_{\rmf}^{\rmT})$], [$\tr(h_{\rmsym}^{\phantom{a}}A_{\rmf}^{*\vphantom{\dagger}}A_{\rmf}^{\dagger})$]           & $h_{\rmsym}$ & ${N(N-1)}$ & $|\tilde\lambda | v^{2}$ \cr
 & 0 & $\tr F^{2}$        &   & [$\tr F_{\rma}^{2}$], [$\tr(F_{\rmf}^{\dagger}F_{\rmf}^{\vphantom{\dagger}})$], [$\tr(A_{\rmf}^{\dagger}F_{\rma}^{\vphantom{\dagger}}A_{\rmf}^{\vphantom{\dagger}})$]           &  &  &   \cr
 & 0 & $\tr (\phid F^{2}\phi)$        &    & [$\tr(F_{\rmf}^{\dagger}F_{\rmf}^{\phantom{\,}})$]           &  &  &  \cr
\cline{2-5}
  & $1/\bar{1}$ & $\det \phi$/$\det \phi^{\dagger}$        & $\det h_{\rmsym}$/$\det h_{\rmsym}^{\dagger}$    &            &  &  &  \cr
\hline
$1^-$  & 0 & $\tr(\phid D^{\mu}\phi)$             & $A_{\rms}^{\mu}$  & [$\tr(h_{\rmsym}^{\dagger} D^{\mu} h_{\rmsym}^{\vphantom{\dagger}})$]              & $A_{\rma}^{\mu}$ & $(N-1)^{2}-1$ & 0 \cr
 & 0 & $\tr(\phid F^{\mu\nu}D_{\nu}\phi)$             &    & [$\tr(F_{\rmf}^{\dagger \mu\nu}A_{\rmf \nu}^{\vphantom{\dagger}})$], [$\tr(F_{\rmf}^{\dagger \mu\nu}h_{\rmsym}^{\vphantom{\dagger}}A_{\rmf \nu}^{*\vphantom{\dagger}})$], $[\tr(F_{\rmf}^{\rmT \mu\nu}h_{\rmsym}^{*\vphantom{\mu}}A_{\rmf \nu}^{\vphantom{\mu}})$]             & $A_{\rms}^{\mu}$ & $1$ & $\frac{2(N-1)}{N}g^{2}v^{2}$  \cr
\cline{2-5}
 & $1/\bar{1}$ & see Op.~\eqref{eq:vectorNonSinglet}            & see main text    &              & $A_{\rmf}^{\mu}$ & $2(N-1)$  & $\frac{g^{2}v^{2}}{2}$   \cr
 & $1/\bar{1}$ & see Op.~\eqref{eq:vectorNonSinglet2}             &   & [see main text]             &  &  &  \cr
 \toprule
\end{tabular}
\caption{Particle content of an $\SU(N)$ gauge theory ($N>2$) with a scalar field in the symmetric second-rank tensor representation and gauge-variant breaking pattern $\SU(N) \to \SU(N-1)$.  
Left: Comparison between operators/states that are strict invariant with respect to $\SU(N)$ transformations, i.e., observables, and operators/states that are considered from the conventional but gauge-variant viewpoint of spontaneous gauge symmetry breaking ($\SU(N-1)$ singlets). Trivial scattering states are ignored. In case the contraction is obvious, we suppress Lorentz indices for better readability. We assign a $\U(1)$ charge of $1/N$ to the scalar field $\phi$. Right: Properties of the elementary building blocks obtained from the standard multiplet decomposition after gauge fixing which are used to construct $\SU(N-1)$ singlets.}

\label{tab:SUsymneg}
\end{center}
\end{table*}


The gauge-invariant $\U(1)$-nonsinglet operator~\eqref{eq:scalarNonSinglet} that contains only scalar fields expands in leading order to the $\SU(N-1)$ invariant state $\det(h_{\rmsym})$,
\begin{align}
 \det \phi = \frac{1}{\sqrt{2}} (v + h) \det(h_{\rmsym}).
\end{align}
Further operators might be investigated for a complete analysis of the mass spectrum within this channel but all local operators that can be build from the elementary fields will generically expand to operators with larger field content than $N-1$ scalar fields.

The simplest gauge-invariant vector operator with vanishing $\U(1)$ quantum number expands to the heaviest vector boson which is an $\SU(N-1)$ singlet,
\begin{align}
 \I\, \tr (\phid D^{\mu}\phi) = - \frac{gv^2}{2} A_{\rms}^{\mu} + \mathcal{O}(\varphi).
\end{align}
This is equivalent to the fundamental case which creates the same breaking pattern. 
Further, the vector operator defined in Eq.~\eqref{eq:SUNvectorSinglet2} expands in leading order to a superposition of the vector hybrid $\tr(F_{\rmf}^{\dagger \mu\nu}A_{\rmf \nu}^{\phantom{\,}})$ and the scattering state $\tr(F_{\rms}^{\mu\nu}A_{\rms \nu})$. Thus, the former state belongs to the second class.

To investigate the bound state structure of vector particles charged under the global $\U(1)$ symmetry, we first study the operator~\eqref{eq:vectorNonSinglet}. 
The leading order contribution forms an intricate $\SU(N)$-invariant hybrid created from $N-2$ different components of $h_{\rmsym}$ and one element of $D^{\mu}h_{\rmsym}$, i.e., $\epsilon^{\dot{\bar{a}}_1 \cdots \dot{\bar{a}}_{N-1}} \epsilon^{\dot{\bar{b}}_1 \cdots \dot{\bar{b}}_{N-1}}  (D^\mu h_{\rmsym})^{\dot{a}_1\dot{b}_1} h_{\rmsym}^{\dot{a}_2\dot{b}_2} \cdots h_{\rmsym}^{\dot{a}_{N-1}\dot{b}_{N-1}}$ where dotted indices run only over those components defining the nontrivial $\SU(N-1)$ subspace. As this hybrid operator follows in a unique way from the FMS expansion, we assign it to the first class. Furthermore, we obtain trivial scattering states, e.g., given by $A_{\rms}^{\mu}$ and $\det h_{\rmsym}$. 
The operator defined in Eq.~\eqref{eq:vectorNonSinglet2} expands in leading order to scattering states as well as another $\SU(N-1)$-invariant hybrid, $\epsilon^{\dot{\bar{a}}_1 \cdots \dot{\bar{a}}_{N-1}} \epsilon^{\dot{\bar{b}}_1 \cdots \dot{\bar{b}}_{N-1}}  (F_{\rmf}^{\mu\nu}A_{\rmf \nu}^{\rmT})^{\dot{a}_1\dot{b}_1} h_{\rmsym}^{\dot{a}_2\dot{b}_2} \cdots h_{\rmsym}^{\dot{a}_{N-1}\dot{b}_{N-1}}$, containing elements of $h_{\rmsym}$ as well as the nonsinglet massive vector multiplet. As we can disentangle this hybrid and the scattering states only via the conventional multiplet decomposition, we assign the hybrid to the second class. At higher orders, we obtain scattering states such as $\tr(F_{\rmf}^{\dagger \mu\nu}A_{\rmf \nu}^{\phantom{\,}})$ and $\det h_{\rmsym}$ or $\tr(F_{\rms}^{\mu\nu}A_{\rms \nu})$ and $\det h_{\rmsym}$.

\subsection{Antisymmetric second-rank tensor representation}
\label{sec:SUN2asym}

We have to consider two different breaking patterns depending on the sign of the non-isotropic coupling $\tilde\lambda$ also for the antisymmetric second-rank tensor representation.

\subsubsection{\texorpdfstring{$\tilde\lambda > 0$}{\tilde\lambda > 0}}
First, we consider the case $\tilde\lambda > 0$. 
As the breaking pattern reads $\SUN \to \Sp(2\lfloor N/2 \rfloor)$ where $\lfloor x \rfloor$ is the floor function \cite{Li:1973mq,Jetzer:1983ij}, it is useful to consider the cases for even and odd argument separately as in case of the adjoint representation for $\SO(N)$ gauge theories. 
The direction of the field configuration that minimizes the potential~\eqref{eq:potential2ndrank} can be transformed into the following block diagonal form
\begin{align}
 \phi_{0} = \frac{1}{\sqrt{2K}} \begin{pmatrix} \varepsilon & & \\ & \ddots & \\ & & \varepsilon   \end{pmatrix}
 \label{eq:SUNasymMin}
\end{align}
for $\SU(N=2K)$ with $\varepsilon = \begin{pmatrix} 0 & 1 \\ -1 & 0 \end{pmatrix}$ and all off-diagonal elements vanish \cite{Li:1973mq}. Note that also this field configuration breaks the global $\U(1)$ symmetry. 

The mass matrix for the vector fields given in Eq.~\eqref{eq:massmatrix2ndrank} has $2K^2+K$ vanishing eigenvalues. The corresponding massless vector fields are gauge bosons of the unbroken $\Sp(2K)$ Yang-Mills sector in the gauge-fixed version of the model. The remaining vector fields, $A_{\rmasym} = A\phi_{0} + \phi_{0}A^{\rmT}$ form an antisymmetric second-rank tensor of $\Sp(2K)$ with mass parameter $m_{\rmA_{\rmasym}}^{2} = \frac{1}{K}g^{2}v^{2}$.

Accordingly, the scalar fluctuation field $\varphi$ contains would-be Goldstone bosons forming an antisymmetric tensor which we remove from the elementary spectrum by the unitary gauge. The remaining scalar degrees of freedom form another antisymmetric $\Sp(2K)$ tensor with mass parameter $m_{\rmh_{\rmasym}} = \frac{1}{2K} \tilde\lambda v^{2}$ and two real-valued singlets being the real and imaginary part proportional to $\phi_0$. While the real part is the radial Higgs excitation $h_{1} = \sqrt{2} \re\big( \tr(\phid_{0}\varphi) \big)$ and thus massive, $m_{\rmh_{1}} = (\lambda + \frac{1}{2K}\tilde\lambda)v^{2}$, the imaginary part $h_{2} = \sqrt{2} \im\big( \tr(\phid_{0}\varphi) \big)$ is massless but a BRST singlet. This field is a real Goldstone degree of freedom associated to the broken generator of the global $\U(1)$ symmetry group of the model.


\begin{table*}[t!]
\begin{center}

\begin{tabular}{c|| cl | ll || lcc}

\toprule

&   \multicolumn{2}{c|}{$\SU(2K)$ invariant} & \multicolumn{2}{c||}{\quad $\Sp(2K)$ singlets \quad} & \multicolumn{3}{c}{$\Sp(2K)$ multiplets} \\
\hline
$J^P$ & $\U(1)$ &  Operator                &  1. Class  & [2. Class]                 & Field  & DOF &  $m_{\mathrm{Field}}^{2}$ \\
\hline

$0^+$  & 0 & $\tr(\phid\phi)$  & $h_{1}$		  & [$\tr (h_{\rmasym}^{\dagger}h_{\rmasym}^{\phantom{\dagger}})$]    &		 $h_{1}$ & 1 & $(\lambda + \frac{1}{2K}\tilde\lambda)v^{2}$  \cr
 & 0 & $\tr(\phid D^{2}\phi)$        & $\tr (A_{\rmasym}^{\dagger}A_{\rmasym}^{\phantom{\dagger}})$, [$h_{1}$]   & [$\tr(h_{\rmasym}^{\dagger}D_{\mu}A_{\rmasym}^{\mu})$], [$\tr(A_{\rmasym}^{\dagger\mu}D_{\mu}h_{\rmasym}^{\vphantom{\dagger}})$]            & $h_{2}$ & 1 & 0 \cr
 & 0 & $\tr F^{2}$        &   & [$\tr F_{\rma}^{2}$], [$\tr (F_{\rmasym}^{\dagger}F_{\rmasym}^{\phantom{\dagger}})$], [$\tr (A_{\rmasym}^{\dagger}F_{\rma}^{\vphantom{\dagger}}A_{\rmasym}^{\vphantom{\dagger}})$]           & $h_{\rmasym}$ & $2K^{2}-K-1$ & $\frac{\tilde\lambda v^{2}}{2K}$  \cr
 & 0 & $\tr (\phid F^{2}\phi)$        &    & [$\tr F_{\rma}^{2}$], [$\tr (F_{\rmasym}^{\dagger}F_{\rmasym}^{\vphantom{\dagger}})$], [$\tr (A_{\rmasym}^{\dagger}F_{\rma}^{\phantom{\dagger}}A_{\rmasym}^{\phantom{\dagger}})$]          &  &  &  \cr
\cline{2-5}
  & $1/\bar{1}$ & $\det(\phi)$        & $h_{1}$, $h_{2}$    & [$\det h_{\rmasym}$], [see main text]          &  &  &  \cr
  & $1/\bar{1}$ & $\epsilon^{\bar{a}_1\cdots \bar{a}_{2K}}\phi^{a_{1}a_{2}}\cdots \phi^{a_{2K-1}a_{2K}}$        & $h_{1}$, $h_{2}$    & [see main text]         &  &  &  \cr
\hline
$1^-$  & 0 & $\tr(\phid D^{\mu}\phi)$             & $\tr (h_{\rmasym}^{\dagger}A_{\rmasym}^{\mu})$  & [$\tr (h_{\rmasym}^{\dagger}D_{\mu}h_{\rmasym}^{\phantom{\dagger}})$]             & $A_{\rma}^{\mu}$ & $K(2K+1)$ & 0 \cr
 & 0 & $\tr(\phid F^{\mu\nu}D_{\nu}\phi)$             & $\tr ( F_{\rmasym}^{\dagger\mu\nu} A_{\rmasym \nu}^{\phantom{\,}})$   & [$\tr(F_{\rmasym}^{\dagger\mu\nu}D_{\nu}h_{\rmasym}^{\vphantom{\dagger}})$], [$\tr(h_{\rmasym}^{\dagger}F_{\rma\vphantom{2}}^{\mu\nu}A_{\rmasym\nu}^{\vphantom{\dagger}})$]              & $A_{\rmasym}^{\mu}$ & $2K^{2}-K-1$ & $\frac{g^{2}v^{2}}{K}$ \cr
\cline{2-5}
 & $1/\bar{1}$ & see Op.~\eqref{eq:vectorNonSinglet}            & see main text    & [see main text]             &  &  &  \cr
 & $1/\bar{1}$ & see Op.~\eqref{eq:vectorNonSinglet2}             &     & [see main text]             &  &  &  \cr
 \toprule
\end{tabular}
\\[0.5cm]

\begin{tabular}{c|| cl | ll || lcc}
\toprule

&   \multicolumn{2}{c|}{$\SU(2K+1)$ invariant} & \multicolumn{2}{c||}{\quad $\Sp(2K)$ singlets \quad} & \multicolumn{3}{c}{$\Sp(2K)$ multiplets} \\
\hline
$J^P$ & $\U(1)$ &  Operator                &  1. Class  & [2. Class]                 & Field  & DOF &  $m_{\mathrm{Field}}^{2}$ \\
\hline

$0^+$  & 0 & $\tr(\phid\phi)$  & $h$		  & [$\tr (h_{\rmasym}^{\dagger}h_{\rmasym}^{\phantom{\dagger}})$]    &		 $h$ & 1 & $(\lambda + \frac{1}{2K}\tilde\lambda)v^{2}$  \cr
 & 0 & $\tr(\phid D^{2}\phi)$        & [$h$]    & [$\tr (A_{\rmasym}^{\dagger}A_{\rmasym}^{\phantom{\dagger}})$], [$\tr (A_{\rmf}^{\dagger}A_{\rmf}^{\phantom{\dagger}})$], [$\tr(h_{\rmasym}^{\dagger}D_{\mu}A_{\rmasym}^{\mu})$],             & $h_{\rmasym}$ & $2K^{2}-K-1$ & $\frac{\tilde\lambda v^{2}}{2K}$ \vphantom{$\frac{\tilde K^{2^{2}}}{\tilde K}$} \cr
&  &         &    &  [$\tr(A_{\rmasym}^{\dagger\mu}D_{\mu}h_{\rmasym}^{\vphantom{\dagger}})$]            &  &  &  \cr
 & 0 & $\tr(\phid\phi(D^{\mu}\phi)^{\dagger}\phi\phid D_{\mu}\phi)$        &    & [$\tr (A_{\rmasym}^{\dagger}A_{\rmasym}^{\phantom{\dagger}})$]           &  &  &  \cr
 & 0 & $\tr F^{2}$        &   & [$\tr F_{\rma}^{2}$], [$\tr (F_{\rmasym}^{\dagger}F_{\rmasym}^{\phantom{\dagger}})$], [$\tr (F_{\rmf}^{\dagger}F_{\rmf}^{\phantom{\dagger}})$],            &  &  &  \cr
  &  &         &   & [$\tr (A_{\rmasym}^{\dagger}F_{\rma}^{\phantom{\,}}A_{\rmasym}^{\phantom{\dagger}})$], [$\tr (A_{\rmf}^{\dagger}F_{\rma}^{\phantom{\,}}A_{\rmf}^{\phantom{\dagger}})$]            &  &  &  \cr
 & 0 & $\tr (\phid F^{\mu\nu}\phi\phid F_{\mu\nu}\phi)$        &    & [$\tr F_{\rma}^{2}$], [$\tr (F_{\rmasym}^{\dagger}F_{\rmasym}^{\phantom{\dagger}})$], [$\tr (A_{\rmasym}^{\dagger}F_{\rma}^{\phantom{\,}}A_{\rmasym}^{\phantom{\dagger}})$],           &  &  &  \cr
 &  &         &    & [$\tr (A_{\rmf}^{\dagger}F_{\rma}^{\phantom{\,}}A_{\rmf}^{\phantom{\dagger}})$]          &  &  &  \cr
\cline{2-5}
  & $1/\bar{1}$ & see Op.~\eqref{eq:scalarNonSinglet3}        &    \multicolumn{2}{c||}{only trivial scattering states}               &  &  &  \cr
\hline
$1^-$  & 0 & $\tr(\phid D^{\mu}\phi)$             & $A_{\rms}^{\mu}$  & [$\tr (h_{\rmasym}^{\dagger}A_{\rmasym}^{\mu})$], [$\tr (h_{\rmasym}^{\dagger}D^{\mu}h_{\rmasym}^{\phantom{\dagger}})$]            & $A_{\rma}^{\mu}$ & $K(2K+1)$ & 0 \cr
 & 0 & $\tr(\phid F^{\mu\nu}D_{\nu}\phi)$             &    & [$\tr ( F_{\rmasym}^{\dagger\mu\nu} A_{\rmasym \nu}^{\phantom{\dagger}})$], [$\tr(A_{\rmf\nu}^{\dagger} F_{\rmf}^{\mu\nu})$],              & $A_{\rmasym}^{\mu}$ & $2K^{2}-K-1$ & $\frac{g^{2}v^{2}}{K}$ \cr
&  &         &    & [$\tr ( F_{\rmasym}^{\dagger\mu\nu} D_{\nu}h_{\rmasym}^{\phantom{\dagger}})$], [$\tr (h_{\rmasym}^{\dagger}F_{\rma}^{\phantom{\,}}A_{\rmasym}^{\phantom{\dagger}})$]           & $A_{\rmf}^{\mu}$ & $4K$ & $\frac{g^{2}v^{2}}{4K}$\vphantom{$\frac{\tilde K^{2^{2}}}{\tilde K}$}  \cr
\cline{2-5}
 & $1/\bar{1}$ & see Op.~\eqref{eq:vectorNonSinglet2}            &    & [see main text]             & $A_{\rms}^{\mu}$  & 1  & $\frac{g^{2}v^{2}}{K(2K+1)}$\vphantom{$\frac{\tilde K^{2^{2}}}{\tilde K}$}  \cr
 \toprule
\end{tabular}

\caption{Particle content of an $\SU(N)$ gauge theory ($N>2$) with a scalar field in the antisymmetric second-rank tensor representation and gauge-variant breaking pattern $\SU(N) \to \Sp(2\lfloor N/2 \rfloor)$. The upper table contains the spectrum for even $N$ while the lower table summarizes the spectrum for odd $N$.
Left: Comparison between operators/states that are strict invariant with respect to $\SU(N)$ transformations, i.e., observables, and operators/states that are considered from the conventional but gauge-variant viewpoint of spontaneous gauge symmetry breaking ($\Sp(2K)$ singlets). Trivial scattering states are ignored. In case the contraction is obvious, we suppress Lorentz indices for better readability. Albeit the global $\U(1)$ symmetry is broken for $N=2K$, we formally divide the $\SU(2K)$-invariant operators into $\U(1)$ singlets and nonsinglets. We assign a $\U(1)$ charge of $1/N$ to the scalar field $\phi$. Right: Properties of the elementary building blocks obtained from the standard multiplet decomposition after gauge fixing which are used to construct $\Sp(2K)$ singlets.}

\label{tab:SUantisym}
\end{center}
\end{table*}


Albeit the VEV of the scalar field $\phi_0$ is gauge-dependent, we are able to formulate the breaking of the global symmetry group in a gauge-invariant manner due to the gauge-invariant order operator $\langle \det\phi \rangle$. Similar to the case of a symmetric second-rank tensor, this order parameter is nonvanishing for the field configuration~\eqref{eq:SUNasymMin} that minimizes the potential~\eqref{eq:potential2ndrank} for $\tilde\lambda >0$ and even $N$. Likewise, the $\U(1)$ charged but $\SU(N)$-invariant operators $\det \phi$ and $\det \phid$ can be translated into two strict gauge-invariant real-valued modes which expand to the elementary fields $h_{1}$ and $h_{2}$, implying that the scalar channel has no mass gap. Thus, we find indeed an $\SU(N)$-invariant composite Goldstone degree of freedom which is connected to the elementary massless field $h_{2}$ via the FMS expansion. As the global $\U(1)$ symmetry is spontaneously broken, we have transitions between $\U(1)$-singlet and -nonsinglet states.

As usual, the FMS expansion of the operator~\eqref{eq:scalarSinglet} predicts that the gauge-invariant bound state spectrum in the scalar channel contains the elementary radial Higgs field which belongs to the first class. The $\Sp(2K)$ meson bound state of two elementary scalar antisymmetric tensor fields $\tr(h_{\rmasym}^{\dagger}h_{\rmasym}^{\phantom{\dagger}})$ belongs to the second class. Further, we obtain the branch cut signaling the decay into two massless Goldstone modes. The analysis of further bound state operators in the scalar as well as the vector channel follows the recipes of the previous sections. We list $\SU(2K)$ gauge-invariant operators and the corresponding $\Sp(2K)$ singlets obtained from either the FMS decomposition or the multiplet decomposition in Tab.~\ref{tab:SUantisym}. As the occurring $\Sp(2K)$-invariant operators obtained from operators with formally open $\U(1)$ quantum number are lengthy, we briefly discuss their constituent field content here. The leading order contribution of the FMS expansion of $\det\phi$ and its conjugate is given by the radial Higgs excitation $h_{1}$ and the $\U(1)$ Goldstone $h_{2}$. Both belong to the first class. At order $2\leq p \leq N$ we obtain various operators of the second class with increasing number of $h_{\rmasym}$ as constituents, $\epsilon^{\bar{a}_1 \cdots \bar{a}_N} \epsilon^{\bar{b}_1 \cdots \bar{b}_N} h_{\rmasym}^{a_1b_1}\cdots h_{\rmasym}^{a_pb_p}\phi_{0}^{a_{p+1}b_{p+1}} \cdots \phi_{0}^{a_Nb_N}$ ($=\det h_{\rmasym}$ for $p=N$). Similarly, the scalar operator $\epsilon^{\bar{a}_1\cdots \bar{a}_{2K}}\phi^{a_{1}a_{2}}\cdots \phi^{a_{2K-1}a_{2K}}$ expands unambiguously to $h_{1}$ and $h_{2}$ at leading. Nontrivial composite operators containing two to $K$ fields $h_{\rmasym}$ can be extracted via the multiplet decomposition.
The leading order contribution vanishes for the Operator~\eqref{eq:vectorNonSinglet} in the vector channel. At next-to-leading order, we obtain $\epsilon^{\bar{a}_1 \cdots \bar{a}_N} \epsilon^{\bar{b}_1 \cdots \bar{b}_N} A_{\rmasym}^{\mu a_1b_1} h_{\rmasym}^{a_2b_2}\phi_{0}^{a_3b_3} \cdots \phi_{0}^{a_Nb_N}$ without requiring the multiplet decomposition, i.e., this meson operator belongs to the first class. At higher order $p$ we extract further composite operators $\epsilon^{\bar{a}_1 \cdots \bar{a}_N} \epsilon^{\bar{b}_1 \cdots \bar{b}_N} A_{\rmasym}^{\mu a_1b_1} h_{\rmasym}^{a_2b_2}\cdots h_{\rmasym}^{a_{p+1}b_{p+1}}\phi_{0}^{a_{p+2}b_{p+2}} \cdots \phi_{0}^{a_Nb_N}$ and $\epsilon^{\bar{a}_1 \cdots \bar{a}_N} \epsilon^{\bar{b}_1 \cdots \bar{b}_N} (D^{\mu}h_{\rmasym})^{a_1b_1} h_{\rmasym}^{a_2b_2}\cdots h_{\rmasym}^{a_pb_p}\phi_{0}^{a_{p+1}b_{p+1}} \cdots$. 
The operator~\eqref{eq:vectorNonSinglet2} does not expand in leading order to a single $\Sp(2K)$-invariant operator. Using the conventional multiplet decomposition, we obtain $\epsilon^{\bar{a}_1 \cdots \bar{a}_N} \epsilon^{\bar{b}_1 \cdots \bar{b}_N} (F_{\rma}^{\mu\nu}A_{\rmasym\nu})^{a_1b_1}\phi_{0}^{a_2b_2} \cdots \phi_{0}^{a_Nb_N}$, $\epsilon^{\bar{a}_1 \cdots \bar{a}_N} \epsilon^{\bar{b}_1 \cdots \bar{b}_N} (F_{\rmasym}^{\mu\nu}A_{\rmasym\nu})^{a_1b_1}\phi_{0}^{a_2b_2} \cdots \phi_{0}^{a_Nb_N}$, as well as $\epsilon^{\bar{a}_1 \cdots \bar{a}_N} \epsilon^{\bar{b}_1 \cdots \bar{b}_N} ([{A_{\rmasym}^{\mu}A_{\rmasym}^{\dagger\nu}-A_{\rmasym}^{\nu}A_{\rmasym}^{\dagger\mu}}]A_{\rmasym\nu})^{a_1b_1}\phi_{0}^{a_2b_2} \cdots$. Further operators of the second class can be extracted at higher orders.

For $\SU(2K+1)$ the normalized field configuration that minimizes the potential reads
\begin{align}
 \phi_{0} = \frac{1}{\sqrt{2K}} \begin{pmatrix} \varepsilon & & & \\ & \ddots & & \\ & & \varepsilon & \\ & & & 0 \end{pmatrix}
\end{align}
causing the breaking pattern ${\SU(2K+1) \to \Sp(2K)}$ \cite{Li:1973mq}. The elementary scalar sector remains unchanged in the unitary gauge apart from the fact that the previous BRST singlet Goldstone $h_{2}$ becomes now a would-be Goldstone, mixing with the gauge boson associated to the additional Cartan generator. Consequently, we get additional massive vector fields from the decomposition of $A^{\mu}$. These are a massive singlet $A_{\rms} = \tr(\phid_{0}A\phi_{0}^{\phantom{\dagger}})$ and a complex $2K$ component field $A_{\rmf}^{\mu} = 2K\phi_{0}^{\phantom{\,}}\phid_{0} (D^{\mu}\phi_{0} - 2K (D^{\mu}\phi_{0}) \phi_{0}^{\phantom{\,}}\phid_{0})$ transforming under the fundamental representation of $\Sp(2K)$. The antisymmetric tensor field $A_{\rmasym}^{\mu}$ is now described in an $\Sp(2K)$ covariant form via $4K^{2} \phi_{0}^{\phantom{\,}}\phid_{0} (D^{\mu}\phi_{0}) \phi_{0}^{\phantom{\,}}\phid_{0} - 2K A_{\rms}\phi_{0}^{\phantom{\,}}\phid_{0}$.

Investigating the strict $\SU(2K+1)$-invariant spectrum of the model, we first note that $\det \phi = 0$. Any other nonvanishing operator in the scalar open $\U(1)$ channel does not develop a homogeneous condensate and generates only massive modes. Thus, we conclude that the global $\U(1)$ symmetry is intact. Nonetheless, the investigation of the spectrum in the scalar open $\U(1)$ channel is intricate. All examined operators contain only scattering states. For instance, one of the nonvanishing operators with least field content is given by
\begin{align}
\epsilon^{\bar{a}_1 \cdots \bar{a}_N} \epsilon^{\bar{b}_1 \cdots \bar{b}_N} (F^{\mu\nu}D_{\nu}\phi)^{a_1b_1}(D_{\mu}\phi)^{a_2b_2}\phi^{a_3b_3} \cdots \phi^{a_Nb_N}.
 \label{eq:scalarNonSinglet3}
\end{align}
The first nontrivial lowest order term of the FMS expansion is a scattering state of the $\Sp(2K)$-invariant vector hybrid $\epsilon^{\bar{a}_1 \cdots \bar{a}_N} \epsilon^{\bar{b}_1 \cdots \bar{b}_N} A_{\rmasym}^{\mu a_1b_1} h_{\rmasym}^{a_2b_2}\phi_{0}^{a_3b_3} \cdots \phi_{0}^{a_Nb_N}$, which we already discussed for $N=2K$, and an $\Sp(2K)$-invariant vector hybrid formed by $A_{\rmf}$ and $F_{\rmf}$. At higher orders in the FMS expansion, we obtain further scattering states, e.g., including the vector singlet $A_{\rms}^{\mu}$ and other $\Sp(2K)$-invariant operators.

In the vector channel, the $\U(1)$-charged operator~\eqref{eq:vectorNonSinglet} vanishes for $N=2K+1$. 
The operator~\eqref{eq:vectorNonSinglet2} expands in leading order merely to operators of the second class, e.g., the $\Sp(2K)$-invariant hybrid operator created by $F_{\rmf}$ and $A_{\rmf}$ which we already encountered in the scalar channel. At higher orders in the FMS expansion, we find scattering states of this particular hybrid with $h$ as well as more sophisticated $\Sp(2K)$-singlets. Only in case $N=3$, we find that the $\SU(2)$-invariant operator containing $A_{\rmf}^{\nu}$ and $D_{\nu}A_{\rmf}^{\mu}$ belongs to the first class. The mass of this hybrid can be approximated by $2m_{\rmA_{\rmf}}$ in the simple constituent model. That we obtain the same result as in the fundamental case, c.f. \cite{Maas:2017xzh} or App~\ref{app:SUNfun}, is not a surprise for $N=3$ because in this case the antisymmetric second-rank tensor can be mapped on the (anti)fundamental representation. The analysis of $\U(1)$-singlet scalar and vector operators is straightforward. A summary can be found in Tab.~\ref{tab:SUantisym}.


\begin{table*}[t!]
\begin{center}

\begin{tabular}{c|| cl | ll || lcc}
\toprule

&   \multicolumn{2}{c|}{$\SU(N)$ invariant} & \multicolumn{2}{c||}{\quad $\Sp(2)\times\SU(N-2)$ singlets \quad} & \multicolumn{3}{c}{$\Sp(2)\times\SU(N-2)$ multiplets} \\
\hline
$J^P$ & $\U(1)$ &  Operator                &  1. Class  & [2. Class]                 & Field  & DOF &  $m_{\mathrm{Field}}^{2}$ \\
\hline

$0^+$  & 0 & $\tr(\phid\phi)$  & $h$		  & [$\tr (h_{\rms\otimes\rmasym}^{\dagger}h_{\rms\otimes\rmasym}^{\phantom{\dagger}})$]    &		 $h$ & 1 & $(\lambda + \frac{1}{2} \tilde\lambda)v^{2}$  \cr
 & 0 & $\tr(\phid D^{2}\phi)$        & [$h$]    & [$\tr (A_{\rmf\otimes\rmf}^{\dagger}A_{\rmf\otimes\rmf}^{\phantom{\dagger}})$]            & $h_{\rms\otimes\rmasym}$ & $(N-2)(N-3)$ & $\frac{\tilde\lambda v^{2}}{4}$ \vphantom{$\frac{\tilde K^{2^{2}}}{\tilde K}$} \cr
 & 0 & $\tr F^{2}$        &   & [$\tr F_{\rma\otimes\rms}^{2}$], [$\tr F_{\rms\otimes\rma}^{2}$], [$\tr (F_{\rmf\otimes\rmf}^{\dagger}F_{\rmf\otimes\rmf}^{\phantom{\dagger}})$]           &  &  &  \cr
 &  &         &   & [$\tr (A_{\rmf\otimes\rmf}^{\dagger}F_{\rma\otimes\rms}^{\phantom{\dagger}}A_{\rmf\otimes\rmf}^{\phantom{\dagger}})$], [$\tr (A_{\rmf\otimes\rmf}^{\dagger}F_{\rms\otimes\rma}^{\phantom{\dagger}}A_{\rmf\otimes\rmf}^{\phantom{\dagger}})$]           &  &  &  \cr
 & 0 & $\tr (\phid F^{2}\phi)$        &    & [$\tr F_{\rma\otimes\rms}^{2}$], [$\tr (F_{\rmf\otimes\rmf}^{\dagger}F_{\rmf\otimes\rmf}^{\phantom{\dagger}})$], [$\tr (A_{\rmf\otimes\rmf}^{\dagger}F_{\rma\otimes\rms}^{\phantom{\dagger}}A_{\rmf\otimes\rmf}^{\phantom{\dagger}})$]          &  &  &  \cr
 & 0 & $\tr (\phid F^{\mu\nu}\phi\phid F_{\mu\nu}\phi)$        &    & [$\tr F_{\rma\otimes\rms}^{2}$], [$\tr (A_{\rmf\otimes\rmf}^{\dagger}F_{\rma\otimes\rms}^{\phantom{\dagger}}A_{\rmf\otimes\rmf}^{\phantom{\dagger}})$]         &  &  &  \cr

\cline{2-5}
  & $1/\bar{1}$ & $\det \phi$\footnote[1]{Vanishes for $N=2K+1$.}        & $h$ & [$\det h_{\rms\otimes\rmasym}$]\footnotemark[1]         &  &  &  \cr
  & $1/\bar{1}$ & see Op.~\eqref{eq:scalarNonSinglet3}        & \multicolumn{2}{c||}{only trivial scattering states}           &  &  &  \cr
\hline
$1^-$  & 0 & $\tr(\phid D^{\mu}\phi)$             & $A_{\rms}^{\mu}$  & [$\tr (h_{\rms\otimes\rmasym}^{\dagger}D^{\mu}h_{\rms\otimes\rmasym}^{\phantom{\dagger}})$]             & $A_{\rma\otimes\rms}^{\mu}$ & $3$ & 0 \cr
 & 0 & $\tr(\phid F^{\mu\nu}D_{\nu}\phi)$             &    & [$\tr ( F_{\rmf\otimes\rmf}^{\dagger\mu\nu} A_{\rmf\otimes\rmf \nu}^{\phantom{\,}})$]             & $A_{\rms\otimes\rma}^{\mu}$ & $(N-2)^{2}-1$ & 0 \cr
\cline{2-5}
 & $1/\bar{1}$ & see Op.~\eqref{eq:vectorNonSinglet}\footnotemark[1]          & see main text    &              & $A_{\rmf\otimes\rmf}^{\mu}$ & $4(N-2)$ & $\frac{g^{2}v^{2}}{4}$\vphantom{$\frac{\tilde K^{2^{2}}}{\tilde K}$}   \cr
 & $1/\bar{1}$ & see Op.~\eqref{eq:vectorNonSinglet2}            &     & [see main text]            & $A_{\rms\otimes\rms}^{\mu}$  & 1  & $\frac{N-2}{N}g^{2}v^{2}$ \cr
 \toprule
\end{tabular}

\caption{Particle content of an $\SU(N)$ gauge theory ($N>2$) with a scalar field in the antisymmetric second-rank tensor representation and gauge-variant breaking pattern $\SU(N) \to \Sp(2)\times\SU(N-2)$.  
Left: Comparison between operators/states that are strict invariant with respect to $\SU(N)$ transformations, i.e., observables, and operators/states that are considered from the conventional but gauge-variant viewpoint of spontaneous gauge symmetry breaking ($\Sp(2)\times\SU(N-2)$ singlets) . Trivial scattering states are ignored. In case the contraction is obvious, we suppress Lorentz indices for better readability. We assign a $\U(1)$ charge of $1/N$ to the scalar field $\phi$. Right: Properties of the elementary building blocks obtained from the standard multiplet decomposition after gauge fixing which are used to construct $\Sp(2)\times\SU(N-2)$ singlets.}

\label{tab:SUantisym2}
\end{center}
\end{table*}


\subsubsection{\texorpdfstring{$\tilde\lambda < 0$}{\tilde\lambda < 0}}
For negative $\tilde\lambda$, the field configuration which minimizes the potential~\eqref{eq:potential2ndrank} can be transformed to the following block-diagonal form \cite{Li:1973mq},
\begin{align}
 \phi_{0} = \frac{1}{\sqrt{2}} \begin{pmatrix} \varepsilon &  &  &  \\ & 0 &  & \\  &  & \ddots &  \\  &  &  & 0   \end{pmatrix},
\end{align}
and we have $\SUN \to \Sp(2) \times \SU(N-2)$ \cite{Jetzer:1983ij,Elias:1975yd}.

In the unitary gauge, the fluctuation field $\varphi$ can be parametrized by a real-valued singlet $h = \sqrt{2}\re\big(\tr(\phid_{0}\varphi)\big)$ and the remaining  components live in the subspace orthogonal to $\phi_{0}$. They are singlets with respect to $\Sp(2)$ but transform as a complex antisymmetric second-rank $\SU(N-2)$ tensor which we denote by $h_{\rms\otimes\rmasym}$ using the notation introduced in Sec.~\ref{sec:ON2sym}. 
The mass parameters of the multiplets are $\mh^2 = (\lambda + \frac{1}{2}\tilde\lambda)v^{2}$ and $m_{\rmh_{\rms\otimes\rmasym}}^{2} = - \frac{1}{4}\tilde\lambda v^{2}$.

As we have $4N-7$ broken generators, we get ${4N-7}$ massive vector fields while the remaining gauge bosons are massless and are in the adjoint representation of either $\Sp(2)$, $A_{\rma\otimes\rms}$, or $\SU(N-2)$, $A_{\rms\otimes\rma}$. 
The massive vector bosons can be divided into a singlet $A_{\rms\otimes\rms}$ with mass $m_{\rmA_{\rms\otimes\rms}}^{2} = \frac{N-2}{N}g^{2}v^{2}$ and $4(N-2)$ degenerated vector fields transforming as a fundamental vector with respect to both subgroups, $A_{\rmf\otimes\rmf}$, with mass parameter $m_{\rmA_{\rmf\otimes\rmf}}^{2} = \frac{1}{4}g^{2}v^{2}$.

The gauge-invariant composite scalar and vector states of this theory can be classified according to the global $\U(1)$ symmetry of the model. 
The FMS expansion for most of the $\SU(N)$-invariant operators introduced in Sec.~\ref{sec:SUNpre} is straightforward and listed in Tab.~\ref{tab:SUantisym2}. 
Particularities appear in the $\U(1)$ open scalar channel for odd $N$. In this case, $\det \phi$ vanishes and all other investigated operators in this channel, e.g., the operator defined in~\eqref{eq:scalarNonSinglet3}, contain only scattering states. In the vector channel, the $\SU(N)$-invariant operator~\eqref{eq:vectorNonSinglet} vanishes for odd $N$. For $N=2K$, it expands to the $\SU(N-2)$ hybrid $\epsilon^{\dot{\bar{a}}_1 \cdots \dot{\bar{a}}_{N-2}} \epsilon^{\dot{\bar{b}}_1 \cdots \dot{\bar{b}}_{N-2}}  (D^\mu h_{\rms\otimes\rmasym})^{\dot{a}_1\dot{b}_1} h_{\rms\otimes\rmasym}^{\dot{a}_2\dot{b}_2} \cdots h_{\rms\otimes\rmasym}^{\dot{a}_{N-2}\dot{b}_{N-2}}$ where dotted indices run only over those components defining the $\SU(N-2)$ subspace as well as scattering states with $h$ at higher orders. The vector operator~\eqref{eq:vectorNonSinglet2} is nonvanishing for all $N$ and the leading order term of the FMS expansion is given by an $\Sp(2)$- and $\SU(N-2)$-invariant hybrid created by $A_{\rmf\otimes\rmf}$, $F_{\rmf\otimes\rmf}$, and $N-3$ fields $h_{\rmasym}$ belonging to the second class as well as scattering states. Only for $N=3$ we obtain a state of the first class at $\approx 2m_{\rmA_{\rmf}}$ as we can map the antisymmetric tensor to the (anti-)fundamental representation. At higher orders in the expansion, we obtain only scattering states.

\section{Implications for GUTs}
\label{sec:GUT}

So far, we found that merely a few single $\mathcal{H}$-invariant operators can be extracted solely via the FMS decomposition of $\mathcal{G}$-invariant composite operators. In case lattice studies will accumulate further evidence that only states belonging to the first class are contained in the $\mathcal{G}$-invariant spectrum of a gauge theory with BEH mechanism, the viewpoint on many proposed BSM models will change. In case states of the second class would also be included in the $\mathcal{G}$-invariant spectrum, a strict $\mathcal{G}$-$\mathcal{H}$ duality could be established at least for those theories for which the $\mathcal{H}$ multiplets carry only non-Abelian gauge charges and the tensor rank of the scalar field is sufficiently low depending on the representation of other nontrivial $\mathcal{G}$ multiplets (gauge fields, fermions, further scalars) and the conventional gauge-dependent breaking pattern. 
However, this is not the case for a proper BSM model which has to be constructed such that one obtains the standard model gauge group (and finally $\SU(3)\times\U(1)$ as a low-energy effective theory). Thus, a description of $\U(1)$ charged objects is necessary in the context of BSM models. 

In our above analysis, we have seen examples which imply that this will turn out to be problematic for GUTs or any other scenario which uses some extended gauge sector which does not include an explicit $\U(1)$ part and is 'broken' to the standard model by some extended Higgs sector. 
While $\U(1)$ charged states are well defined observables from the perspective of any $\mathcal{H} = \tilde{\mathcal{H}}\times\U(1)$ gauge theory,\footnote{As long as the state is invariant with respect to $\tilde{\mathcal{H}}$.} there is no $\mathcal{G}$-invariant operator from which we can extract such a $\U(1)$ charged state, neither via the FMS nor the multiplet decomposition. Thus, these particular states belong to the third class. For an explicit example, see our toy model studies in Sec.~\ref{sec:ONfun} for $N=3$ or Sec.~\ref{sec:ON2anti} for $\tilde\lambda <0$.\footnote{For the $\SO(N)$ antisymmetric tensor and $\tilde\lambda >0$ as well as for the $\SU(N>2)$ adjoint case, we obtained Abelian charged multiplets as well. However, the corresponding multiplets carried also a non-Abelian charge in such a way that $\tilde{\mathcal{H}}$-invariant operators are always $\U(1)$ singlets. Thus, there is no $\U(1)$ charged $\mathcal{H}$-invariant state that can be assigned to the third class in these cases.} 

In order to examine a more sophisticated scenario, consider the $\SU(5)$ GUT. None of the $\SU(5)$-invariant operators containing only one $\SU(5)$ gauge field  expand to a $W^{\pm}$. For instance, $\phid \Sigma^{n} D^{\mu} \phi \sim Z^{\mu}$ ($n\geq 0$), $\partial_{\nu}\tr(\Sigma F^{\mu\nu}) \sim Z^{\mu}$, $\partial_{\nu}\tr(\Sigma^{n} F^{\mu\nu}) \sim Z^{\mu} + A_{\U(1)}^{\mu}$ ($n>1$), where $\phi$ and $\Sigma$ are the scalar fields in the fundamental and adjoint representation, respectively. At least the $Z$ boson belongs to the first class. The photon can only be disentangled from the $Z$ if we use the multiplet decomposition for the latter operator. Thus, there is not only no $\SU(5)$-invariant description of $W^{\pm}$ but also none of the massless photon in case states of the second class are indeed not contained in the strict gauge-invariant spectrum of the model.\footnote{In a previous work, we classified the photon as a potential observable of the $\SU(5)$-invariant spectrum \cite{Maas:2017xzh}. However, we did not take $v_{\mathrm{SM}}/v_{\mathrm{GUT}}$ corrections into account within this prior analysis and also did not performed the present classification of states. Taking such effects into account leads to a reclassification of the photon.}

That $\U(1) \subset \mathcal{H}$ charged states are assigned to the third class can also be seen within the fermion sector straightforwardly. Any $\SU(5)$-invariant operator build by one of the fermion multiplets as well as an arbitrary number of Higgs multiplets provides only an $\SU(5)$-invariant description of a neutrino which is consistent as this is the only noncharged particle with respect to $\SU(3)\times\U(1)$. However, there is no $\SU(5)$-invariant description of an electron or a quark (which is fine for the latter as quarks are additionally color charged). 
The only states that can be extracted via the FMS or the multiplet decomposition from $\mathcal{G}$-invariant objects are $\mathcal{H}$-invariant. Thus, we obtain at most $\U(1)$-singlet bound states as $W^{+}W^{-}$ or positronium but not their charged constituents. Also the additional global $\U(1)_{\mathrm{global}}$ symmetry cannot be used to describe electrically charged particles as a careful distinction between the different $\U(1)$ groups is necessary. Any operator obtained from an $\SU(5)$-invariant object carrying an open global $\U(1)_{\mathrm{global}}$ quantum number is still invariant with respect to $\U(1)_{\mathrm{em}}$ transformations as the latter is the central $\U(1)$ of $\mathrm{S}(\U(3)\times\U(2))$. Indeed, one finds only $\SU(5)$-invariant operators charged with respect to $\U(1)_{\mathrm{global}}$ that expand to bound states, e.g., containing a $W^{\pm}$ and three leptoquarks with charge $\mp \frac{1}{3}$, such that the net electric charge vanishes.

From this perspective, the conventional formulation of $\SU(5)$ GUT fails to describe the experimentally observed particles in a gauge-invariant manner. This point of view can be generalized to any other currently investigated GUT scenario with a single gauge group as well as further BSM models with extended gauge sector where $\mathcal{G}$ does not include an explicit $\U(1)$, e.g., $\SO(10)$, $\mathrm{E}_{6}$, Pati-Salam, or trinification. As to whether similar constraints also extend to models where $\mathcal{G}$ contains an explicit $\U(1)$ factor, $\mathcal{G} = \tilde{\mathcal{G}}\times \U(1)$, e.g., flipped $\SU(5)$, is an open issue and currently under investigation. Nonetheless, the FMS analysis of the spectrum of gauge theories with a BEH mechanism formulates new field theoretical constraints on the construction of GUTs and BSM model building. Of course, we assume here that the $\mathcal{G}$-invariant spectrum is only analyzed via the FMS approach. If novel strict gauge-invariant formulations beyond the FMS prescription could be developed, the additional restrictions might be circumvented. But even if it would be possible to find $\mathcal{G}$-invariant formulations of the complete $\mathcal{H}$-invariant spectrum, the FMS approach still provides additional insights and possible constraints in terms of $\mathcal{G}$-invariant bound states. Furthermore, the FMS approach could allow for new routes towards model building. For instance consider the $\SO(N)$ antisymmetric tensor. Albeit there is no state charged with respect to the Abelian part of either $\SU(K)\times\U(1)$ ($\tilde\lambda > 0$) or $\U(1)\times\SO(N-2)$ ($\tilde\lambda < 0$), we found hints of an additional emergent Abelian gauge structure at the level of the $\mathcal{G}$-invariant bound states which could mimic the electric charge.

\section{Conclusions}
\label{sec:Conclusions}

We investigated the spectrum of non-Abelian gauge theories with a BEH mechanism in a strict gauge-invariant manner without using the misleading notion of spontaneous gauge symmetry breaking. In order to perform these studies, we systematically extended and adapted the method developed by Fr\"ohlich, Morchio, and Strocchi who formulated the electroweak standard model in a gauge-invariant language for the first time. We demonstrated that properties of some $\mathcal{G}$-invariant operators can be read off from properties of $\mathcal{H}$-invariant operators in those gauges where the gauge-depended VEV of the scalar field is identified with the field configuration that minimizes the action. Within these commonly chosen gauges, a $\mathcal{G}$-$\mathcal{H}$ duality is established between some but not all states where $\mathcal{H}$ is the stability group obtained from the usual perspective on the BEH mechanism. Of course, the properties of $\mathcal{G}$-invariant objects do not change once a different gauge condition is chosen. It is rather that they are probably easiest accessible within these special gauges but the $\mathcal{G}$-invariant nature of the spectrum is always intact even in the presence of a BEH mechanism in accordance to Elitzur's theorem.

That the spectra of a $\mathcal{G}$ gauge theory and the corresponding $\mathcal{H}$ gauge theory are in most cases not identical is in contrast to the assumption of the standard treatment of the BEH mechanism via spontaneous gauge symmetry breaking. The FMS expansion in terms of the split~\eqref{eq:split} reveals that only a subset of the spectra coincides. Therefore, a reinterpretation of the BEH mechanism is mandatory for a strict gauge-invariant and thus field theoretical well-defined formulation of observables in gauge theories with BEH mechanism. In fact, the BEH mechanism does not lead to spontaneous gauge symmetry breaking but provides a $\mathcal{G}$-$\mathcal{H}$ duality between the states of these theories. Which states are actually related via this duality can be extracted via the FMS expansion. From a bottom up perspective this analysis shows which states of an $\mathcal{H}$ gauge theory can be embedded in the context of a theory with gauge group $\mathcal{G}$ with suitable scalar sector. This latter point leads to new constraints for BSM model building.

In particular, we investigated the FMS expansion of various operators of $\SO(N)$ and $\SU(N)$ gauge theories with a BEH mechanism induced by a single scalar field in a low dimensional representation up to second rank tensors. Further, we categorized the possible states of the remaining unbroken $\mathcal{H}$ gauge theory in a gauge-fixed formulation according to the definitions of Sec.~\ref{sec:FMS}. 
We put a special emphasis on the distinction between states of the first and second class. Operators/States of both classes can formally be obtained from $\mathcal{G}$-invariant operators. However, states of the first class are distinguished by the fact that the FMS expansion directly project on these $\mathcal{H}$ singlets. No unique FMS projection exists for states of the second class. Thus, they can only be obtained via the standard gauge-dependent multiplet decomposition. 
This classification and the assumption that only states of the first class are present in the $\mathcal{G}$-invariant spectrum is based on present available lattice simulations \cite{Maas:2016ngo,Maas:2018xxu,Maas:2017pcw,Maas:2013aia,Maas:2014pba,Shrock:1985ur,Shrock:1985un}. In particular, states of the first class can be treated within the usual BEH framework at leading order in the FMS prescription. 
Slight modifications of this assumption might be conceivable once further theories are simulated. In particular, larger operator bases are necessary to decide the indeterminate role of operators of the second class. Furthermore, we ignored possible nontrivial analytic structures of the propagators which might become more involved due to the various interactions. These could give rise to additional bound states or resonances.

In case $\mathcal{H}$ is non-Abelian, we always obtain operators of the second class, e.g., the $\mathcal{H}$-glueballs, showing that the $\mathcal{G}$-invariant spectrum is modified compared to the conventional treatment. But even if it turns out that some or maybe even all operators of the second class are present in the $\mathcal{G}$-invariant spectrum, the spectrum of possible standard model extensions is constrained by the FMS analysis. Clearly, the standard model contains fields with an Abelian charge. Thus, any BSM model has to be able to describe $\U(1)$ charged particles at least as some low-energy effective model. However, this cannot be a $\U(1)$ charge being part of $\mathcal{H}$ as we can only extract $\mathcal{H}$-invariant operators from any $\mathcal{G}$-invariant operator via the FMS or the multiplet decomposition, i.e., the operator is a $\U(1)$ singlet. 
Therefore, the FMS mechanism provides an additional field theoretical tool to examine the validity of proposed BSM models similar to constraints from anomalies. Only if the strict gauge-invariant spectrum of the model is at least compatible with the standard model, or to be more precise with experiment, it can be viewed as a valid standard model extension. Otherwise the proposed model has to be rejected or alternative ways beyond the FMS description in terms of bound states have to be developed for a proper definition of observables.

From the FMS viewpoint, the electroweak structure of the standard model is special as a duality between gauge-invariant bound states and elementary fields can be established at leading order in the FMS approach. This can be traced back to the following two facts. First, $\mathcal{H}$ does not contain a non-Abelian substructure. Second, gauge-invariant operators can be assigned to multiplets of an additional global $\SU(2)$ symmetry within the Higgs sector, although this symmetry is broken by the hypercharge. That the global symmetry of the model is the same as the weak gauge structure provides a one-to-one mapping at leader order in the expansion. Precisely this circumstance explains why the perspective of electroweak symmetry breaking within the standard model is successful albeit not well defined. As to whether similar constructions are possible for general $\mathcal{G}$ gauge theories with enlarged global symmetry group will be discussed in subsequent work. Further, it is important to investigate the phenomenological implications of the FMS approach also beyond the leading order contributions within the standard model (and beyond). Otherwise modifications arising from the gauge-invariant formulation of the standard model could be misinterpreted as BSM signals. Current exploratory studies in this direction show that these effects are suppressed but might get observed by future experiments \cite{Maas:2017wzi,Egger:2017tkd,Maas:2017swq,Maas:2018ska,Maas:2019dwd}. Moreover, the FMS mechanism might lead to novel model building strategies. Although, it constrains theories which try to embed the standard model within some larger gauge group and an extended Higgs sector, the duality between states of different gauge theories provides various new possibilities for dark matter phenomenology \cite{Buttazzo:2019mvl,Buttazzo:2019iwr}. Note that the authors of the latter two papers assume that also states which we assign to the second class are related via the duality relation and that the conventional analysis of the BEH mechanism can be used to study the spectrum of the investigated models. Furthermore, we observed hints that also new Abelian gauge structures can emerge at the level of the $\mathcal{G}$ bound states. This emergent gauge structure was induced by a nontrivial interplay of the BEH and FMS mechanism.


\section*{Acknowledgments}

It is a pleasure to thank A. Maas, P. T\"orek, and A.~Wipf for valuable discussions and especially A. Maas and P. T\"orek for comments on the manuscript. 
This work received funding by the DFG under Grant No. \mbox{SO1777/1-1}.

\appendix

\section{Fundamental representation of \texorpdfstring{$\SU(N)$}{SU(N)}}
\label{app:SUNfun}

In this appendix, we summarize the FMS analysis of an $\SU(N)$ gauge theory with scalar field in the fundamental representation. The main results can be found in \cite{Maas:2017xzh}. Here, we perform a classification of the states introduced in Sec.~\ref{sec:FMS} for the first time and express the results in terms of the more general viewpoint of the present paper.

For the fundamental representation, the scalar field transforms as a complex vector with respect to the $\SUN$ gauge symmetry, $\phi \to U \phi$ with $U\in \SUN$. 
We are always able to perform an $\SUN$ transformation such that the field configuration which minimizes the potential, 
\begin{align*}
 V(\phi) = -\mu^{2} \phid\phi + \frac{\lambda}{2} (\phid\phi)^{2},
\end{align*}
takes the simple form $\phi_{0}^{a} = \delta^{aN}$, where $\mu^{2} = \frac{1}{2}v^{2}\lambda$. 
In this case, the radial Higgs excitation, ${h \equiv \frac{1}{\sqrt{2}}(\phid_{0}\varphi + \varphi^{\dagger}\phi_{0})}$, is located in the real part of the $N$th component of the scalar field. The imaginary part of $\phid_{0}\varphi$ and the other components are populated by the would-be Goldstone bosons which mix with those gauge bosons that acquire a nonvanishing mass term. Thus in the unitary gauge, the scalar field contains only one real-valued degree of freedom which is a BRST singlet, $\phi(x) = \frac{1}{\sqrt{2}}\big(v+h(x)\big)\phi_{0}$. The mass of $h$ is given by $\mh^2 = \lambda v^2$.

The mass matrix of the gauge bosons can be derived from the kinetic term of the scalar field with covariant derivative $D_{\mu}\phi = \partial_{\mu}\phi + \I g A_{\mu}\phi$ and is defined via $\frac{1}{2} {(\MA^{2})}_{ij}A_{\mu i}^{\phantom{\mu}}A_{j}^{\mu} = \frac{1}{2} g^{2}v^{2} \phid_{0} A_{\mu}A^{\mu}\phi_{0}$ and $A^{\mu} = A_{i}^{\mu}T_{i}^{\phantom{\,}}$ where $T_{i}$ are the generators of $\SU(N)$. 
The gauge field $A^{\mu}$ decomposes into 
a massive $\SU(N-1)$-singlet, $ \phid_{0}A^{\mu}\phi_{0}^{\phantom{\dagger}} \equiv A_{\rms}^{\mu} $, a massive field transforming as a complex fundamental vector of $\SU(N-1)$, $A^{\mu}\phi_{0}-A_{\rms}^{\mu}\phi_{0} \equiv A_{\rmf}^{\mu} $, and the remaining degrees of freedom are the massless gauge fields of  the unbroken gauge group $SU(N-1)$, $A_{\rma}^{\mu}$. The mass parameters are listed in Tab.~\ref{tab:SUfun}.

The considered model obeys a global $\U(1)$ symmetry for $N\geq 3$. Thus, we can distinguish the states of the theory as $\U(1)$ singlets and states with an open $\U(1)$ quantum number. For the $\U(1)$ singlet channel, we investigate the $\SU(N)$-invariant scalar operator $\phid\phi$. Its FMS expansion reveals a duality to the states generated by the elementary Higgs field $h$ and contains the trivial scattering state $h^{2}$ as well. It is also possible to find an $\SU(N)$-invariant operator that maps on the only other elementary $\SU(N-1)$ singlet in the gauge fixed formulation which is the vector singlet $A_{\rms}^{\mu}$.  
This conclusion can be drawn by investigating the following operator in the $\U(1)$-singlet vector channel 
\begin{align}
 \I \phid D^{\mu} \phi &= - \frac{g v^{2}}{2} A_{\rms}^{\mu} + \frac{\I v}{2} \partial^{\mu}h \notag\\
 &\quad - gv\, h  A_{\rms}^{\mu}  - \frac{g}{2} h^{2} A_{\rms}^{\mu}  + \frac{\I}{2} h \partial_{\mu} h
 \label{eq:derivativeMixing}
\end{align}
implying that the bound state can be mapped to the elementary vector $A_{\rms}^{\mu}$. Several trivial scattering states of this vector boson and the Higgs are included as well. These are separated in the second line. Further, a pole at the mass of the elementary Higgs will appear at the level of the correlator of this composite operator. However, this does not necessarily give rise to a new vector particle in the $1^{-}$ singlet channel as the pole appears only in the longitudinal component. This is expected because the derivative acting on the scalar field transforms as a vector and thus mixes with operators in this channel.

Further $\U(1)$ singlet operators in the $0^{+}$ and $1^{-}$ channel as $\phid D^{2}\phi$, $\phid F^{2} \phi$, $\tr F^{2}$, and $\phid F^{\mu\nu}D_{\nu}\phi$ contain $\SU(N-1)$-invariant composite objects as the meson operator $A_{\rmf}^{\dagger}A_{\rmf}^{\vphantom{f}}$, the hybrids $F_{\rmf}^{\dagger}F_{\rmf}^{\vphantom{f}}$, $F_{\rmf}^{\dagger}A_{\rmf}^{\vphantom{f}}$, and $A_{\rmf}^{\dagger}F_{\rma}^{\vphantom{f}}A_{\rmf}^{\vphantom{f}}$, as well as the glueball $\tr F_{\rma}^{2}$. As the FMS expansion of the $\SU(N)$-invariant operators does not provide a direct mapping on these $\SU(N-1)$-invariant operators and we need the multiplet decomposition to extract them, they belong to the second class.


\begin{table*}[t!]
\begin{center}

\begin{tabular}{c|| cl | ll || ccc}
\toprule

&   \multicolumn{2}{c|}{$\SU(N)$ invariant} & \multicolumn{2}{c||}{\quad $\SU(N-1)$ singlets \quad} & \multicolumn{3}{c}{$\SU(N-1)$ multiplets} \\
\hline
$J^P$ & $\U(1)$ &  Operator                &  1. Class  & [2. Class]                 & Field  & DOF &  $m_{\mathrm{Field}}^{2}$ \\
\hline

$0^+$  & 0 & $\phid\phi$  & $h$		  &     		& $h$ & 1 & $\lambda v^{2}$  \cr
 & 0 & $\phid D^{2}\phi$        & $h$    & [$ A_{\rmf}^{\dagger}A_{\rmf}^{\vphantom{f}}$]            &  &  &  \cr
 & 0 & $\tr F^{2}$        &   & [$\tr F_{\rma}^{2}$], [$F_{\rmf}^{\dagger}F_{\rmf}^{\vphantom{f}}$], [$A_{\rmf}^{\dagger}F_{\rma}^{\vphantom{f}}A_{\rmf}^{\vphantom{f}}$]           &  &  &  \cr
 & 0 & $\phid F^{2}\phi$        &    & [$F_{\rmf}^{\dagger}F_{\rmf}^{\vphantom{f}}$]           &  &  &  \cr
\cline{2-5}
  & $1/\bar{1}$ & see Op.~\eqref{eq:SUfunNonSingletScalar}        &  & [see main text]          &  &  &  \cr
& $1/\bar{1}$ & see Op.~\eqref{eq:SUfunNonSingletScalar2}        &  & [see main text]          &  &  &  \cr
\hline
$1^-$  & 0 & $\phid D^{\mu}\phi$             & $A_{\rms}^{\mu}$  &              & $A_{\rma}^{\mu}$ & $(N-1)^{2}-1$ & 0 \cr
 & 0 & $\tr(\phid F^{\mu\nu}D_{\nu}\phi)$             &    & [$F_{\rmf}^{\dagger\mu\nu} A_{\rmf \nu}^{\vphantom{f}}$]             & $A_{\rmf}^{\mu}$ & $2(N-1)$ & $\frac{g^{2}v^{2}}{4}$ \cr
\cline{2-5}
 & $1/\bar{1}$ & see Op.~\eqref{eq:SUfunNonSingletVector}\footnote{For $N=3$, one may consider the operator $\epsilon^{\bar{a}_{1}\bar{a}_{2}\bar{a}_{3}} \phi^{a_1} (D^{\nu_1}\phi)^{a_2}(D_{\nu_1}D^{\mu}\phi)^{a_3}$ which contains an operator of the first class with mass $\approx 2m_{\rmA_{\rmf}}$.}          &    & [see main text]             & $A_{\rms}^{\mu}$ & $1$ & $\frac{N-1}{2N}g^{2}v^{2}$   \cr
 \toprule
\end{tabular}

\caption{Particle content of an $\SU(N)$ gauge theory ($N>2$) with scalar field in the fundamental representation and gauge-variant breaking pattern $\SU(N) \to \SU(N-1)$. 
Left: Comparison between operators/states that are strict invariant with respect to $\SU(N)$ transformations, i.e., observables, and operators/states that are considered from the conventional but gauge-variant viewpoint of spontaneous gauge symmetry breaking ($\SU(N-1)$ singlets). Trivial scattering states are ignored. In case the contraction is obvious, we suppress Lorentz indices for better readability. We assign a global $\U(1)$ charge of $1/N$ to the scalar field $\phi$. Right: Properties of the elementary building blocks obtained from the standard multiplet decomposition after gauge fixing which are used to construct $\SU(N-1)$ singlets.}

\label{tab:SUfun}
\end{center}
\end{table*}


In addition to the $\U(1)$-singlet operators, operators with an open global $\U(1)$ quantum number can be constructed in the vector and the scalar channel. The lightest ground state of these channels is necessarily stable as the $\U(1)$ charge is conserved. In order to build such an operator, we have to contract the indices with the aid of the $\epsilon$ tensor. A realization of a $\U(1)$-charged scalar operator was proposed in \cite{Maas:2017xzh} and is given by
\begin{align}
 & \epsilon^{\bar{a}_{1}\cdots\bar{a}_{N}} \phi^{a_1} (F_{\nu_1}^{\phantom{\mu}\nu_2}\phi)^{a_2}(F_{\nu_2}^{\phantom{\mu}\nu_3}\phi)^{a_3} \cdots (F_{\nu_{N-1}}^{\phantom{\nu_{N-1}}\nu_{1}}\phi)^{a_N} 
\label{eq:SUfunNonSingletScalar}
\end{align}
for $3 < N \leq \frac{1}{2}d(d-1)+1$. At leading order, the hybrid $\epsilon^{\dot{\bar{a}}_{1}\cdots\dot{\bar{a}}_{N-1}}(F_{\rmf\nu_1}^{\phantom{\mu}\nu_2})^{\dot{a}_1} \cdots (F_{\rmf\nu_{N-1}}^{\phantom{\nu_{N-1}}\nu_1})^{\dot{a}_{N-1}}$\footnote{Dotted indices run over those components defining the subspace perpendicular to $\phi_{0}$.} can be extracted but only if we additionally use the multiplet decomposition to separate it from scattering stats containing $A_{\rms}^{\mu}$. Therefore, this ${\SU(N-1)}$ hybrid operator belongs to the second class. For the case $N=3$, an operator with minimal field content is given by $\epsilon^{\bar{a}_{1}\bar{a}_{2}\bar{a}_{3}}\phi^{a_1}(D_{\mu}\phi)^{a_2}(F^{\mu\nu}D_{\nu}\phi)$. Performing the FMS expansion, all terms appearing on the right-hand side can be decomposed further via the conventional multiplet decomposition. For $N > d(d-1)/2$ the $\SU(N)$-invariant operator defined on the left-hand side of Eq.~\eqref{eq:SUfunNonSingletScalar} vanishes due to the antisymmetric property of the $\epsilon$ tensor and more involved objects have to be constructed, e.g., including more covariant derivatives or anticommutators thereof. For instance for $N\leq d^{2}+1$, we consider
\begin{align}
 & \epsilon^{\bar{a}_{1}\cdots\bar{a}_{N}} \phi^{a_1} (D_{\nu_1}D^{\nu_2}\phi)^{a_2} \cdots (D_{\nu_{N-1}}D^{\nu_{1}}\phi)^{a_N}. 
\label{eq:SUfunNonSingletScalar2}
\end{align}
Single $\SU(N-1)$-invariant operators can only be extracted via the multiplet decomposition. This is also the case for any further operator with larger field content in the scalar open $\U(1)$ channel. Thus, all states generated by these operators belong to the second class.

A similar construction can be done in the vector channel. For $N \leq \frac{1}{2}d(d-1) + 2$, a vector operator with open $\U(1)$ quantum number reads 
\begin{align}
 & \epsilon^{\bar{a}_{1}\cdots\bar{a}_{N}} \phi^{a_1} (D^{\nu_1}\phi)^{a_2}(F_{\nu_1}^{\phantom{\mu}\nu_2}\phi)^{a_3} \cdots (F_{\nu_{N-2}}^{\phantom{\nu_{N-2}}\mu}\phi)^{a_N}   
\label{eq:SUfunNonSingletVector}
\end{align}
For larger $N$, we may consider the same operator where the field strength tensors are replaced by two covariant derivatives $F_{\mu}^{\phantom{\mu}\nu}\to D_{\mu}D^{\nu}$ which is nonvanishing as long as $N \leq d^{2}+2$.
All single $\SU(N-1)$-invariant operators contained in these $\U(1)$-charged vector operators are assigned to the second class as we have to use the multiplet decomposition to obtain them. The only possibility to construct an $\SU(N)$-invariant operator whose FMS expansion provides a projection on an operator of the first class is given for $N=3$. There, $\epsilon^{\bar{a}_{1}\bar{a}_{2}\bar{a}_{3}} \phi^{a_1} (D^{\nu_1}\phi)^{a_2}(D_{\nu_1}D^{\mu}\phi)^{a_3} = -g^{2}v^{3}\epsilon^{\dot{\bar{a}}_{1}\dot{\bar{a}}_{2}} (A_{\rmf}^{\nu_1})^{\dot{a}_1}(D_{\nu_1}A_{\rmf}^{\mu})^{\dot{a}_2} + \mathcal{O}(\varphi)$. 
Applying the naive constituent model to the $\SU(2)$ hybrid, the mass can be approximated by $2m_{\rmA_{\rmf}}$. This is in agreement with lattice simulations for $N=3$ \cite{Maas:2018xxu}.\footnote{Interestingly, a different operator, $\epsilon^{\bar{a}_{1}\bar{a}_{2}\bar{a}_{3}} \phi^{a_1} (D^{\mu}\phi)^{a_2}(D^{2}\phi)^{a_3}$, was considered in the lattice analysis where the state with mass $\approx 2m_{\rmA_{\rmf}}$ can only be extracted via the multiplet decomposition. Thus, we conjecture that a $\mathcal{G}$-invariant operator can have overlap with states of the first class even if these states of the first class can only be extracted via the multiplet decomposition from the considered $\mathcal{G}$-invariant operator.}

\subsection*{\texorpdfstring{$N=2$}{N=2} and the electroweak sector}
The case $N=2$ is of particular interest as it describes the weak-Higgs subsector of the standard model. Furthermore, it is special regarding the above analysis as the fundamental representation of $\SU(2)$ is pseudo-real. This causes that the global symmetry is not $\U(1)$ but $\SU(2)$ and we are able to assign the gauge-invariant operators to multiplets of the global $\SU(2)$ group. This can be realized by using the charge conjugated scalar field $\tilde\phi = \epsilon \phi^{*}$ to construct gauge-invariant operators \cite{Frohlich:1980gj,Frohlich:1981yi}.

The scalar singlet $\phid\phi + \tilde\phi^{\dagger}\tilde\phi$ expands in leading order to the radial Higgs excitation while the triplet vanishes, $(\tilde\phi^{\dagger}\phi,\phid\tilde\phi, \phid\phi - \tilde\phi^{\dagger}\tilde\phi) = (0,0,0)$. In the vector channel the $\SU(2)_{\mathrm{gauge}}$-invariant $\SU(2)_{\mathrm{global}}$-triplet $(\tilde\phi^{\dagger}D^{\mu}\phi,\phid D^{\mu}\tilde\phi, \phid D^{\mu}\phi - \tilde\phi^{\dagger}D^{\mu}\tilde\phi)$ expands in leading order to the elementary massive vector bosons. Equally, these form a triplet of a diagonal $\SU(2)_{\mathrm{diag}}$ subgroup explaining the degeneracy of the elementary degrees of freedom from the conventional perspective of gauge symmetry breaking as the gauge-dependent breaking pattern is $\SU(2)_{\mathrm{gauge}} \times \SU(2)_{\mathrm{global}} \to \SU(2)_{\mathrm{diag}}$.\footnote{Often a reformulation of the scalar sector via the field variable $X = (\tilde\phi,\phi)$ is done for convenience, e.g., see \cite{Maas:2017wzi,Egger:2017tkd}. The advantage of this reformulation is that $\SU(2)_{\mathrm{global}}$ transformations act linear on $X$ by multiplication from the right while usual $\SU(2)_{\mathrm{gauge}}$ transformations act via multiplication from the left. In terms of $\phi$ and $\tilde\phi$, $\SU(2)_{\mathrm{global}}$ transformations are realized in a nonlinear way.} The vector $\SU(2)_{\mathrm{global}}$ singlet does not contain an elementary gauge boson but only the standard mixing of a derivative term of the scalar degree of freedom with the vector channel, $\phid D^{\mu}\phi + \tilde\phi^{\dagger}D^{\mu}\tilde\phi = 2\phid\partial^{\mu}\phi$. Thus, in the particular case of $N=2$, we obtain a one-two-one mapping from simple gauge-invariant bound state operators with least field content to the elementary fields of the model. This is due to the fact that the structure of the global symmetry group coincides with the gauge structure of the model as well as the diagonal subgroup of both such that we obtain a mapping from the global to the local multiplets. Further note that the non-Abelian gauge group is fully broken in the conventional picture implying that no gauge multiplets are left in the elementary spectrum.

A generalization to the full electroweak sector is straightforward as the additional $\U(1)_{\mathrm{Y}}$ hypercharge group is Abelian. Of course, the additional hypercharge breaks the global $\SU(2)_{\mathrm{global}}$ explicitly at the level of the Lagrangian. Nonetheless, we have a sufficiently large operator basis as we can still use the scalar doublet as well as its dual field to construct $\SU(2)_{\mathrm{gauge}}$-invariant operators. The only manifestation of the explicitly broken $\SU(2)_{\mathrm{global}}$ symmetry is that the corresponding multiplets split into nondegenerate degrees of freedom. In order to describe observable states also in a $\U(1)_{\mathrm{Y}}$-invariant manner, we may use suitable dressings via Dirac phase factors \cite{Maas:2017wzi}. Equally, we can use the FMS description for a fully $\SU(2)_{\mathrm{gauge}} \times \U(1)_{\mathrm{Y}}$-invariant description of all experimentally observed particles in the electroweak sector. For instance, we can still use the four operators $\phid\phi + \tilde\phi^{\dagger}\tilde\phi$, $\tilde\phi^{\dagger}\phi$, $\phid\tilde\phi$, and $\phid\phi - \tilde\phi^{\dagger}\tilde\phi$ in the scalar channel. These form a mass eigenbasis at the level of the gauge-invariant bound state operators as the latter three operators vanish and the formal $\SU(2)_{\mathrm{global}}$-singlet is dual to the Higgs boson according to the FMS framework. In the vector channel, the conjugated operators $\tilde\phi^{\dagger}D^{\mu}\phi$ and $\phid D^{\mu}\tilde\phi$ expand to the elementary $W^{+\mu}$ and $W^{-\mu}$, respectively. Note that these two gauge-invariant operators are the $\SU(2)$ versions of the $\SU(N>2)$-invariant operators with open $\U(1)$ quantum number, see above. The remaining operator of the formal triplet, $\phid D^{\mu}\phi - \tilde\phi^{\dagger}D^{\mu}\tilde\phi$, provides an unambiguous mapping on the elementary neutral $Z^{\mu}$. Finally, the operator $\phid D^{\mu}\phi - \tilde\phi^{\dagger}D^{\mu}\tilde\phi$ provides an electroweak gauge-invariant description of the photon field $A^{\mu}$. Furthermore, the generalization to the fermion sector of the standard model is straightforward \cite{Frohlich:1980gj,Frohlich:1981yi}.


\begin{table*}[t!]
\begin{center}

\begin{tabular}{c|| ll | ll || lcc}

\toprule

& \multicolumn{2}{c|}{\quad $\SU(N)$ invariant \quad} & \multicolumn{2}{c| |}{\quad $\mathrm{S}(\U(P){\times}\U(N-P))$ singlets \quad} & \multicolumn{3}{c}{$\mathrm{S}(\U(P){\times}\U(N-P))$ multiplets} \\
\hline
$J^P$ & $\Ztwo$ \, &  Operator                &  1. Class  & [2. Class]                 & Field  & DOF &  $m_{\mathrm{Field}}^{2}$ \\
\hline

$0^+$ & $+$ & $\tr\phi^{2}$ &		 $h$ & [$\tr h_{\rms\otimes\rma}^{2}$],[$\tr h_{\rma\otimes\rms}^{2}$]    & $h$ & 1 & see Eq.~\eqref{eq:masshsymSO} \cr
& $+$ & $\tr(\phi D^{2}\phi)$        & $\tr A_{\rmf\otimes\rmf}^{2}$, [$h$]  & [$\tr (A_{\rmf\otimes\rmf}^{2}h_{\rma\otimes\rms}^{\vphantom{2}})$], [$\tr (A_{\rmf\otimes\rmf}^{2}h_{\rms\otimes\rma}^{\vphantom{2}})$]            & $h_{\rma \otimes \rms}$ & $P^{2}-1$ & see Eq.~\eqref{eq:masshsymSO} \cr
& $+$ & $\tr F^{2}$        &  & [$\tr(F_{\rma\otimes\rms}^{2})$], [$\tr(F_{\rms\otimes\rma}^{2})$], [$\tr(F_{\rmf\otimes\rmf}^{2})$],           & $h_{\rms \otimes \rma}$ & $(N-P)^{2}-1$ & see Eq.~\eqref{eq:masshsymSO} \cr
&  &        &   & [$\tr(A_{\rmf\otimes\rmf}F_{\rma\otimes\rms}A_{\rmf\otimes\rmf})$], [$\tr(A_{\rmf\otimes\rmf}F_{\rms\otimes\rma}A_{\rmf\otimes\rmf})$]            & & & \cr
& $+$ & $\tr(\phi^{2}F^{2})$        &   & [$\tr(F_{\rma\otimes\rms}^{2})$], [$\tr(F_{\rms\otimes\rma}^{2})$], [$\tr(F_{\rmf\otimes\rmf}^{2})$],           & & & \cr
&  &        &   & [$\tr(A_{\rmf\otimes\rmf}F_{\rma\otimes\rms}A_{\rmf\otimes\rmf})$], [$\tr(A_{\rmf\otimes\rmf}F_{\rms\otimes\rma}A_{\rmf\otimes\rmf})$]            & & & \cr
&  &        &   & [$\tr(F_{\rma\otimes\rms}^{2}h_{\rma\otimes\rms}^{\phantom{2}})$], [$\tr(F_{\rms\otimes\rma}^{2}h_{\rms\otimes\rma}^{\phantom{2}})$],            & & & \cr
&  &        &   & [$\tr(F_{\rmf\otimes\rmf}^{2}h_{\rma\otimes\rms}^{\phantom{2}})$], [$\tr(F_{\rmf\otimes\rmf}^{2}h_{\rms\otimes\rma}^{\phantom{2}})$]            & & & \cr
\cline{2-5}
& $-$ & $\tr\phi^{3}$        & $h$   & [$\tr h_{\rms\otimes\rma}^{2}$], [$\tr h_{\rma\otimes\rms}^{2}$], [$\tr h_{\rms\otimes\rma}^{3}$]\footnote{Vanishes for $P=2$}, [$\tr h_{\rma\otimes\rms}^{3}$]\footnote{Vanishes for $N-P=2$}            & & & \cr
& $-$ & $\tr(\phi^{2} D^{2}\phi)$        & $\tr A_{\rmf\otimes\rmf}^{2}$\footnote[3]{This state is not present for ${P=N-P}$}, [$h$]  & [$\tr (A_{\rmf\otimes\rmf}^{2}h_{\rma\otimes\rms}^{\vphantom{2}})$], [$\tr (A_{\rmf\otimes\rmf}^{2}h_{\rms\otimes\rma}^{\vphantom{2}})$]            &  &  &  \cr
& $-$ & $\tr(\phi F^{2})$        &   & [$\tr(F_{\rma\otimes\rms}^{2})$], [$\tr(F_{\rms\otimes\rma}^{2})$], [$\tr(F_{\rmf\otimes\rmf}^{2})$]\footnotemark[3],            & & & \cr
&  &        &   & [$\tr(A_{\rmf\otimes\rmf}F_{\rma\otimes\rms}A_{\rmf\otimes\rmf})$], [$\tr(A_{\rmf\otimes\rmf}F_{\rms\otimes\rma}A_{\rmf\otimes\rmf})$]            & & & \cr
&  &        &   & [$\tr(F_{\rma\otimes\rms}^{2}h_{\rma\otimes\rms}^{\phantom{2}})$], [$\tr(F_{\rms\otimes\rma}^{2}h_{\rms\otimes\rma}^{\phantom{2}})$],            & & & \cr
&  &        &   & [$\tr(F_{\rmf\otimes\rmf}^{2}h_{\rma\otimes\rms}^{\phantom{2}})$], [$\tr(F_{\rmf\otimes\rmf}^{2}h_{\rms\otimes\rma}^{\phantom{2}})$]            & & & \cr
\hline
$1^-$ & $+$ & $\partial_{\nu} \tr (F^{\mu\nu}\phi^{2})$        & $A^{\mu}_{\perp \U(1)}$ &             & $A_{\rma \otimes \rms}^{\mu}$ & $P^{2}-1$ & 0 \cr
& $+$ & $\tr(\phi F^{\mu\nu}D_{\nu}\phi)$             & $\tr (F_{\rmf\otimes\rmf} A_{\rmf\otimes\rmf})$  & [$\tr(F_{\rma\otimes\rms}^{\mu\nu}D_{\nu}h_{\rma\otimes\rms}^{\vphantom{\mu}})$], [$\tr(F_{\rms\otimes\rma}^{\mu\nu}D_{\nu}h_{\rms\otimes\rma}^{\vphantom{\mu}})$],   	& $A_{\rms \otimes \rma}^{\mu}$ & $(N-P)^{2}-1$ & $0$ \cr
&  &              &   & [$\tr(F_{\rmf\otimes\rmf}^{\mu\nu}A_{\rmf\otimes\rmf\nu}^{\vphantom{\mu}}h_{\rma\otimes\rms}^{\vphantom{\mu}})$], [$\tr(F_{\rmf\otimes\rmf}^{\mu\nu}h_{\rma\otimes\rms}^{\vphantom{\mu}}A_{\rmf\otimes\rmf\nu}^{\vphantom{\mu}})$],             & $A_{\rmf\otimes\rmf}^{\mu}$ & $2P(N-P)$ & $\frac{N}{2P(N-P)}g^2v^2$ \cr
 &  &              &   & [$\tr(F_{\rmf\otimes\rmf}^{\mu\nu}A_{\rmf\otimes\rmf\nu}^{\vphantom{\mu}}h_{\rms\otimes\rma}^{\vphantom{\mu}})$], [$\tr(F_{\rmf\otimes\rmf}^{\mu\nu}h_{\rms\otimes\rma}^{\vphantom{\mu}}A_{\rmf\otimes\rmf\nu}^{\vphantom{\mu}})$]              & $A_{\U(1)}^{\mu}$ & $1$ & $0$ \cr
\cline{2-5}
& $-$ & $\partial_{\nu} \tr (F^{\mu\nu}\phi)$        & $A^{\mu}_{\perp \U(1)}$  &              &  &  &  \cr
 & $-$ & $\tr(F^{\mu\nu}D_{\nu}\phi)$             & $\tr (F_{\rmf\otimes\rmf} A_{\rmf\otimes\rmf})$   & [$\tr(F_{\rma\otimes\rms}^{\mu\nu}D_{\nu}h_{\rma\otimes\rms}^{\vphantom{\mu}})$], [$\tr(F_{\rms\otimes\rma}^{\mu\nu}D_{\nu}h_{\rms\otimes\rma}^{\vphantom{\mu}})$],               &  &  &  \cr
 &  &              &   & [$\tr(F_{\rmf\otimes\rmf}^{\mu\nu}A_{\rmf\otimes\rmf\nu}^{\vphantom{\mu}}h_{\rma\otimes\rms}^{\vphantom{\mu}})$], [$\tr(F_{\rmf\otimes\rmf}^{\mu\nu}h_{\rma\otimes\rms}^{\vphantom{\mu}}A_{\rmf\otimes\rmf\nu}^{\vphantom{\mu}})$],              &  &  &  \cr
 &  &              &   & [$\tr(F_{\rmf\otimes\rmf}^{\mu\nu}A_{\rmf\otimes\rmf\nu}^{\vphantom{\mu}}h_{\rms\otimes\rma}^{\vphantom{\mu}})$], [$\tr(F_{\rmf\otimes\rmf}^{\mu\nu}h_{\rms\otimes\rma}^{\vphantom{\mu}}A_{\rmf\otimes\rmf\nu}^{\vphantom{\mu}})$]              &  &  &  \cr

 \toprule
\end{tabular}
\caption{Particle content of an $\SU(N)$ gauge theory with scalar field in the adjoint representation and gauge-variant breaking pattern $\SU(N)\to\mathrm{S}(\U(P){\times}\U(N-P))$. 
Left: Comparison between operators/states that are strict invariant with respect to $\SU(N)$ transformations, i.e., observables, and operators/states that are considered from the conventional but gauge-variant viewpoint of spontaneous gauge symmetry breaking ($\mathrm{S}(\U(P){\times}\U(N-P))$ singlets). Trivial scattering states are ignored. We suppress Lorentz indices for better readability in the $0^{+}$ channel. Right: Properties of the elementary building blocks obtained from the standard multiplet decomposition after gauge fixing which are used to construct $\mathrm{S}(\U(P){\times}\U(N-P))$ singlets.}

\label{tab:SUadj}
\end{center}
\end{table*}


\section{Adjoint representation of \texorpdfstring{$\SU(N)$}{SU(N)}}
\label{app:SUNadj}

The FMS analysis of an $\SU(N)$ gauge theory with one scalar field in the adjoint representation and a $\Ztwo$ symmetric scalar sector was discussed in Ref.~\cite{Maas:2017xzh}. For $\SU(3)$ also a cubic interaction term was considered such that the global $\Ztwo$ symmetry was explicitly broken. These results will be generalized to an arbitrary $\SU(N)$ gauge theory with nonsymmetric $\Ztwo$ scalar sector in the following. 

The transformation property of the scalar field reads $\phi \to U \phi U^{\dagger}$ ($U\in \SU(N)$) and the scalar potential has structurally the same form as the symmetric tensor in the $\SO(N)$ group, Eq.~\eqref{eq:potSONsym}. Thus, the direction of the field configuration that minimizes the potential is given in Eq.~\eqref{eq:minSONsym} and its modulus $v$ is determined by the parameters of the potential via $\mu^{2} = \frac{1}{2}\lambda v^{2}+2\tilde\lambda v^{2}\tr\phi_{0}^{4} + \gamma v \tr\phi_{0}^{3}$. Equation~\eqref{eq:potSONsym} implies a breaking pattern $\SU(N)\to\mathrm{S}(\U(P)\times\U(N-P))$ with $P<N$. Without loss of generality, we assume $P\leq N/2$. Depending on the sign of the nonisotropic couplings $\gamma$ and $\tilde\lambda$ different breaking patterns are favored, see the discussion below Eq.~\eqref{eq:minSONsym}. The decomposition of the elementary multiplets is also similar to the symmetric tensor of $\SO(N)$ and summarized in Tab.~\ref{tab:SUadj}. We just highlight the presence of an additional Abelian gauge boson $A_{\U(1)}^{\mu} = \tr(A^{\mu}\phi_{0})$ being invariant with respect to the non-Abelian subgroups $\SU(P)\times\SU(N-P)$. Further note that we use $A_{\rmf\otimes\rmf} = [A,\phi_{0}]$ for convenience. Alternatively, one could define $A'_{\rmf\otimes\rmf} =  \mathbb{P}_{P} A \mathbb{P}_{N-P}$ where $\mathbb{P}_{P} = \frac{P}{N}\mathbbm{1}+ \frac{\sqrt{2NP(N-P)}}{N}\phi_{0}$ and $\mathbb{P}_{N-P} = \mathbbm{1} - \mathbb{P}_{P}$ such that $A_{\rmf\otimes\rmf} = A^{\prime\dagger}_{\rmf\otimes\rmf} - A^{\prime\phantom{\dagger}}_{\rmf\otimes\rmf}$. Further, we have $A_{\rma\otimes\rms} =  \mathbb{P}_{P} A  \mathbb{P}_{P}$ and $A_{\rms\otimes\rma} =  \mathbb{P}_{N-P} A  \mathbb{P}_{N-P}$.

In order to compute the strict gauge-invariant spectrum of the $\SU(N)$ gauge theory without using the misleading notion of spontaneous gauge symmetry breaking, we use the same operators as for the symmetric tensor in the $\SO(N)$ case. As the $\Ztwo$-odd scalar operator $\tr\phi^{3}$ will acquire a nonvanishing VEV, we conclude that the global $\Ztwo$ symmetry will be either spontaneously or, depending on $\gamma$, explicitly broken. Thus, the distinction in $\Ztwo$ odd and even operators in Tab.~\ref{tab:SUadj} is merely introduced to organize the operators and the analysis but does not lead to two separate ground states in the vector and the scalar channel. Note that such a breaking of the global symmetry was not considered in \cite{Maas:2017xzh}.

Finally, we would like to emphasize that the case $N=2$ is special. For $\SU(2)$, the scalar $\Ztwo$-odd operators listed in Tab.~\ref{tab:SUadj} vanish and all nonvanishing operators that can be constructed contain only scattering states. As this theory has further only one invariant Casimir operator, the cubic term vanishes such that the global $\Ztwo$ symmetry is always manifest in this particular model. Furthermore, the $\Ztwo$ even vector operator $\partial_{\nu} \tr (F^{\mu\nu}\phi^{2})$ vanishes and we obtain only one vector operator that expands to the massless $\U(1)$ vector field $A_{\U(1)}$, implying that the model contains a massless vector in the $\SU(2)$-invariant spectrum. From the conventional perspective of spontaneous gauge symmetry breaking, one would also expect that the other two elementary vector fields charged with respect to the unbroken Abelian subgroup are part of the physical spectrum. However, we find only $\U(1)$-invariant objects as dictated by the FMS framework. Thus these states belong to the third class which is not a surprise because the results of the $\SU(2)$ adjoint model can be mapped to the $\SO(3)$ fundamental theory. Furthermore, it is proposed that the $\mathcal{G}$-$\mathcal{H}$ duality in this model can be used to investigate the confinement of Yang-Mills theories \cite{Shibata:2019ygp,Kondo:2018qus,Kondo:2016ywd}.

\bibliography{bibliography}

\end{document}